\documentclass[12pt]{article}

\usepackage{graphicx,times,amsmath,natbib}
\usepackage{amsfonts,amssymb,latexsym,textcomp}
\usepackage{epsfig}
\usepackage{color}
\usepackage{subfigure}
\usepackage{url,theorem}
\usepackage{rotating,multirow}

\newcommand{\bfalpha} {\boldsymbol{\alpha}}
\newcommand{\bfbeta} {\boldsymbol{\beta}}
\newcommand{\bfgamma} {\boldsymbol{\gamma}}

\newcommand{\bfxi} {\boldsymbol{\xi}}

\newcommand{\bfeta} {\boldsymbol{\eta}}
\newcommand{\bfzeta} {\boldsymbol{\zeta}}

\newcommand{\bflambda} {\boldsymbol{\lambda}}

\newcommand{\bfpi} {\boldsymbol{\pi}}
\newcommand{\bfphi} {\boldsymbol{\phi}}

\newcommand{\bfnu} {\boldsymbol{\nu}}

\newcommand{\bfTheta} {\boldsymbol{\Theta}}

\newcommand{\bfvarsigma} {\boldsymbol{\varsigma}}
\newcommand{\bfvarpi} {\boldsymbol{\varpi}}

\newcommand{\dd}{\mbox{d}}

\newcommand{\bfw} {\mathbf{w}}

\newcommand{\bfy} {\mathbf{y}}

\newcommand{\bfY} {\mathbf{Y}}

\renewcommand{\Pr}{\mathsf{Pr}}
\newcommand{\E}{\mathsf{E}}
\newcommand{\V}{\mathsf{Var}}
\newcommand{\Cov}{\mathsf{Cov}}
\newcommand{\Cor}{\mathsf{Cor}}

\newcommand{\Gam}{\mathsf{Gam}}

\newcommand{\Poi}{\mathsf{Poi}}
\newcommand{\bet}{\mathsf{Beta}}
\newcommand{\Uni}{\mathsf{Uni}}

\newcommand{\Ber}{\mathsf{Ber}}

\newtheorem{lemma}{Lemma}
\theoremstyle{definition}

\title{Stochastic blockmodels for exchangeable collections of networks}
\author{Perla E. Reyes and Abel Rodr\'{\i}guez \footnote{Perla E. Reyes is Postdoctoral Associate, Department of Applied Mathematics and Statistics, University of California, Mailstop SOE2, Santa Cruz, CA 95064, {\tt perla@soe.ucsc.edu}.  Abel Rodriguez is Associate Professor, Department of Applied Mathematics and Statistics, University of California, Mailstop SOE2, Santa Cruz, CA 95064, {\tt abel@soe.ucsc.edu}.  This material is based upon work supported by DARPA Award No. N66001-10-1-4016. Any opinions, findings, and conclusions or recommendations expressed in this publication are those of the authors and do not necessarily reflect the views of DARPA.}
\\
\\
Department of Applied Mathematics and Statistitics\\
\\
University of California, Santa Cruz}

\date{}
\begin{document}
%\thanks{}
%\keywords{Dependent Dirichlet process; Nonparametric Bayes; Random probability measure; Travel Costs; Insurance Claim Distributions}
\maketitle
\begin{abstract}
We construct a novel class of stochastic blockmodels using Bayesian nonparametric mixtures.  These model allows us to jointly estimate the structure of multiple networks and explicitly compare the community structures underlying them, while allowing us to capture realistic properties of the underlying networks.  Inference is carried out using MCMC algorithms that incorporates sequentially allocated split-merge steps to improve mixing.  The models are illustrated using a simulation study and a variety of real-life examples.
%Multiple networks, defined as networks where a set of actors are connected by more than one type of relationships, arise in wide variety of areas of application, such as psychology, sociology, epidemiology and finance. We construct a novel class of hierarchical blockmodels for collections of networks using Bayesian nonparametric mixtures.  The model allows us to jointly estimate the structure of multiple networks and 
% structures whose properties are investigated, targeting their sensitivity to prior specification. On the other, the hierarchical specification has the feature of not only detecting community structures for each network, but also compare them across networks. An efficient MCMC algorithm is developed for posterior estimation, combining sequentially allocation and regular Gibbs steps to enhance the exploration of the posterior space and reduce the computational load.
\end{abstract}

\section{Introduction}

%A network consists of a set of nodes (which we call actors) along with measurements of the interactions between them.  
Network data consists of measurements associated with the interactions among a set of nodes (which we call actors), and is often visualized in the form of a (weighted) graph.  Network data has become quite ubiquitous in fields as diverse as sociology, bioinformatics, finance and physics.  In fact, it is often the case that multiple relationships are observed among a given set of actors, in which case it is of interest to model them jointly.  Indeed, since the actors are the same, we might expect similarities across the networks, which means that performing independent analyses on each network is potentially wasteful.  On the other hand, since different networks might reflect slightly different modes of interaction, just collapsing all observations into a single ``concensus'' network, or assuming that all networks arise from the same underlying stochastic process, might yield misleading results.

Popular statistical approaches for modeling networks include the class of exponentially weighted random graphs of \cite{FrSt86}, the class of $p_1$ models introduced in \cite{HoLe75}, and the latent social space models of \cite{HoRaHa02} and \cite{HaRaTa07}.  This paper is concerned with building hierarchical Bayesian models for an exchangeable collection of networks, with a particular emphasis on procedures that allow for network comparison. The methods we discuss build on the popular literature on stochastic blockmodels \citep{Wi76,WaWo87}.  Stochastic blockmodels  extend the notion of model-based clustering to network data.  More specifically, stochastic blockmodels aim at discovering an ``optimal partition'' of the actors in the network into homogenous groups (the factions or communities), which are made of actors that are (approximately) structurally equivalent \citep{LoWh71,WaFa94}.  These communities represent ``social positions'' or ``social roles'' ; members of the same faction are ``substitutable'', in the sense that they are subject to similar opportunities and constraints.  Hence, stochastic blockmodels are appealing because of their interpretability: factions are meaningful social constructs that are often driven by unobservable (or unobserved) variables.  Of particular interest to us are the subclass of infinite-dimensional blockmodels introduced by \cite{KeTeGrYaUe06} and \cite{XuTrYuKr06}, which treat the number of communities as an unknown parameter and place a probability distribution over all possible partitions of the set of actors. 

As a motivating example, consider the study reported by \cite{Kr87}, who collected cognitive social structure data from 21 management personnel in a high-tech, machine manufacturing firm to asses the effects of a recent management intervention program.  Each person indicated not only who he or she believes is their friend, but also his or her perception of others friendships; the result is a set of twenty one networks, each one of size $21 \times 21$.  In this example it is natural to ask how similar the perception of the network is among the subjects involved; the answer to this question provides important insights about the social structures within network (see \citealp{Kr87} for further discussion).  In addition, once we have identified groups of individuals with similar perceptions, it would be natural to try to aggregate information within each  of these groups to improve estimation of the underlying parameters determining the network structure. Another motivating example comes from the study on the interactions among 14 employees in a Western Electric plant reported by \citet{RoDi39}.  The researchers recorded six types of interactions among the employees: friendship, participation in horseplay, helping others with work, antagonistic behavior, arguments about open windows and number of times workers traded job assignments. Again, it would be natural to compare the social structure associated with the different relationships; however, this task is particularly challenging because of the different data types involved in the analysis (which includes directed and undirected networks, as well as binary and count data).  

Recently, the literature on blockmodels, and more generally, community identification algorithms, has been very active.  \cite{GiNe02}, \cite{Ne04}, \cite{Cl05} and \cite{MiScStTa08} are some recent examples within the physics and machine learning literature.  However, most of these approaches are algorithmic and it is unclear how uncertainty estimates can be obtained.  On the statistics side, two recent publications are particularly noteworthy.  On one hand, \cite{AiBlFiXi06} develop mixed-membership blockmodels, where subjects can be members of more than one community simultaneously.  On the other, \cite{BiCh09} show that likelihood-based community detection algorithms are asymptotically consistent, while algorithms based on certain modularity scores (such as the one described in \citealp{GiNe02}) are not.  In any case, all the approaches discussed above deal with single network problems rather than with the analysis of multiple networks, which is the focus of this manuscript.  

Indeed, although network comparison and aggregation are issues that seem to appear often in practice, we are not aware of work on model-based statistical methods to address them. Contingency tables have been used for modeling multi-relational data at least since \cite{GaMa78}.  However, this type of ``macroanalysis'' focuses solely on the relations and ignores individual actors.  \cite{FiMeWa85} extend the class of $p_1$ models to multi-relational data.  However, their approach relies on a partition of the actors into subgroups that is generated from external (extra-relational) information, which is assumed to be common to all relationships.  Moreover, these approaches are restricted binary networks and cannot be applied to problems with mixed data types.

The dearth of formal statistical procedures for the comparison of network structures might be explained in part by the lack of standard asymptotic results. To ascertain the similarity among networks, practitioners usually contrast summary statistics (such as indegree, outdegree, or betweenness distributions) across networks using simple tests.  However, this approach is often inappropriate, specially when networks correspond to different data types and there is no obvious way to construct a common summary that applies across all networks (as is the case, for example, of the Western Electric example mentioned before). In the case of aggregation, commonly employed methods assign a  ``consensus'' value to the interaction between two subjects across several networks (for an example, see \citealp{Kr87}). A typical aggregate would be a weighted average of all observed networks.  An obvious drawback of this approach is that choosing the weights is not straightforward, and assigning the same weight to all networks will often be inappropriate.  Furthermore, this type of approach cannot be applied if networks correspond to different data types.
 %Similarly, most procedures to generate ``consensus'' networks use simple thresholding rules that often weight all networks equally, no matter how %In this type of situations, Bayesian methods are particularly appealing since they allow for exact inference in the absence of asymptotic theory.

One important feature of the models developed in this paper is their ability to account for complex features of the networks such as assortative / disassortative mixing, and to incorporate prior information about the degree distribution.  Indeed, existing approaches to Bayesian inference in stochastic blockmodels pay little attention to prior elicitation and the effect of the prior on inference.  In addition to discussing models for multiple networks, this paper explores the {\it a priori} properties of networks generated by stochastic blockmodels and discusses hierarchical specifications that can be used to include prior information about the topology of the network.

The remaining of the paper is organized as follows: Section~2 describes a general framework for modeling a single network and reviews the use of nonparametric mixture priors in the context of network models. Section~3 presents generalization of this class of models that allow for more flexible prior specification.  Section ~4 extends the single-network model to a multiple networks model that simultaneously identifies community structure per network and establishes similarities across networks.  This section also discusses the MCMC algorithm that we use for posterior inference. Key features of the multiple networks model are showcase in Section~5 using first a simulation study and then datasets of employee relationships. Finally, Section~6 presents our conclusions.

\section{Stochastic blockmodels}\label{sec:blockmodel}

For the purpose of this paper, a network is a $I \times I$ matrix $\bfY = [y_{i,j}]$, where $y_{i,j}$ measures the strength of the relationship between actor $i$ and actor $j$.  The network is called undirected if $\mathbf{Y}$ is symmetric (e.g., if it is irrelevant who initiates the interaction), and directed otherwise. Further, the network is called acyclic if subjects do not interact with themselves, in which case the diagonal elements of $\bfY$ are treated as structural zeros\footnote{We deviate slightly from the standard definition of acyclic networks, which typically applies only to directed networks and precludes any sort closed loops.}.  

%This paper will focus most of its attention on binary and acyclic networks, which are arguably the most common type of network encountered in practice.  In a binary network, $y_{ij} \in \{ 0, 1\}$, so that $y_{ij} = 1$ if actor $i$ interacts with subject $j$, and $y_{ij} = 0$ otherwise.  Hence, binary networks only capture the presence or absence of a relation between actors.  However, in spite of our focus, most of our discussion applies to more general data types, including situations where different types of networks are present (see our illustration in Section \ref{se:wiring}).

Stochastic blockmodels \citep{Wi76,WaWo87} extend the notion of model-based clustering to networks.  In the case of acyclic, directed networks, Bayesian stochastic blockmodels are hierarchical models that for $i,j = 1,\ldots,I$ and $j \ne i$.
\begin{align}\label{eq:blockmodel}
y_{i,j} | \xi_i, \xi_j, \bfTheta &\sim  \psi(y_{ij} | \theta_{\xi_i, \xi_j}, \bfnu),  & \xi_i | \bfw & \sim \sum_{k=1}^{N} w_k \delta_k, & (\bfw, \bfTheta, \bfnu | \bfvarsigma, \bflambda) &\sim p(\bfw| \bfvarsigma) p(\bfTheta| \bflambda) p(\bfnu),
\end{align}
where $N$ is the maximum number of potential factions, $\bfvarsigma$ and $\bflambda$ are vectors of hyperparameters (which will be assigned hyperpriors $p(\bfvarsigma)$ and $p(\bflambda)$), $\bfTheta = [\theta_{k,l}]$ is a $N \times N$ matrix, $\bfw = (w_1,\ldots,w_N)'$ is such that $\sum_{k=1}^{N} w_k =1$, $\delta_k$ denotes the degenerate probability distribution placing probability 1 on $k$, and $\psi(y | \theta, \bfnu)$ is parametric kernel indexed by the parameters $\theta$ and $\bfnu$, where $\theta$ is a random effect, and $\bfnu$ is a vector of fixed effects.  In the case of undirected networks, a similar definition applies with the added constrains $y_{i,j} = y_{j,i}$ and $\theta_{k,l}=\theta_{l,k}$.  %For example, a simple version of the model is obtained by assuming $N$ known and letting $\bfw \sim \Dir(1,1,\ldots,1)$ and $\theta_{k,l} \sim_{iid} H$ for some distribution distribution $H$.  

The formulation in \eqref{eq:blockmodel} is extremely flexible and easily interpretable.  The latent variables $\bfxi = (\xi_1, \ldots, \xi_I)'$ act as (unobserved) faction indicators; the prior probability that any two subjects are assigned to the same cluster (i.e., the share the same social role) is given by $\E\left\{ \sum_{k=1}^{N} w_k^2 \right\}$.  Binary networks (i.e., those where $y_{i,j} \in \{ 0, 1\}$, so that $y_{i,j} = 1$ if actor $i$ interacts with actor $j$, and $y_{i,j} = 0$ otherwise) can be accommodated by taking $y_{i,j} |\xi_i, \xi_j, \bfTheta \sim \Ber(\theta_{\xi_i, \xi_j})$ and selecting (for computational convenience) $\theta_{k,l} \sim \bet(a, b)$.  In this case, the entries $\theta_{k,l}$ give the probability that an interaction occurs between actor from factions $k$ and $l$.  On the other hand, count data could be incorporated by taking $y_{i,j} | \xi_i, \xi_j, \bfTheta \sim \Poi(\theta_{\xi_i, \xi_j})$ and $\theta_{k,l} \sim \Gam(a, b)$. In this case, $\theta_{k,l}$ is the intensity of the interaction between factions $k$ and $l$.

\begin{figure}[h]
\begin{center}
\includegraphics[width=6.5in,angle=0]{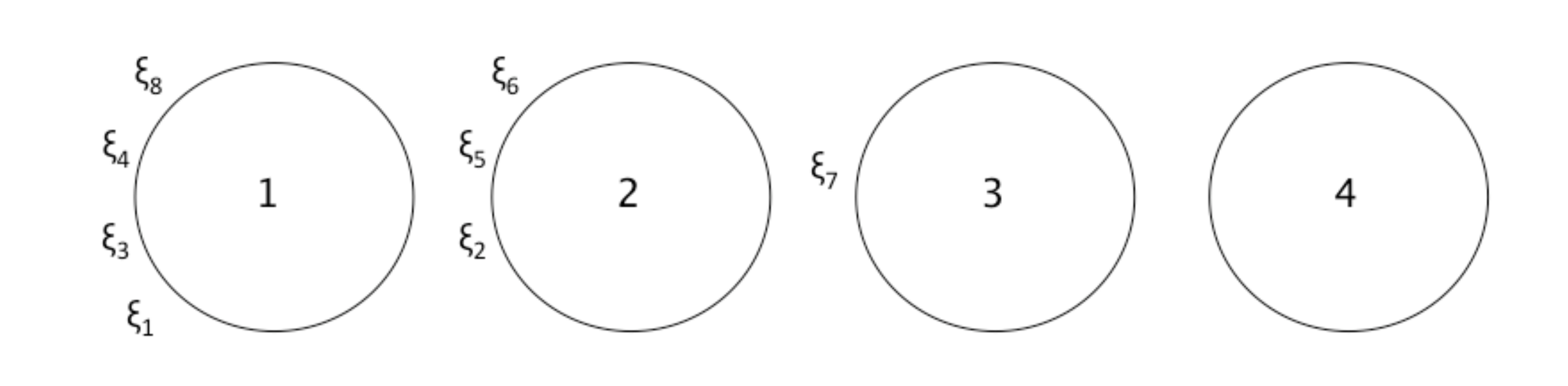} 
\caption{Schematic representation for the Chinesse restaurant process.  In this analogy, observations correspond to customers who plan to dine in a restaurant with an infinite number of tables.  The variable $\xi_i$ indicates which table is occupied by customer $i$.  Customer 1 sits at table 1; customer $i$ sits at any of the occupied tables with probability proportional to the number of customers already sitting in it, while she sits on the next empty table with probability proportional to $\beta$. In this particular example, the next customer (number 9) would sit on table 1 with probability $4/(\beta+8)$, on table 2 with probability $3/(\beta+8)$, on table 3 with probability $1/(\beta+8)$, or on table 4 with probability $\beta/(\beta+8)$.} \label{fi:crp1}
\end{center}
\end{figure}

One example of a stochastic blockmodel is the infinite relational blockmodel (IRM) \citep{KeTeGrYaUe06,XuTrYuKr06}.  In the IRM, $N=\infty$ and %, the $\theta_{k,l}$s form an independent and identically distributed sample from a common distribution $H^{\bflambda}$, and 
\begin{align}\label{eq:DPstickbreaking}
\theta_{k,l} & \sim_{iid} H^{\bflambda}, &  w_k &= v_k \prod_{s < k} (1 - v_s), & v_k &\sim_{iid} \bet(1, \beta), & k,l&=1,2,\ldots,
\end{align}
%$w_k = v_k \prod_{k < l} (1 - v_k)$ with $v_1,v_2,\ldots$ being another independent and identically distributed sequence with $v_l \sim \bet(1, \beta)$. 
where $H^{\bflambda}$ is a parametric distribution indexed by the hyperparmeter $\bflambda$. This structure for the weights, which is strongly connected to the stick-breaking construction of the Dirichlet process \citep{Se94}, implies that the joint distribution of $\bfxi = (\xi_1, \ldots, \xi_I)$ obtained {\it after} integrating  out $\bfw$ can be described by a sequence of predictive distributions with $\xi_1 = 1$ and
\begin{align}\label{eq:polyaurn1}
\xi_{i} | \xi_{i-1}, \ldots, \xi_1 &\sim \sum_{k=1}^{K^{i-1}} \frac{m_k^{i-1}}{\beta+ i - 1}\delta_{k} +  \frac{\beta}{\beta+ i - 1}\delta_{K^{i-1} + 1},   & 2 \le i &\le I,
\end{align}
where $\delta_a$ denotes the degenerate probability distribution on $a$, $K^{i-1} = \max_{j < i} \{ \xi_j \}$ is the number of unique values among $\xi_1,\ldots,\xi_{i-1}$, $m_k^{i-1} = \sum_{j=1}^{i-1} \mathbf{1}_{(\xi_j = k)}$ is the number of indicators equal to $k$ among $\xi_1,\ldots,\xi_{i-1}$, and $\beta > 0$ is a constant.  This sequence of predictive distributions is sometimes called the Chinese restaurant process (CRP) (see Figure \ref{fi:crp1}), and implies that $\Pr(\xi_i = \xi_j) = \sum_{k=1}^{\infty} \E\left\{w_k^2 \right\} = 1/(1 + \beta)$ for all $i$ and $j$.

The CRP places a probability distribution on all possible partitions of $I$ actors, whose shape is controlled by the parameter $\beta$.  For example, the probability that all actors are assigned to a single faction is $\prod_{i=2}^{I} \frac{i-1}{\beta + i - 1}$, in which case the model reduces to an Erd{\"o}s-R{\'e}nyi random graph \citep{ErRe59} with an unknown link frequency.  On the other hand, the probability that each actor is assigned to a different faction is given by $\prod_{i=2}^{I} \frac{\beta}{\beta + i -1}$.  Hence, although the number of potential factions $N$ is infinite, the effective number of factions $K$ actually occupied by actors is treated as a random variable taking values with support on the set $\{ 1, \ldots, I \}$, and is automatically estimated from the data.  This prior distribution for $K$ implied by the CRP is such that $\E(K) \sim \beta \log \{ \frac{\beta + I}{\beta} \}$ for large $I$.

\section{Marginal model properties and prior elicitation}\label{se:modelproperties}

The IRM described in the previous Section has two unappealing properties.  Firstly, note that for a binary network under \eqref{eq:blockmodel} and \eqref{eq:DPstickbreaking}, $\Pr(y_{i,j} = 1 | \xi_i = \xi_j) = \Pr(y_{i,j} = 1 | \xi_i \ne \xi_j)$  a priori, i.e., membership in the same faction does not provide any information about how likely a link is.  This flies in the face of well known empirical facts (for example, most social networks present assortative mixing, where members of the same faction have a higher probability of interacting).  Secondly, note that $\Pr(y_{i,j} | \bflambda) = \E_{H^{\bflambda}}(\theta_{k,l}) = \bar{\theta}$ for every $\beta$.  Moreover, the expected number of links per actor is simply given by $(I-1) \E_{H^{\bflambda}}(\theta_{k,l})$.  Hence, prior information about the community structure provides no information about the expected degree in the network.

To alleviate these issues we consider a more general prior specification such that
\begin{align}\label{eq:hierarchicalprior2}
p(\bfTheta | \bflambda = (\bflambda_O, \bflambda_D)) &= \begin{cases}
\left\{ \prod_{k=1}^{\infty} H^{\bflambda_D}(\theta_{k,k}) \right\} \left\{ \prod_{k=1}^{\infty} \prod_{l=k+1}^{\infty} H^{\bflambda_O}(\theta_{k,l}) \delta_{\theta_{k,l}}(\theta_{l,k}) \right\} & \mbox{$\bfY$ undirected }  \\
\left\{ \prod_{k=1}^{\infty} H^{\bflambda_D}(\theta_{k,k}) \right\} \left\{ \prod_{k=1}^{\infty} \prod_{l=k+1}^{\infty} H^{\bflambda_O}(\theta_{k,l}) H^{\bflambda_{O}}(\theta_{l,k}) \right\}  & \mbox{$\bfY$ directed }  \\
\end{cases},
\end{align}
where $H^{\bflambda}$ is a distribution function indexed by a parameter $\bflambda$, and $p(\bfw)$ is defined by 
\begin{align}\label{eq:weightprior}
w_k &= u_{k} \prod_{s < k} (1 - u_{s}),  & u_{k} &\sim_{iid} \bet (1-\alpha, \beta + \alpha k), & k=&1,2,\ldots.
\end{align}

This Section explores the properties of this new specification, with a particular emphasis on 1) studying the {\it a-priori} properties of the network, 2) determining procedures for prior specification, and 3) generating model-based alternatives to traditional summaries of network topology.  Most of our discussion is focused on binary networks because they are arguably the most common type of networks found in practice.  However, most of the comments can be easily extended to more general classes of networks. %Hence, the discussion on this Section extends naturally to blockmodels for single networks.

\subsection{Exchangeability of actors and communities}

An infinite two-dimensional array $\bfY$ is said to be jointly exchangeable if, for any finite submatrix, the distribution is unchanged when the same permutation is applied to both the rows and the columns of the submatrix.  In the absence of node-specific covariate information, this is often a natural assumption for networks that simply implies that the order in which the actors are observed should have to influence in our probability model.

The blockmodel defined in Equation \eqref{eq:blockmodel} defines a jointly infinitely exchangeable prior on inifnite arrays; this can be most easily seen by exploiting the representation theorem from \cite{Al81}: an array is jointly exchangeable if we can write $y_{i,j} = g( \bfphi, v_i, v_j, u_{i,j})$ where $\bfphi$, $\{ v_i \}$ and $u_{i,j}$ are independent random variables such that $v_i \sim \Uni[0, 1]$ and $u_{i,j} \sim \Uni[0, 1]$.  In the case of blockmodels, we can take $\bfphi = ( \bfw, \bfTheta )$ and write
$$
y_{i,j} = g(\bfphi, v_i, v_j, u_{i,j}) = \begin{cases}
1 & u_{i,j}  \le \theta_{G^{-1}(v_i), G^{-1}(v_j)} \\
0 & u_{i,j}  > \theta_{G^{-1}(v_i), G^{-1}(v_j)}
\end{cases},
$$
where $G(\cdot) = \sum_{k} w_k \delta_k$ and $G^{-1}(\cdot)$ is the generalized inverse of $G$.  Hence, the representation theorem ensures that the actors are exchangeable, no matter what the priors on $\bfw$ and $\bfTheta$ are.  

Although ensuring exchangeability in the matrix $\bfY$ does not impose constrains on our choice of priors for $\bfTheta$ and $\bfw$,  in practice we are often interested in models where not only the individuals but also the unknown communities are exchangeable.  In order to ensure that a blockmodel induces exchangeability among factions, we need the prior on $\bfTheta$ to also be jointly exchangeable.  The following lemma ensures that our generalized model satisfies this constraint.
\begin{lemma}\label{le:exchangeability}
Let $p(\bfTheta | \bflambda)$ be defined as in \eqref{eq:hierarchicalprior2}, and let $\bflambda \sim p(\bflambda)$.  Then, $p(\bfTheta) = \int p(\bfTheta | \bflambda) p(\bflambda) \dd \bflambda$ defines a jointly exchangeable prior on $\bfTheta$.  \hfill $\diamond$
\end{lemma}
The proof can be found in Appendix~\ref{ap:exchangeability}.   One implication of Equation \eqref{eq:hierarchicalprior} is that the a priori marginal probability of a link is given by a weighted average of the probability of a link under $H^{\bflambda_D}$ and $H^{\bflambda_O}$.
\begin{lemma}\label{le:problink}
Let $\bfY$ be an undirected network.  For the blockmodel in \eqref{eq:blockmodel} and the priors in  \eqref{eq:hierarchicalprior2} and \eqref{eq:weightprior}, the a priori marginal probability of a link between any two nodes $i$ and $j$ is given by
\begin{align*}
\bar{\theta} & =  \Pr(y_{ij}=1 | \bflambda, \bfvarsigma) = \left( \frac{1-\alpha}{\beta + 1} \right)  \E_{H^{\bflambda_D}}\left\{ \theta_{k,k} \right\}  + \left( \frac{\beta + \alpha}{\beta + 1} \right) \E_{H^{\bflambda_O}}\left\{ \theta_{k,l} \right\}.
\end{align*}
Under the same circumstances, but for a directed network $\bfY$, we have $\bar{\theta}^{in} = \bar{\theta}^{out} = \bar{\theta}$, where $\bar{\theta}^{in}= \Pr(y_{ji} =1 | \bflambda, \bfvarsigma)$ and $\bar{\theta}^{out} = \Pr(y_{ij}=1 | \bflambda, \bfvarsigma)$ represent the prior probabilities of a link from $j$ to $i$ and a link from $i$ to $j$ respectively.  \hfill $\diamond$
\end{lemma}
For a proof, see Appendix \ref{ap:problink}.  Note that $(1-\alpha)/(\beta + 1)$ represents the percentage of links of node $i$ with members of its same community, and 
$$
\min\{ \E_{H^{\bflambda_D}}\{ \theta_{k,k} \}, \E_{H^{\bflambda_O}}\{ \theta_{k,l} \} \} \le \bar{\theta} \le \max\{ \E_{H^{\bflambda_D}}\{ \theta_{k,k} \}, \E_{H^{\bflambda_O}}\{ \theta_{k,l} \} \}
$$
for any choice of prior on $\bfw$.  On the other hand, if $\bflambda_D = \bflambda_O$ we recover $\Pr(y_{ij} = 1 | \bflambda, \bfvarsigma) = \E_{H^{\bflambda}}(\theta_{k,l})$ as in traditional IRMs.

\subsection{Capturing assortative / disassortative mixing}\label{sec:assortative}

A network is said to be assortative (or have assortative mixing) if interactions among actors in the same faction tend to be more common than interactions among actors in different factions.  Similarly, a network is said to be disassortative if the opposite is true.  It is often the case that we have a-priori knowledge about the mixing pattern associated with a specific network; for example, social networks are often assortative, while ecological prey-predator networks are often disassortative.

One advantage of \eqref{eq:hierarchicalprior2} is that, by allowing different hyperparameters for the diagonal and off-diagonal elements of $\bfTheta$, it allows us to incorporate information about the type of mixing in the network.  For example, for an undirected binary network with $y_{ij} | \xi_i, \xi_j, \bfTheta \sim \Ber(\theta_{\xi_i,\xi_j})$ which is believed to have assortative mixing, we might assign $\theta_{k,k} \sim \bet(a_D, b_D)$ and $\theta_{k,l} \sim \bet(a_O, b_O)$, where $\bflambda_D = (a_D, b_D)$ and $\bflambda_O = (a_O, b_O)$ are given a joint prior $p(a_D, b_D, a_O, b_O)$ such that $\{ a_D/(a_D + b_D) >  a_O/(a_O + b_O) \}$ has high probability.

This discussion suggests that a simple way to summarize the effect of our prior choice on the topology of the network is through the assortativity index
\begin{align*}
\Upsilon = \log \left( \E_{H^{\bflambda_D}}\{ \theta_{k,k} | \bflambda_D \} \right) -  \log \left( \E_{H^{\bflambda_O}}\{ \theta_{k,l}| \bflambda_O \} \right).  %\quad\quad\quad \E\{  \log \E(\bar{\theta}_{D}) - \log \E(  \bar{\theta}_{O} ) | \bfY\}
\end{align*}
Note that $p(\bflambda_D) = p(\bflambda_O)$ implies $\E\{ \Upsilon \} = 0$ a priori, which means that a priori we have no information about whether the network is assortative or disassortative;  $\E\{ \Upsilon \} > 0$ is associated with assortative networks, while $\E \{ \Upsilon \} < 0$ is associated with disassortative networks.  A posteriori, the distribution of $\Upsilon$ provides a simple summary of the type of mixing in the network and a model-based alternative to the assortativity coefficients discussed in \cite{Ne03b}. %and 170, 186 and 

\subsection{Degree distribution}

The degree of a node refers to the number of links associated with it.  Hence, for undirected binary networks, the degree of actor $i$ is simply $D_i = \sum_{j \ne i} y_{ij}$.  For directed networks, we can analogously define the in-degree $D_i^{I}$ and the out-degree $D_i^{O}$ of actor $i$ as the number of links that start or end at actor $i$, $D_i^{I} = \sum_{j \ne i} y_{ji}$ and $D_i^{O}= \sum_{j \ne i} y_{ij}$. The distribution of $D_i$ is often used to compare how well analytical models fit the observed data.  
\begin{lemma}\label{le:expDi}
Consider an undirected network $\bfY$.  For the blockmodel in \eqref{eq:blockmodel} and the priors in  \eqref{eq:hierarchicalprior2} and \eqref{eq:weightprior}, the degree distribution satisfies
\begin{align*}
%\rho &= (I - 1) \left\{  \left( \frac{1 - \alpha}{\beta + 1} \right)  \bar{\theta}_{D} + \left( \frac{\beta + \alpha}{\beta + 1}\right) \bar{\theta}_{O}  \right\} \\
\E \{ D_i | \bflambda, \bfvarsigma \} &= \bar{\rho} = [ I - 1 ] \left[  \left( \frac{1 - \alpha}{\beta+1} \right) \E_{H^{\bflambda_D}} \{ \theta_{k,k} \} + \left( \frac{\beta + \alpha}{\beta+1} \right) \E_{H^{\bflambda_O}} \{ \theta_{k,l} \}  \right ] \\
\V( D_i | \bflambda, \bfvarsigma ) &=  \bar{\kappa} =  \bar{\rho}(1-\bar{\rho}) + \frac{[I-1][I-2]}{2(\beta + 1)(\beta + 2)}  \bigg[ (1-\alpha)(2-\alpha) \E_{H^{\bflambda_D}} \{ \theta_{k,k}^2 \} \bigg. \\
& \qquad \qquad \qquad \quad+  (\beta + \alpha)(\beta + 2\alpha)  \left( \E_{H^{\bflambda_O}} \{ \theta_{k,l} \} \right)^2  \\
& \qquad \qquad \qquad \qquad + (1 - \alpha)(\beta + \alpha) \left[ 2 \E_{H^{\bflambda_D}} \{ \theta_{k,k} \} \E_{H^{\bflambda_O}} \{ \theta_{k,l} \}   +   \E_{H^{\bflambda_O}} \{ \theta_{k,l}^2 \}  \right] \bigg. \bigg], 
\end{align*}
where $I$ is the number of actors in the network.  If $\bfY$ is a directed network, a similar result holds separately for the in-degree and the out-degree, and we also have
$$
\Cor(D_i^{in}, D_i^{out} | \bflambda,\bfvarsigma) = \Cor(y_{ij}, y_{ji} | \bflambda, \bfvarsigma) = \bar{\Delta} = \frac{\bar{\Delta}^{N}}{\bar{\Delta}^{D}},
$$
where
\begin{multline*}
\bar{\Delta}^{N} =  \left( \frac{1 - \alpha}{\beta + 1} \right) \E_{H^{\bflambda_D}}\{ \theta^2_{k,k} \} +  \left( \frac{\beta + \alpha}{\beta + 1} \right) \left( \E_{H^{\bflambda_O}}\{ \theta_{k,l} \} \right)^2 \\
 -  \left[ \left( \frac{1 - \alpha}{\beta + 1} \right) \E_{H^{\bflambda_D}}\{ \theta_{k,k} \} +  \left( \frac{\beta + \alpha}{\beta + 1} \right) \E_{H^{\bflambda_O}}\{\theta_{k,l}\}  \right]^2
\end{multline*}
and
\begin{multline*}
\bar{\Delta}^{D} =  \left( \frac{1 - \alpha}{\beta + 1} \right) \E_{H^{\bflambda_D}}\{ \theta_{k,k} \} + \left( \frac{\beta + \alpha}{\beta + 1} \right) \E_{H^{\bflambda_O}}\{ \theta_{k,l} \} \\
 - \left[  \left( \frac{1 - \alpha}{\beta + 1} \right) \E_{H^{\bflambda_D}}\{ \theta_{k,k} \} + \left( \frac{\beta + \alpha}{\beta + 1} \right) \E_{H^{\bflambda_O}}\{\theta_{k,l}\}  \right]^2
 = \hat{\theta} \left\{ 1 -\hat{\theta} \right\}.
\end{multline*}
\hfill $\diamond$
\end{lemma}
The proof is in Appendix~\ref{ap:expDi}.  %Note that $\bar{\Delta}^{N}$ and $\bar{\Delta}^{D}$ differ exclusively on their second terms.  Hence, the correlation among the in-degree and the out-degree is controlled by $\E_{H^{\bflambda_O}}\{ \theta^2_{k,l} \} - \left( \E_{H^{\bflambda_O}}\{ \theta_{k,l} \} \right)^2 = \V_{H^{\bflambda_O}} (\theta_{k,l})$.  If $\V_{H^{\bflambda_O}} (\theta_{k,l}) \to 0$ then $\bar{\Delta} \to 1$; on the other hand, if $\V_{H^{\bflambda_O}} (\theta_{k,l}) \to \infty$ while $\E_{H^{\bflambda_O}} (\theta_{k,l})$ is kept constant implies that $\bar{\Delta} \to 0$.
%Obtaining more general results for the degree distribution for general blockmodels is extremely challenging.  For example, consider an undirected network where $H^{\bflambda}$ is chosen to be a beta distribution and $\lambda_D = (a_D, b_D)$ and $\lambda_O = (a_O, b_O)$.  In that case, $D_i$ follows a compound Beta-Binomial distribution, i.e.,
%\begin{align*}
%D_i = \sum_{l=1}^{L} X_l
%\end{align*}
%where $X_1,\ldots,X_n$ is an independent sequence of random variables such that $X_1 \sim \betbin (m_1, a_D, b_D)$ and, for $l \ge 2$, $X_l \sim \betbin (m_l, a_O, b_O)$, and the joint distribution of $(L,m_1,\ldots, m_L)$ is given by
%\begin{align}\label{eq:jointPDprocess}
%p(L, m_1, \ldots, m_L | I, \alpha, \beta) = \left\{ \prod_{l=1}^{L-1} (\beta + l \alpha) \right\} \frac{\Gamma(\beta + 1)}{ \Gamma(\beta + I) } \prod_{l=1}^{L} \frac{\Gamma(m_l - \alpha)}{\Gamma(1 - \alpha)}
%\end{align}
Moreover, the moment generating function for $D_i$ can be written as (see Appendix \ref{ap:momgenfuncDi})
\begin{align}\label{eq:jointPDprocess}
\E \left\{ 
\left[ 1 + \sum_{s=1}^{m_1 - 1} {m_1 -1 \choose s} \E_{H^{\bflambda_{D}}} (\theta_{1,1}^s) (e^t -1)^s \right]
\prod_{k=2}^{K} \left[ 1 + \sum_{s=1}^{m_k} {m_k \choose s} \E_{H^{\bflambda_{O}}} (\theta_{1,k}^s) (e^t -1)^s \right]
 \right\},
\end{align}
where the last expectation is taken with respect to the joint distribution of the number of factions and faction sizes, which is given by \citep{Pi95}
%\begin{align*}
%p(K, m_1, \ldots, m_K) &=  \frac{\Gamma(\beta+1)}{\Gamma(\beta+I)} \prod_{k=1}^{K} (\beta + k \alpha) \frac{\Gamma(m_k - \alpha)}{\Gamma(1 - \alpha)} & 1 &\le K \le I  & 1 \le &m_k \le I  & \sum_{k=1}^{K} m_k &= I
%\end{align*}
\begin{align*}%\label{eq:jointPDprocess}
p(K, m_1, \ldots, m_K ) &= \frac{\Gamma(\beta + 1)}{(\beta + K \alpha) \Gamma(\beta + I) } \prod_{k=1}^{K} (\beta + k \alpha) \frac{\Gamma(m_k - \alpha)}{\Gamma(1 - \alpha)}
\end{align*}
where  $1 \le K \le I$,  $1 \le m_k \le I$ and  $ \sum_{k=1}^{K} m_k = I$.  Although obtaining closed-from results for the full degree distribution is challenging, our results on the mean and variance of the degree distributions, together with information about the assortative/disassortive structure in the network and/or the number of communities, can be used to elicit meaningful informative priors for network models.

To explore the effect of diverse values of $\alpha$, $\beta$, and assumptions on $H_D$ and $H_O$ in the degree distribution of the resulting network we conducted a simulation study. For the stick-breaking parameters, we set $\alpha$ in $\{ 0, 0.2, 0.5, 0.8\}$ and $\beta$ in $\{ 0.5, 1.5, 5.0\}$. $H_D$ and $H_O$ are selected such that $(\bar{\theta}_{D}, \bar{\theta}_{O}) = (\E_{H^{\bflambda_D}} \{ \theta_{kk} \}, \E_{H^{\bflambda_O}} \{ \theta_{kl} \})$ are in $\{ (0.2, 0.2), (0.2, 0.8), (0.8, 0.2), (0.8,0.8) \}$ with 3 levels of variability: 1) point-masses at $\bar{\theta}_{D} $ and $  \bar{\theta}_{O}$ (no variability), 2) beta distributions with parameters $a_D+ b_D = a_O+ b_O = 0.5$ (high variability), and 3) beta distributions with parameters $a_D+ b_D = a_O+ b_O = 5$ (low variability). For each of the different combinations of parameters 10,000 networks with $I=100$ actors each were generated. 

\begin{figure}[h]
\begin{center}
{\small
\begin{tabular}{rcccc}
&$\bar{\theta}_{D} =  \bar{\theta}_{O} = 0.2$   &   $\bar{\theta}_{D} = 0.2$, $\bar{\theta}_{O} = 0.8$   &   $\bar{\theta}_{D} = 0.8$, $\bar{\theta}_{O} = 0.2$   &   $\bar{\theta}_{D} =  \bar{\theta}_{O} = 0.8$\\
%%%
\begin{sideways} $\quad \quad \quad \;\; \alpha=0$  \end{sideways}&
\subfigure{\includegraphics[width=1.4in,angle=0]{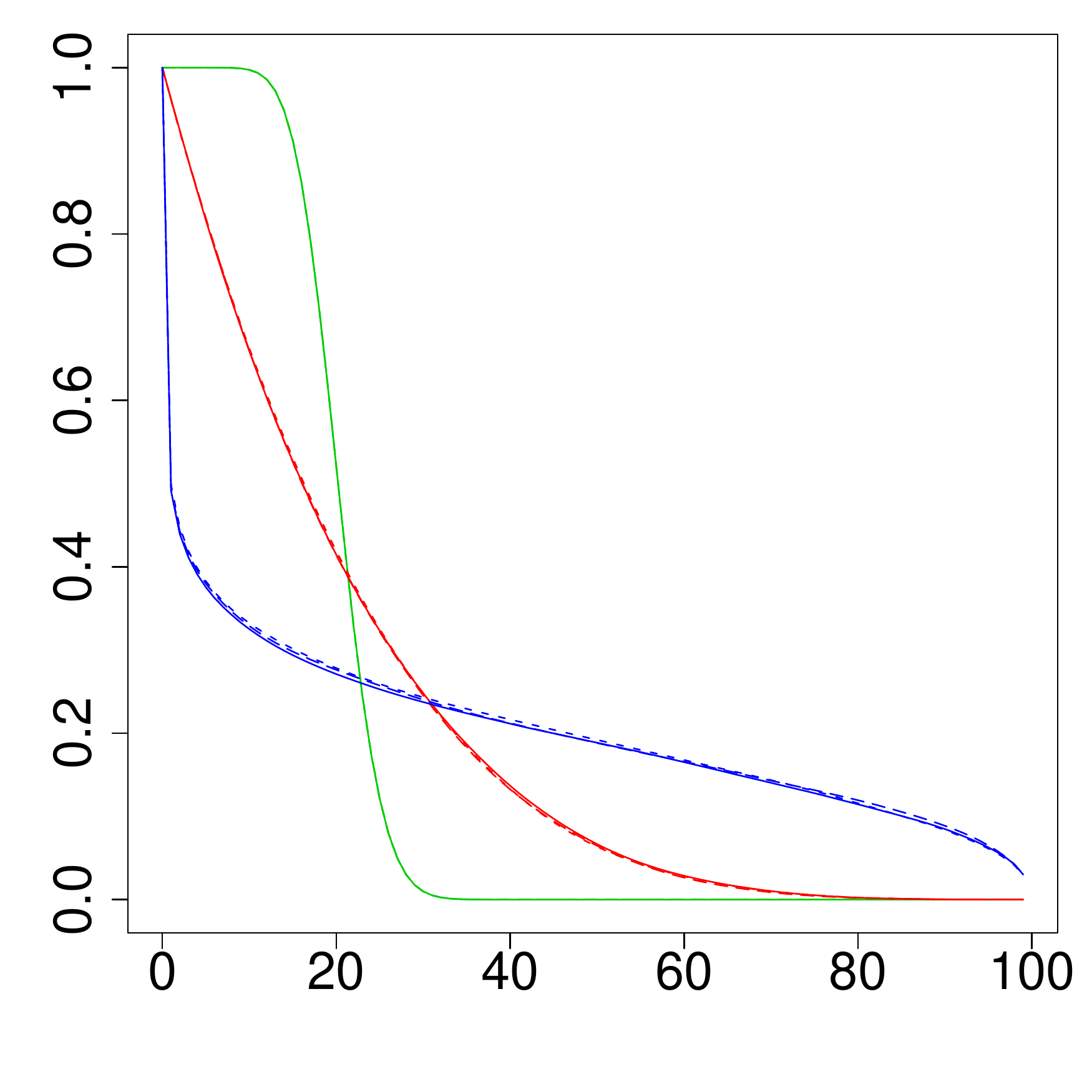}}  &
\subfigure{\includegraphics[width=1.4in,angle=0]{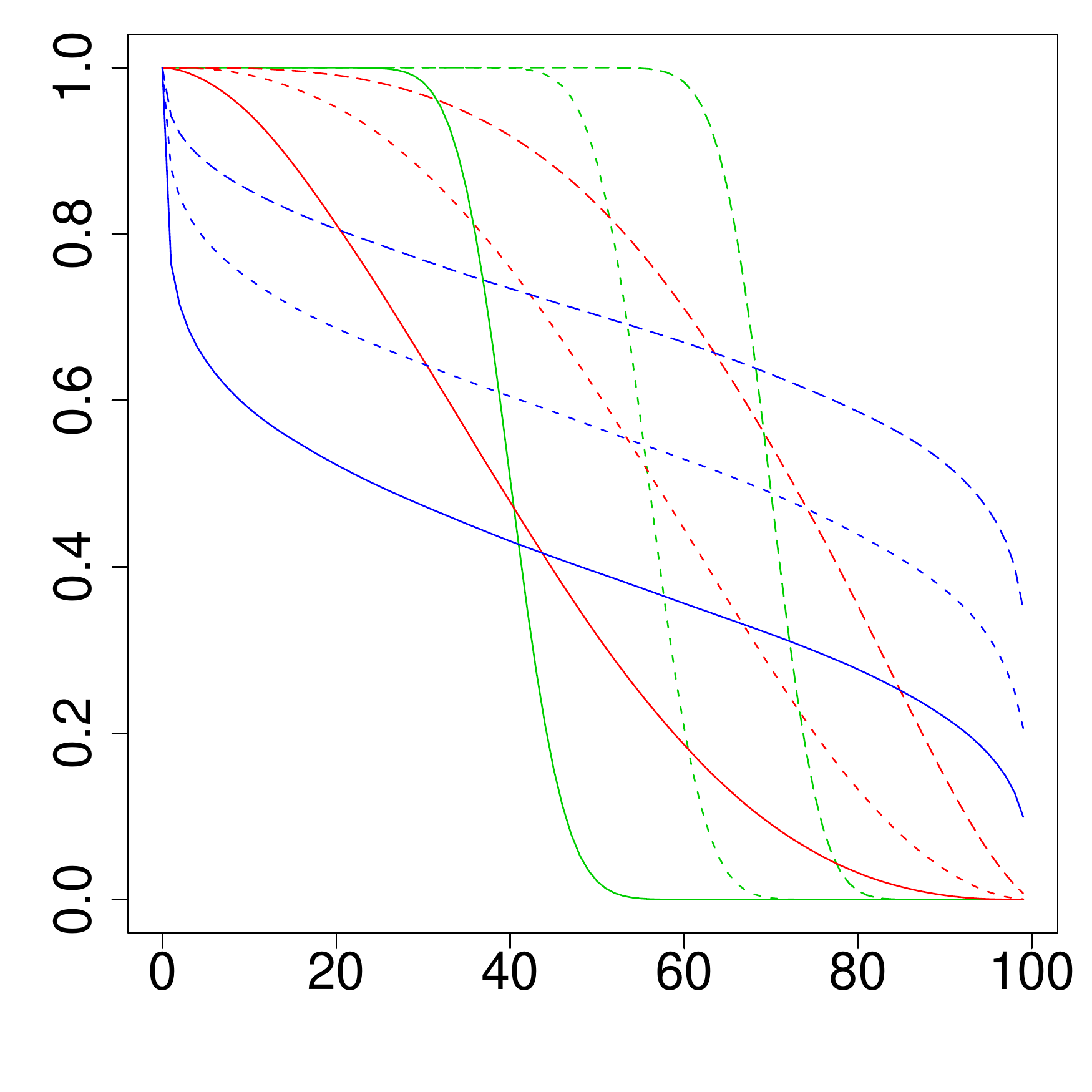}}  &
\subfigure{\includegraphics[width=1.4in,angle=0]{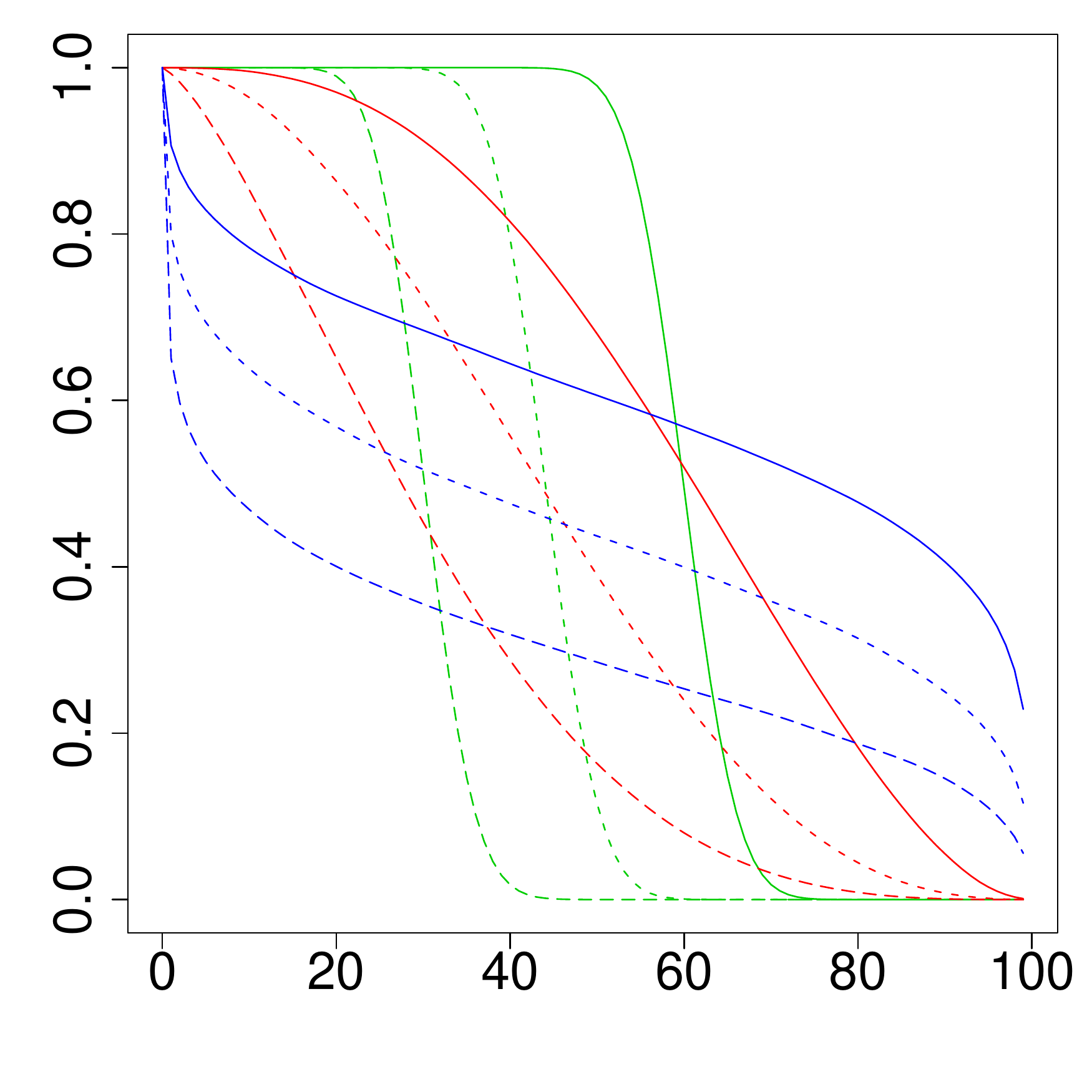}}  &
\subfigure{\includegraphics[width=1.4in,angle=0]{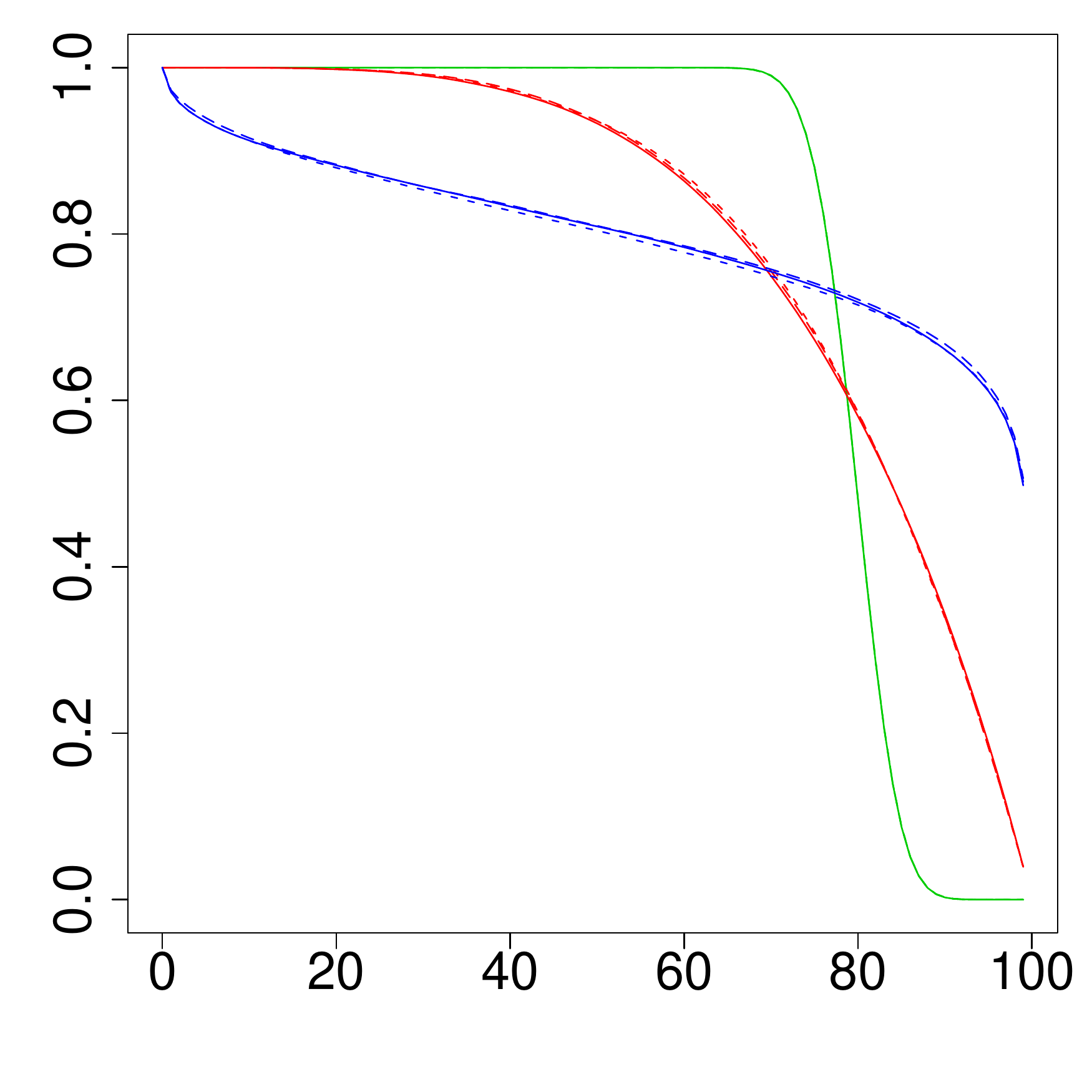}} \\
%%%
\begin{sideways} $\quad \quad \quad \;\; \alpha=0.2$  \end{sideways}&
\subfigure{\includegraphics[width=1.4in,angle=0]{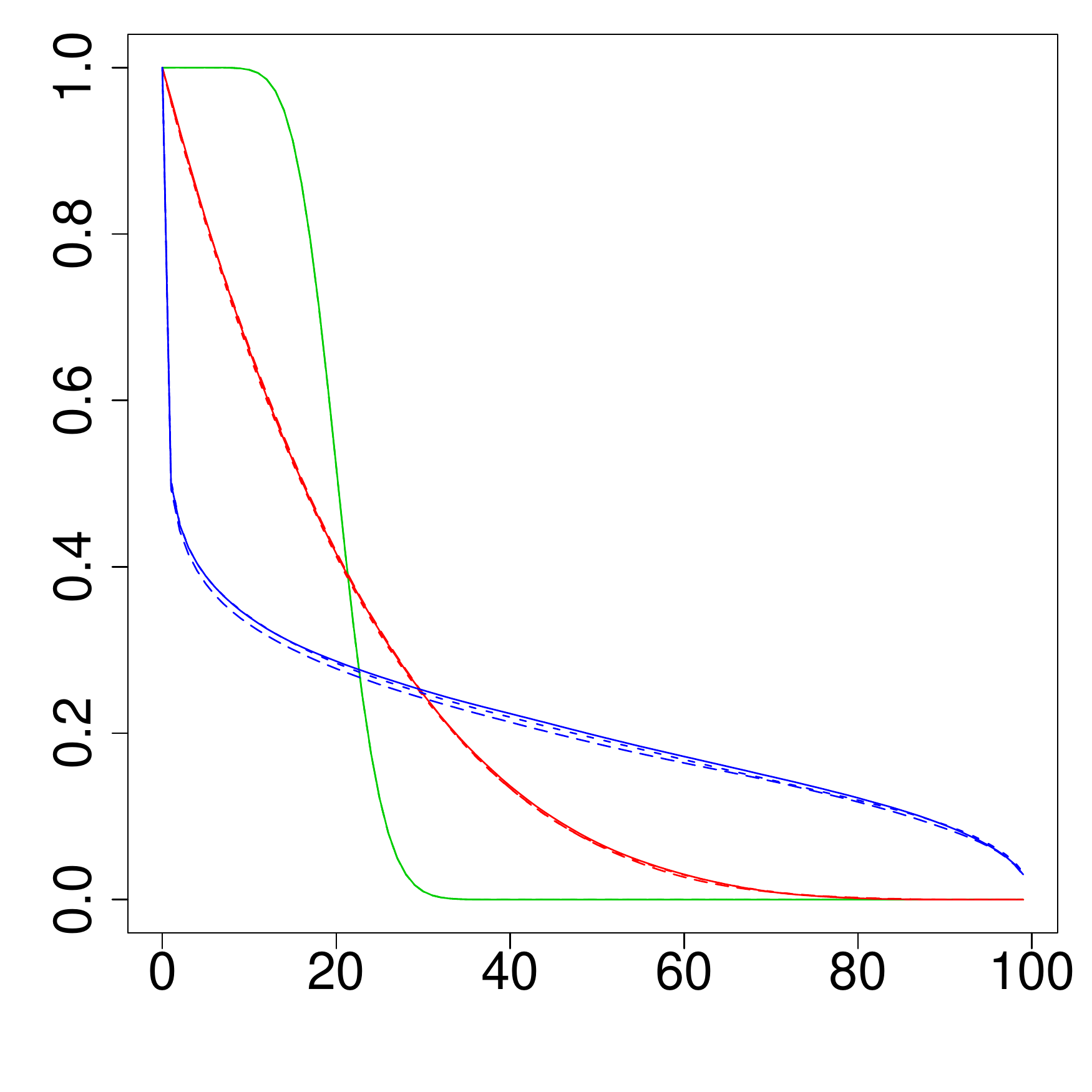}}  &
\subfigure{\includegraphics[width=1.4in,angle=0]{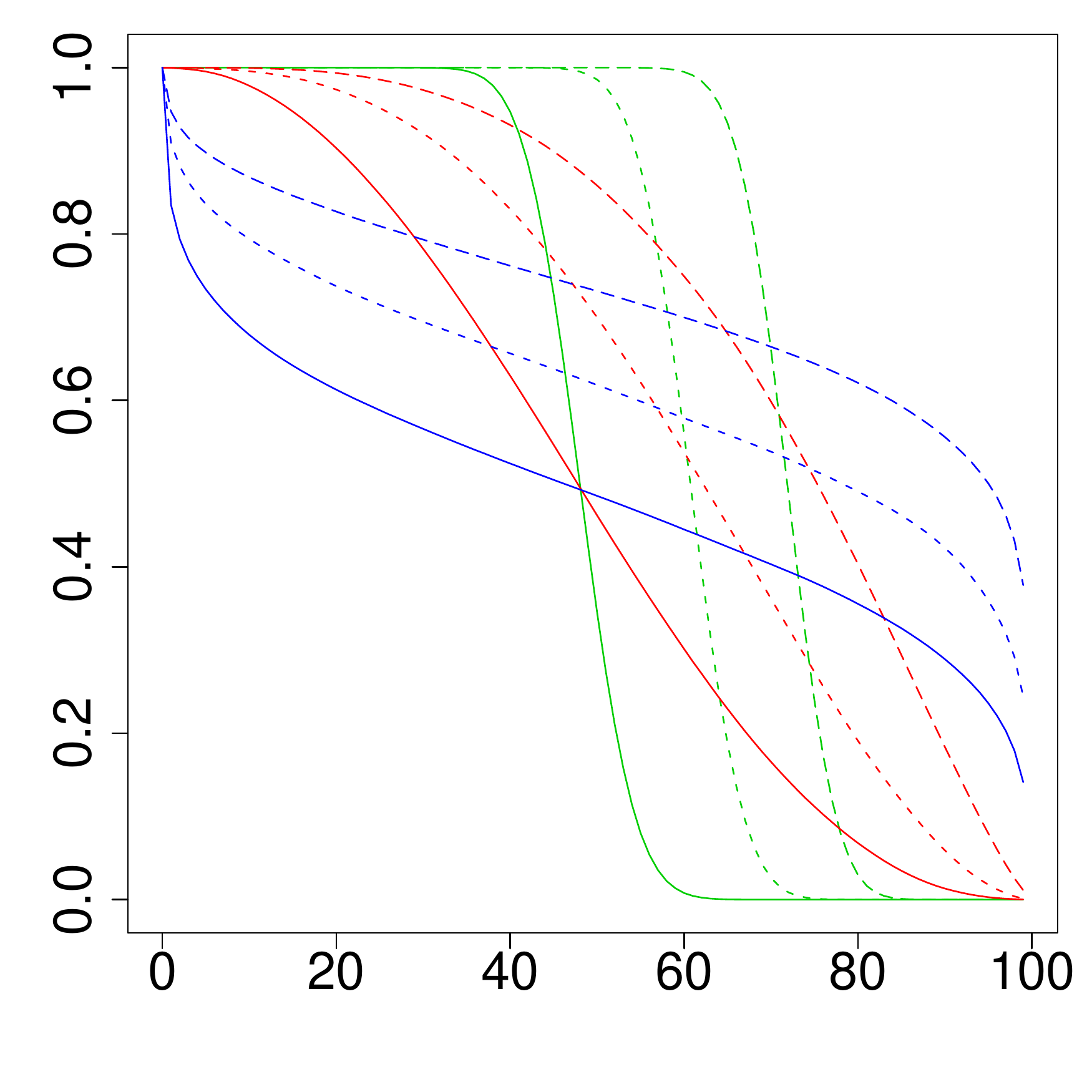}}  &
\subfigure{\includegraphics[width=1.4in,angle=0]{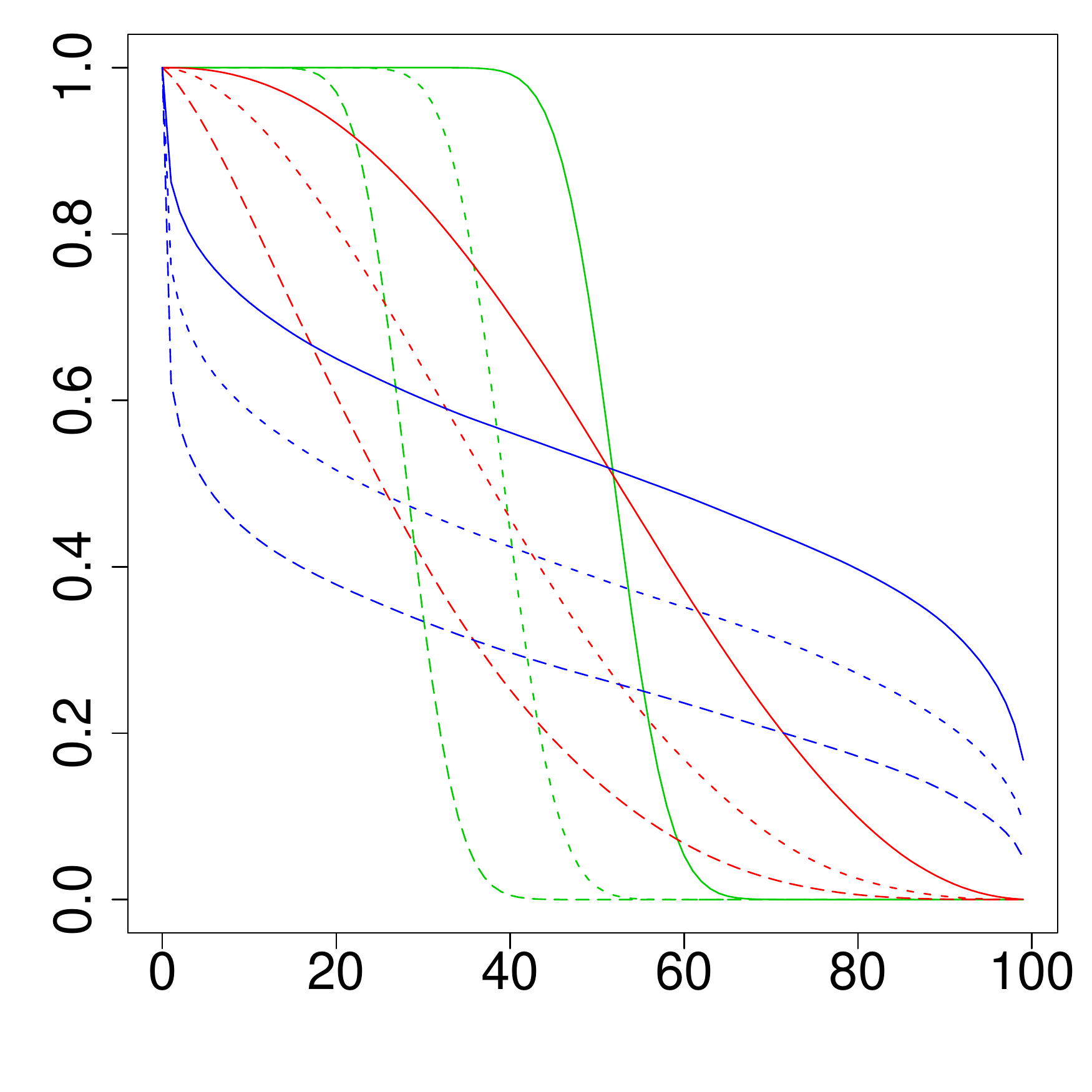}}  &
\subfigure{\includegraphics[width=1.4in,angle=0]{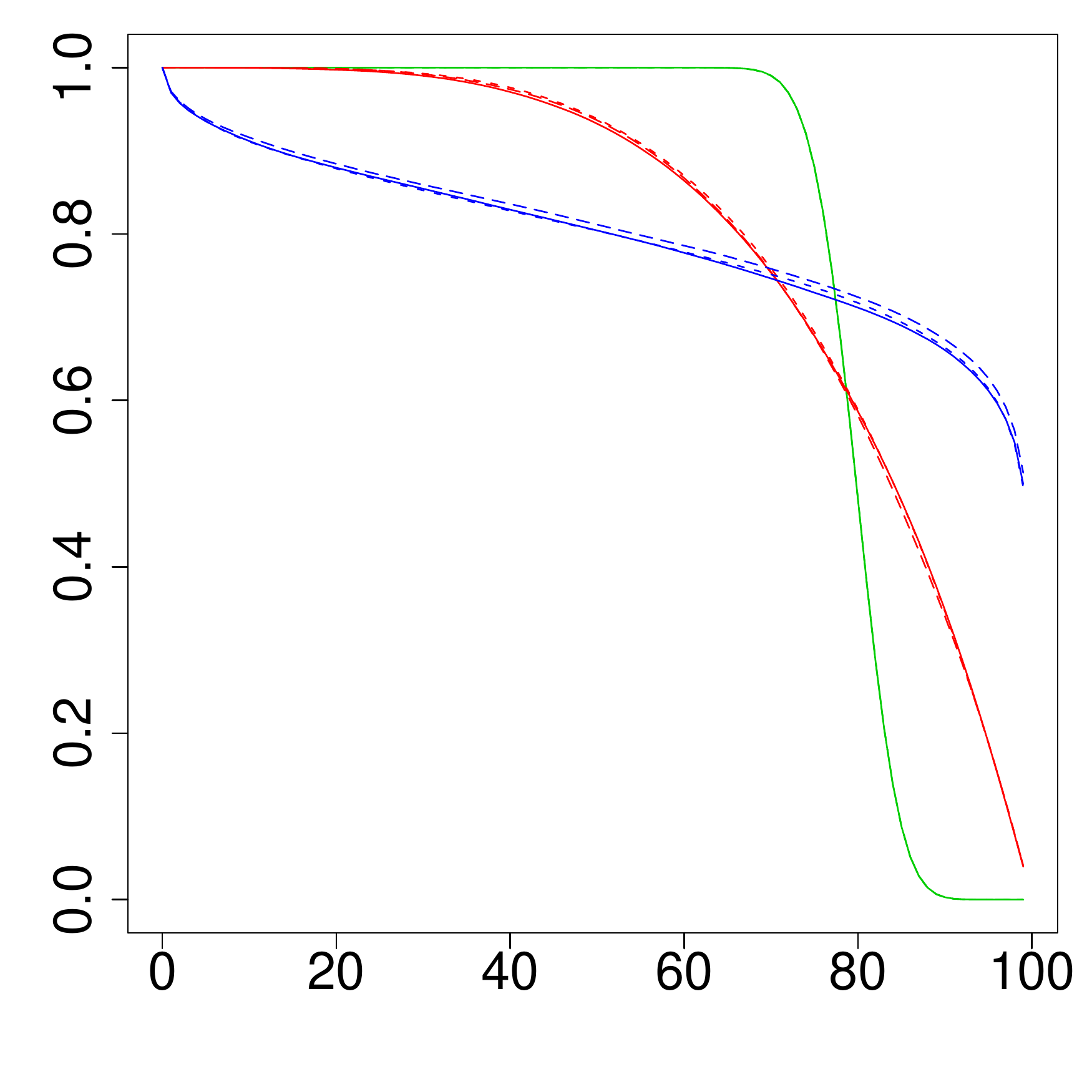}} \\
%%%
\begin{sideways} $\quad \quad \quad \;\; \alpha=0.8$ \end{sideways}&
\subfigure{\includegraphics[width=1.4in,angle=0]{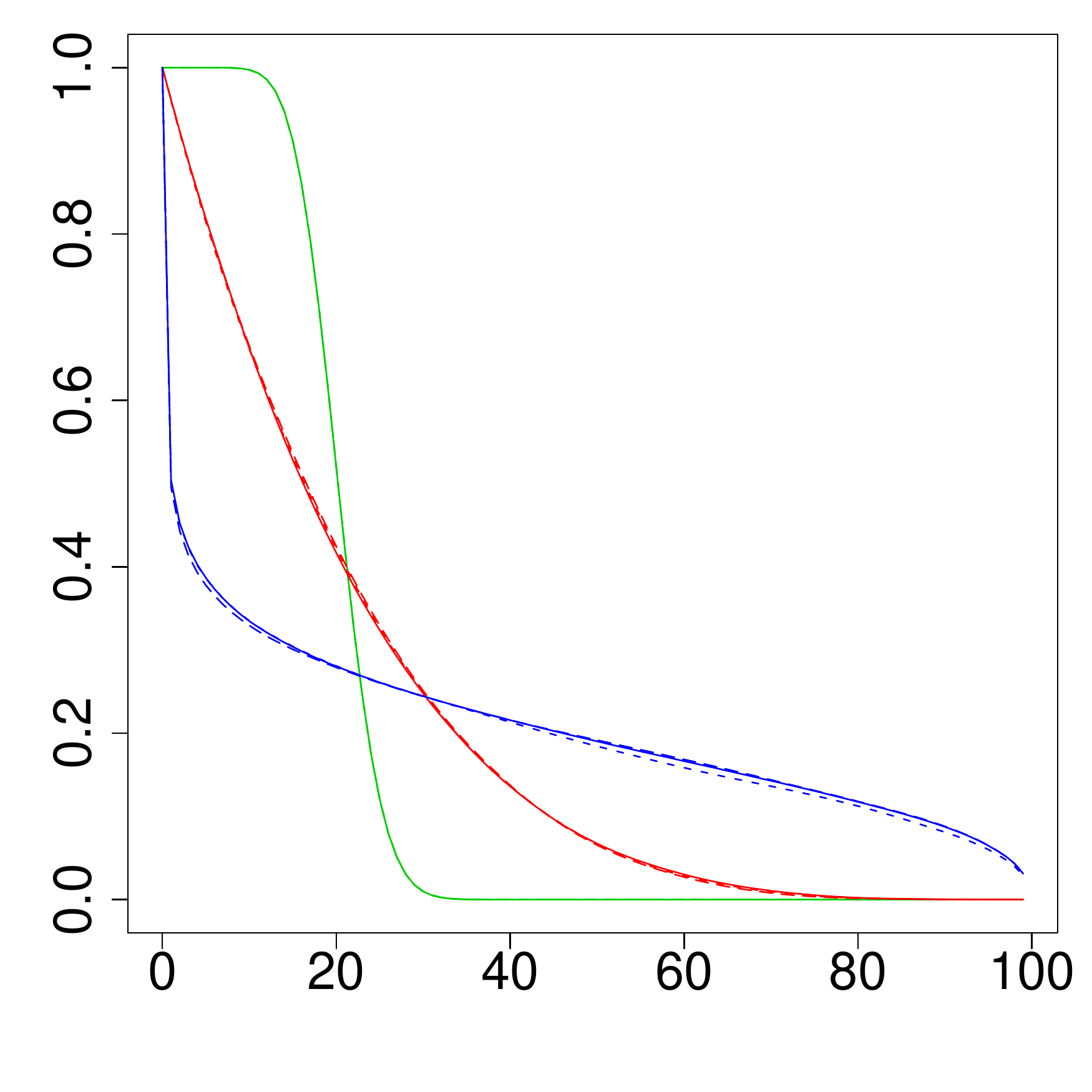}} &
\subfigure{\includegraphics[width=1.4in,angle=0]{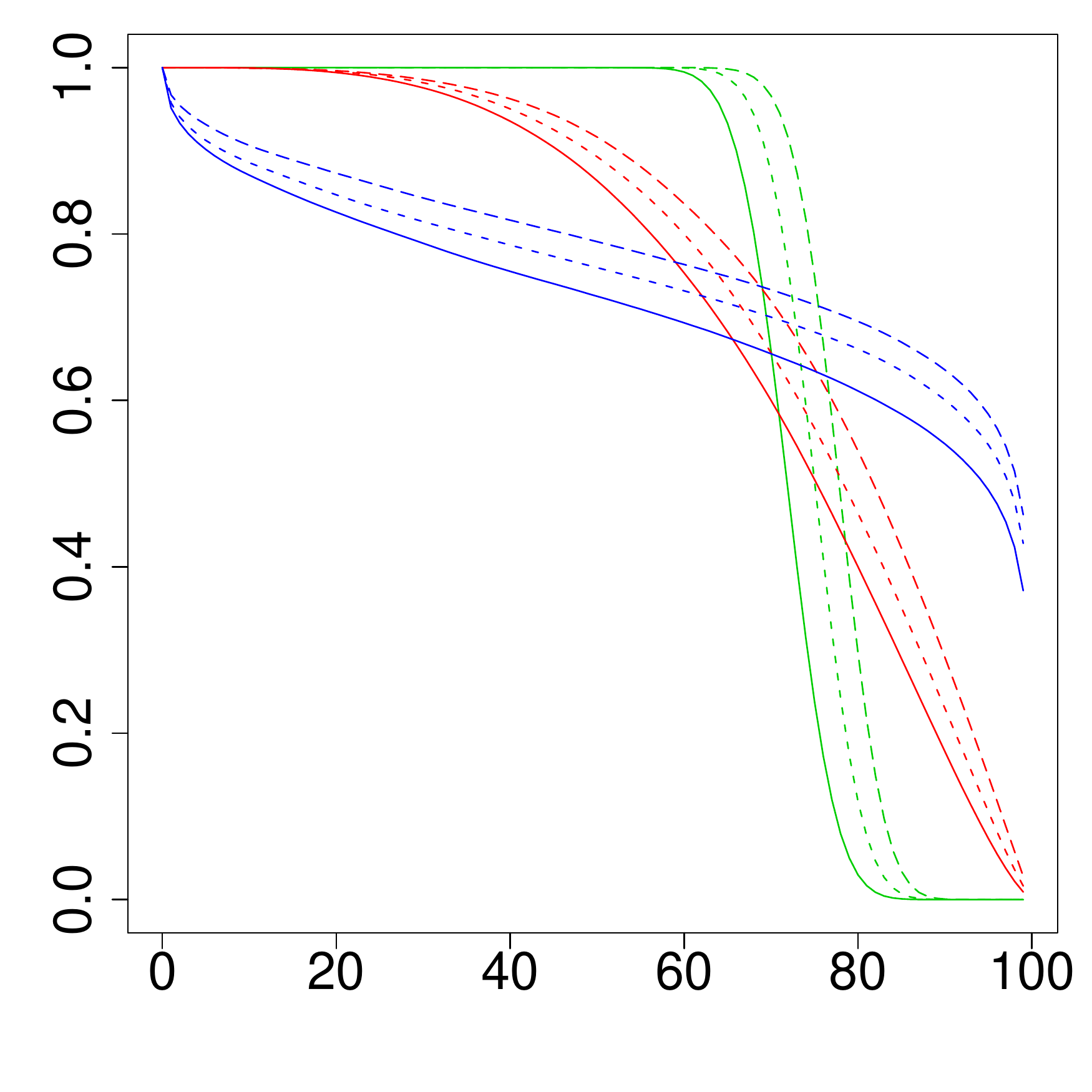}} &
\subfigure{\includegraphics[width=1.4in,angle=0]{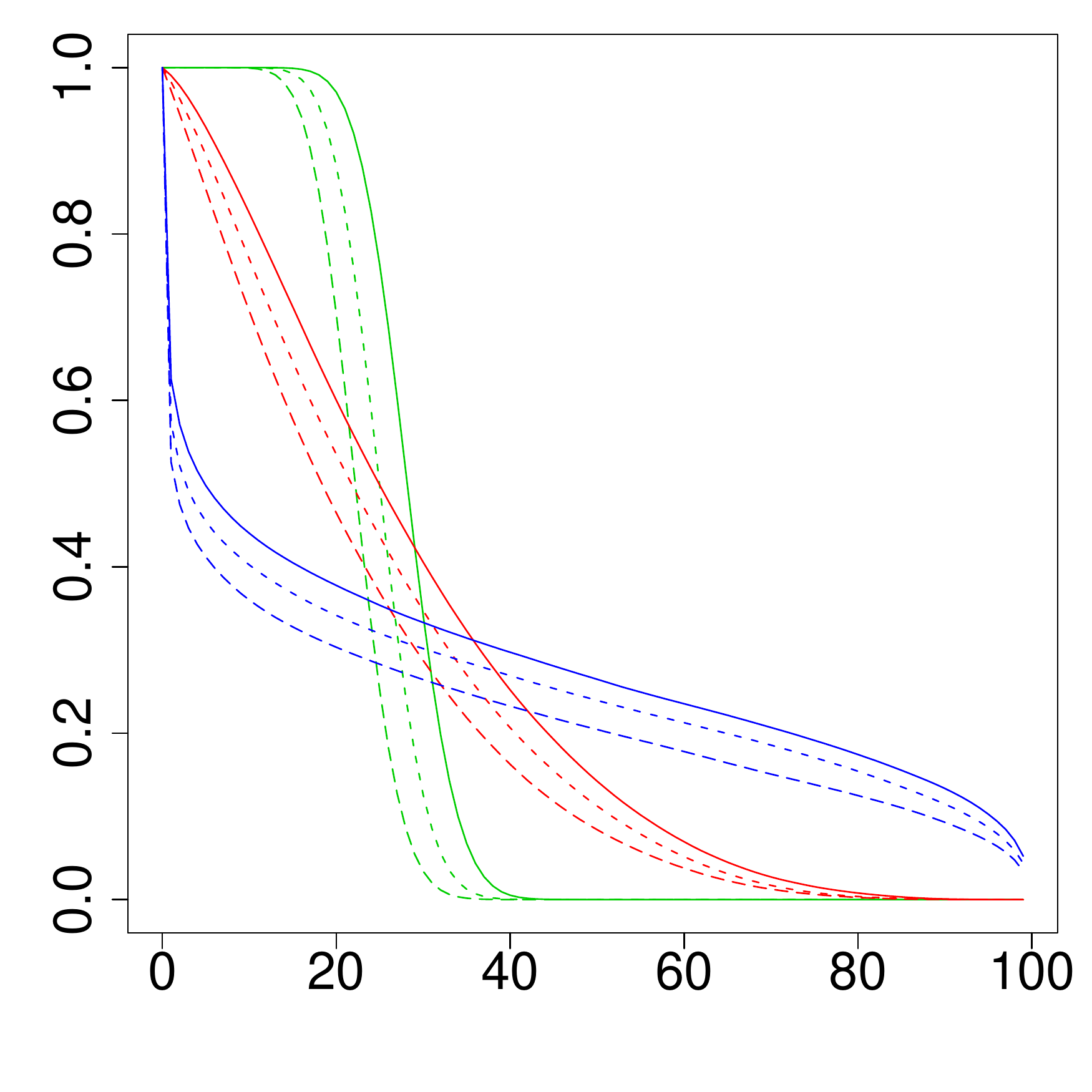}} &
\subfigure{\includegraphics[width=1.4in,angle=0]{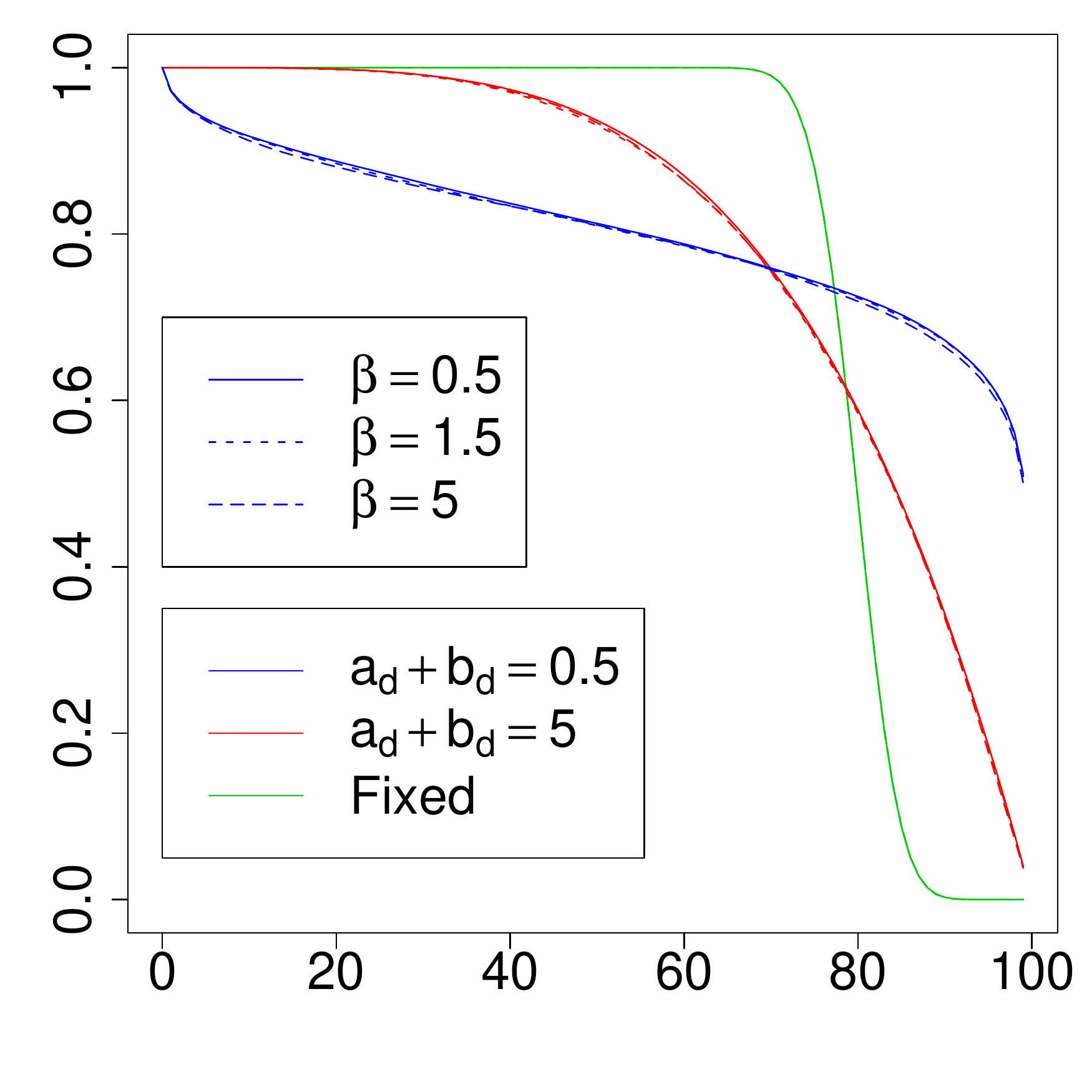}} \\
\end{tabular}
}
\caption{Degree distributions. $\Pr (D_i \geq k)$ is plotted on the vertical axis vs $k$ on the the horizontal axes. Each line represents a combination of diverse values of the stick-breaking parameter $\beta = 0.5$ ({\small \tt solid}), $\beta = 1.5$ ({\small \tt dotted}) and $\beta = 5$ ({\small \tt dashed}); and levels of variability, point-mass ({\small \tt green}), $a_D+ b_D = a_O+ b_O = 5$ ({\small \tt red}) and $a_D+ b_D = a_O+ b_O = 0.5$ ({\small \tt blue}).  } \label{fi:degree}
\end{center}
\end{figure}

The resulting degree distributions are presented in Figure \ref{fi:degree}, each plot shows $\Pr (D_i \geq k)$ for a given $(\bar{\theta}_{D} ,  \bar{\theta}_{O})$ pair and $\alpha$, the first row shows results for $\alpha=0$ which is the blocking model defined on \eqref{eq:DPstickbreaking}, the second and third row use model \eqref{eq:weightprior} with $\alpha=0.2$ (small)  and $\alpha=0.8$ (large), respectively. We observe that 1) when $\bar{\theta}_{D} =  \bar{\theta}_{O}$ the specification of the cluster modeling ($\alpha$ and $\beta$) has no effect on the degree distribution; 2) the stick-breaking parameters have  a bigger influence when assortative or disassortative behavior is present, but the specification of the prior on $\bfTheta$ is still the most influential feature again.

\subsection{Transitivity}

Transitivity is the tendency that makes more likely the connection between two actors when both of them are related to a common third one (the friend of your friend is likely also to be your friend).  Intuitively, the presence of transitivity in the network implies an increased (or decreased) number of triangles (sets of three elements all connected to each other).  More formally, we can define the transitivity coefficient $\chi_{ijh} = \Pr(y_{ij} = 1 | y_{ih}=1, y_{jh}=1, \bflambda, \bfvarsigma)$; $\chi_{ijh} > \Pr(y_{ij} | \bflambda, \bfvarsigma) = \bar{\theta}$ implies positive transitivity, $\chi_{ijh} < \bar{\theta}$ implies negative transitivity, and $\chi_{ijh} = \bar{\theta}$ implies no transitivity in the network.

\begin{lemma}\label{le:transitivity}
For the blockmodel in \eqref{eq:blockmodel} and the priors in  \eqref{eq:hierarchicalprior2} and \eqref{eq:weightprior}, the a priori transitivity index for a binary undirected network is given by 
$$
\Pr(y_{ij} = 1 | y_{ih}=1, y_{jh}=1, \bflambda, \bfvarsigma) = \bar{\chi} = \bar{\chi}^N / \bar{\chi}^D,
$$
where
\begin{multline*}
\bar{\chi}^N = 
\left\{ \frac{(1 - \alpha)(2 - \alpha)}{(\beta + 1)(\beta + 2)}\right\} \E_{H^{\lambda_D}}\{ \theta_{k,k}^3 \}   +
3\left\{ \frac{(1 - \alpha)(\beta+\alpha)}{(\beta + 1)(\beta + 2)}\right\}  \E_{H^{\lambda_D}}\{ \theta_{k,k} \}   \E_{H^{\lambda_O}}\{ \theta_{k,l}^2  \}   \\
+ \left\{ \frac{(\beta + \alpha)(\beta + 2\alpha)}{(\beta + 1)(\beta + 2)}\right\}  \left( \E_{H^{\lambda_O}} \left\{ \theta_{k,l} \right\} \right)^3
\end{multline*}
and
\begin{align*}
\bar{\chi}^D = 
\left( \frac{1 - \alpha}{\beta + 1}\right)  \E_{H^{\lambda_D}}\{ \theta_{k,k}^2 \}   +  \left( \frac{\beta + \alpha}{\beta + 1}\right)
\left( \E_{H^{\lambda_O}} \left\{ \theta_{k,l} \right\} \right)^2
\end{align*}
for any $i$, $j$ and $h$.  \hfill $\diamond$
\end{lemma}
The proof can be seen in Appendix \ref{ap:transitivity}.  Note that $\chi_{ijk} \ne \bar{\theta}$ even under \eqref{eq:hierarchicalprior2}; hence all blockmodels assume a priori transitivity in the relationship among the subjects.

%Even though \eqref{eq:traditionalblockomodels} is more general than \eqref{eq:hierarchicalprior}, the structure of blockmodels is such that to control the clustering coefficient separately from the clustering structure and the reflexivity index we would need to use a prior $H^{\lambda_D}$ that allows for the third moment to be determined separately from the first two moments.  Hence, using beta priors for the elements of the diagonal entries of $\bfTheta$ implies that the prior value of the transitivity index is fixed.

\begin{figure}[h]
\begin{center}
{\small
\begin{tabular}{rcccc}
& $\bar{\theta}_{D} =  \bar{\theta}_{O} = 0.2$   &   $\bar{\theta}_{D} = 0.2$, $\bar{\theta}_{O} = 0.8$   &   $\bar{\theta}_{D} = 0.8$, $\bar{\theta}_{O} = 0.2$   &   $\bar{\theta}_{D} =  \bar{\theta}_{O} = 0.8$\\
%%%
\begin{sideways} $\quad \quad a_D + b_D =  0.5 $  \end{sideways}&
\subfigure{\includegraphics[width=1.4in,angle=0]{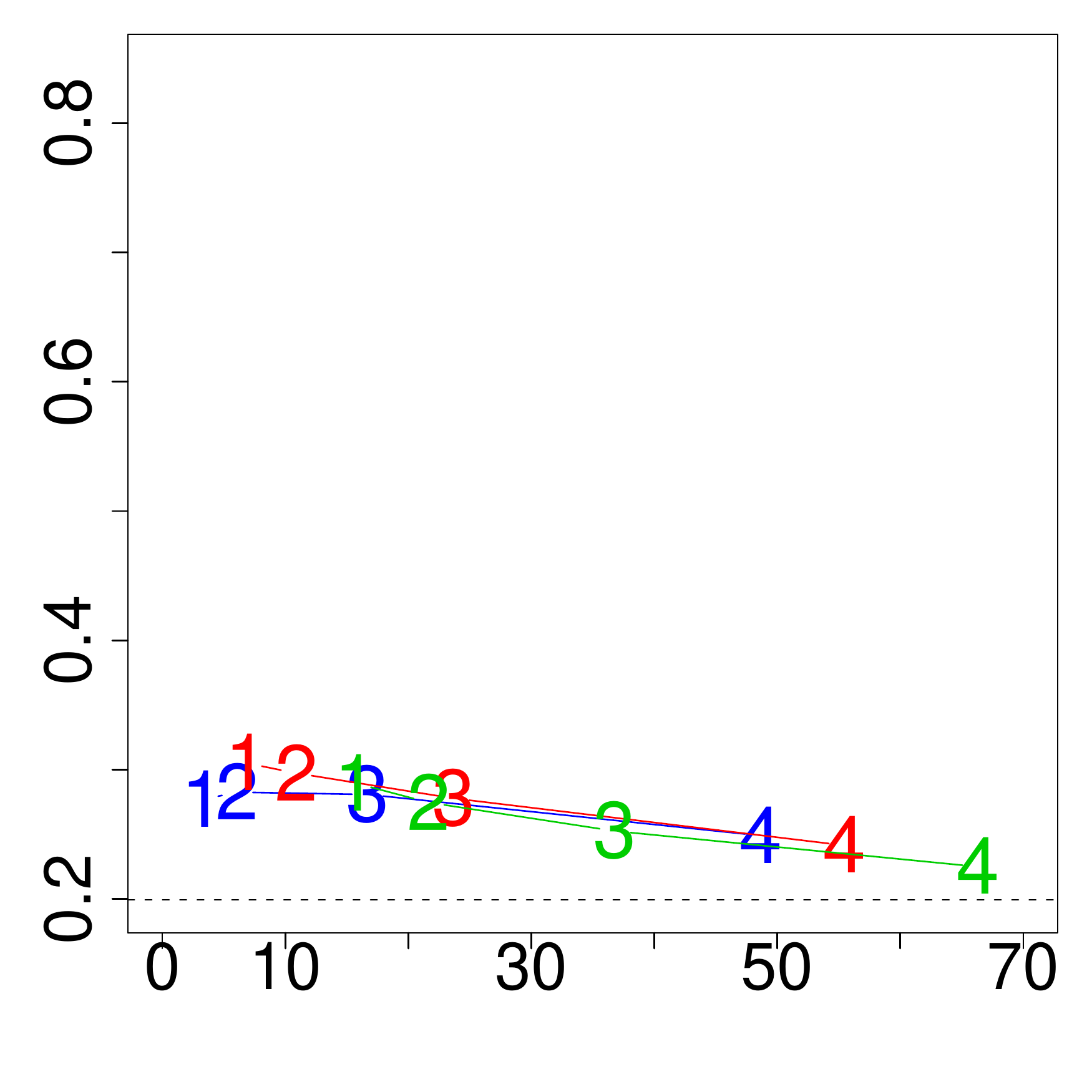}} &
\subfigure{\includegraphics[width=1.4in,angle=0]{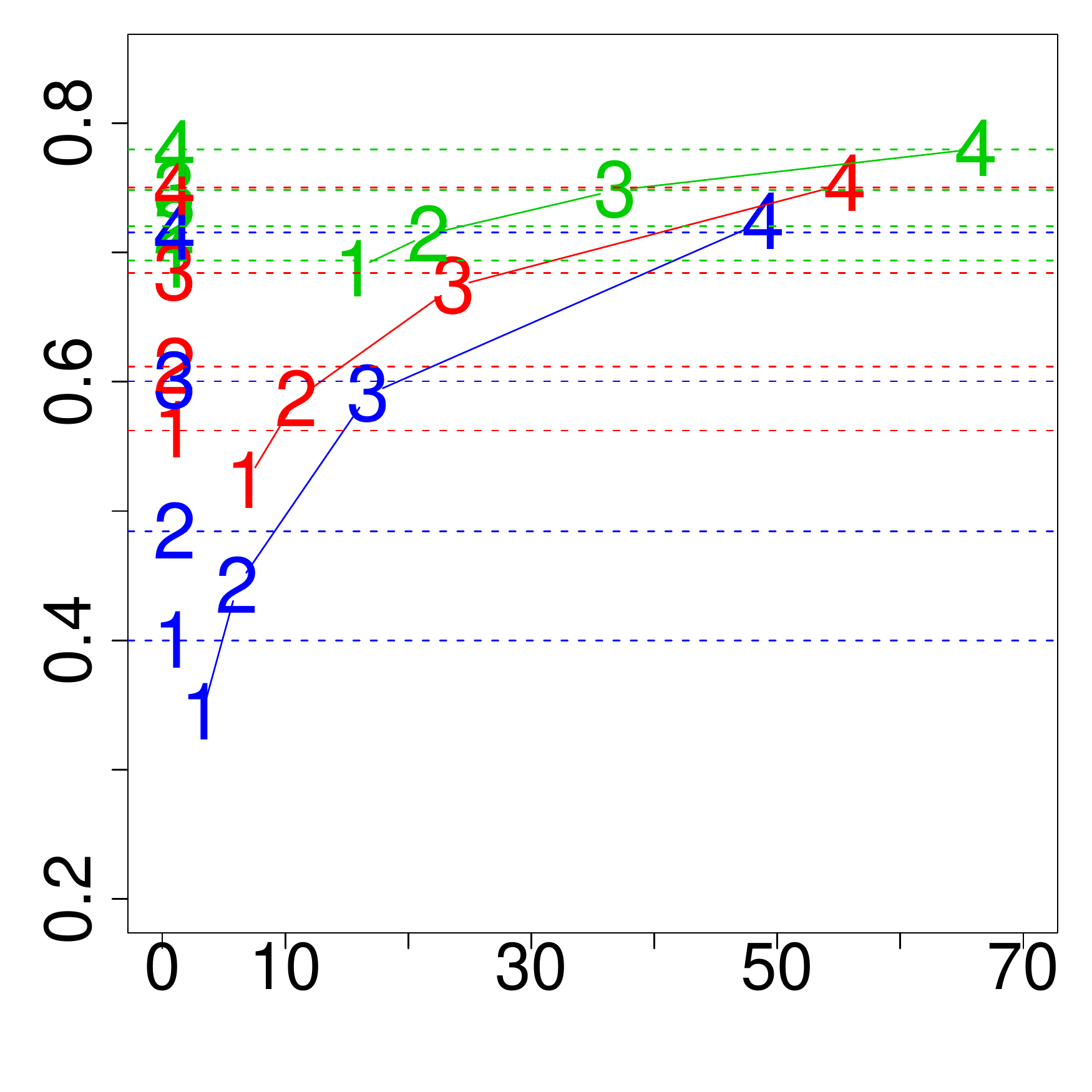}} &
\subfigure{\includegraphics[width=1.4in,angle=0]{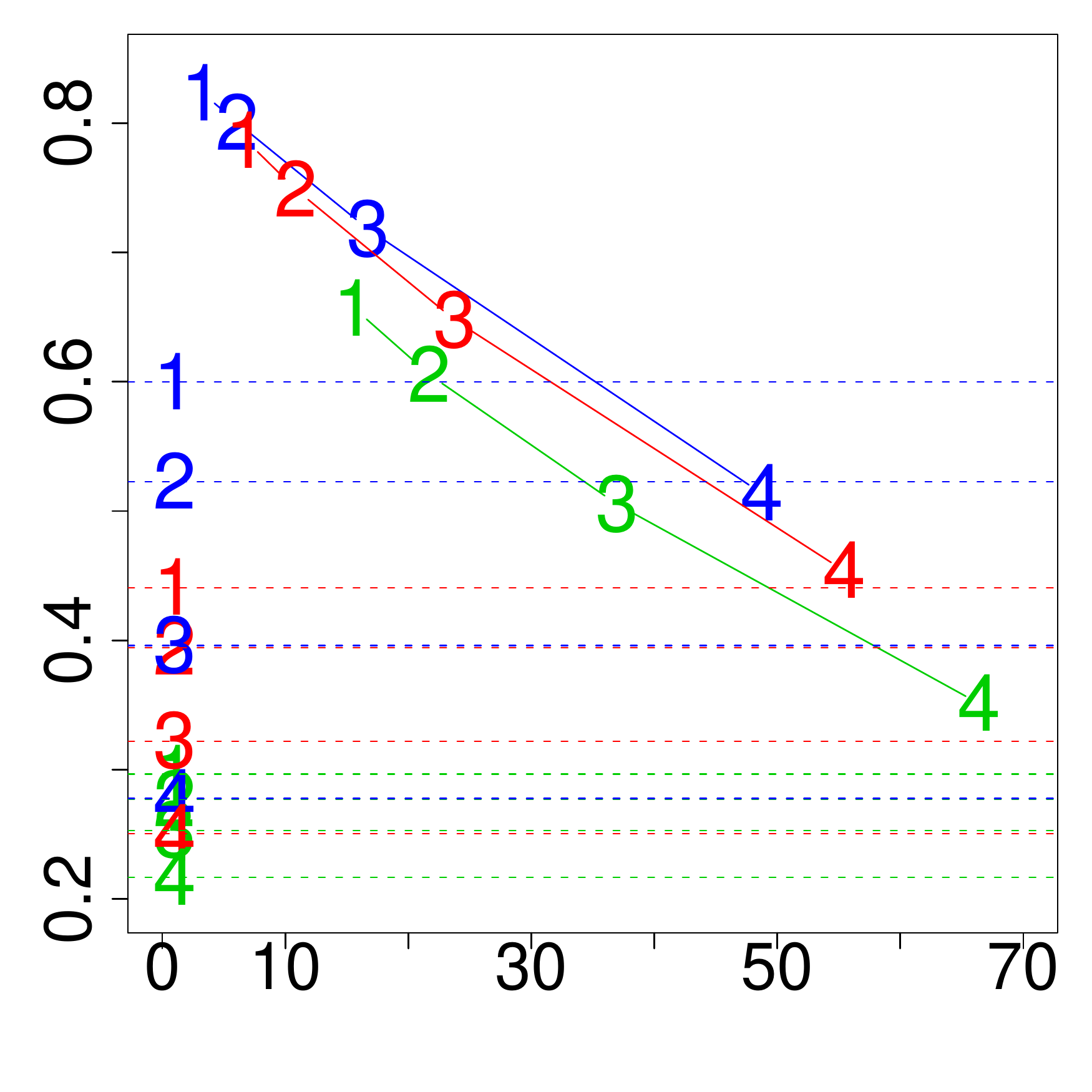}} &
\subfigure{\includegraphics[width=1.4in,angle=0]{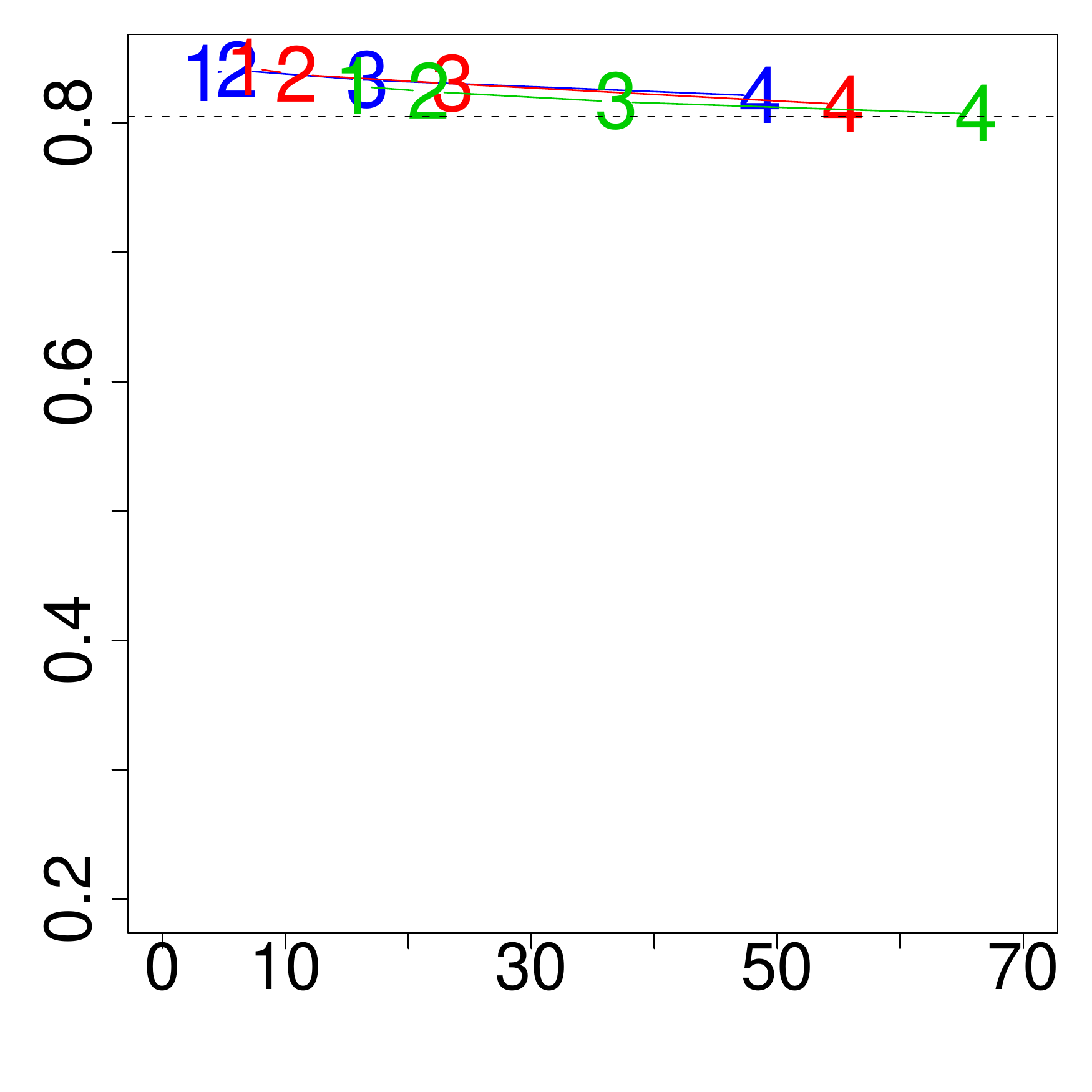}} \\
%%%
\begin{sideways} $\quad \quad a_D + b_D = 5 $  \end{sideways}&
\subfigure{\includegraphics[width=1.4in,angle=0]{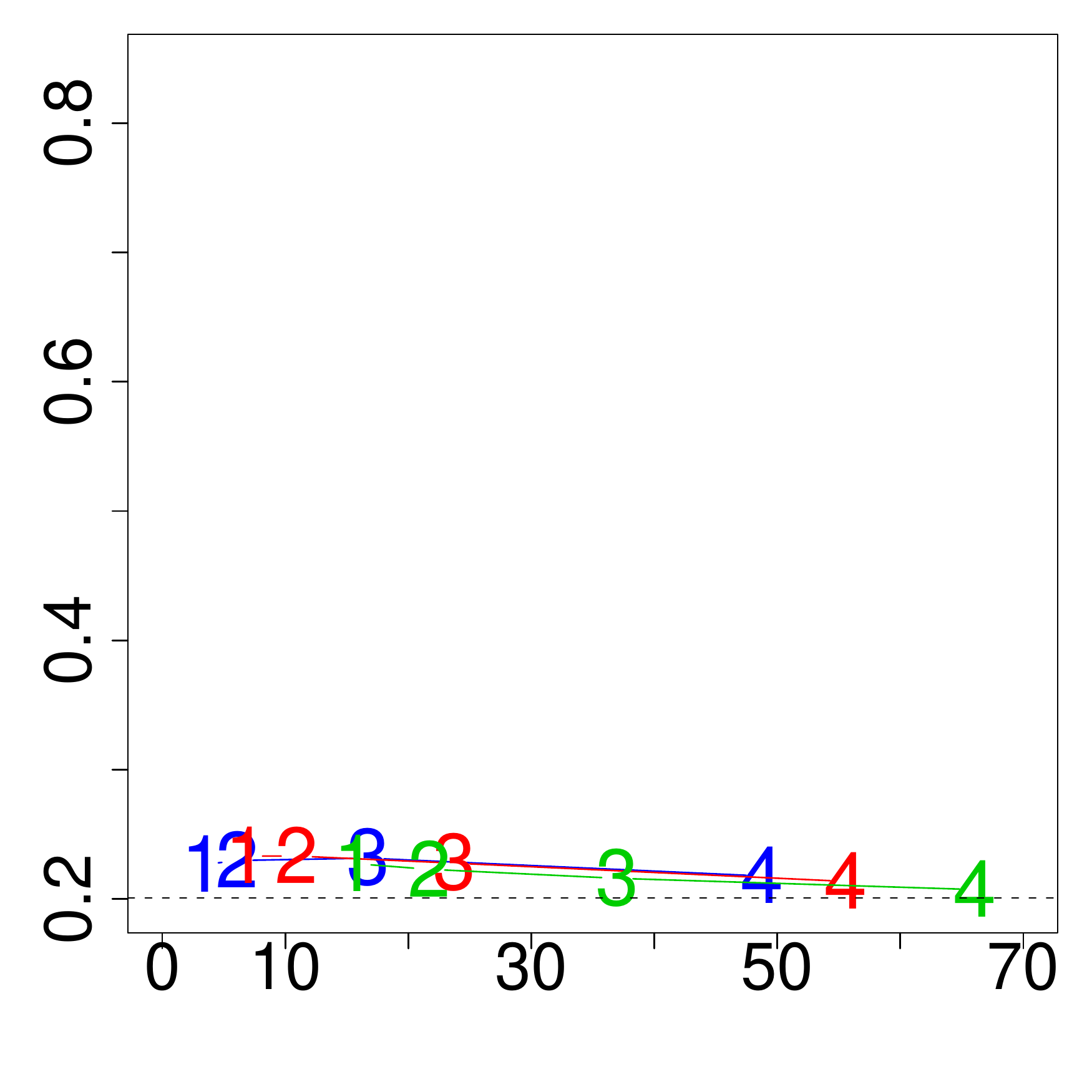}} &
\subfigure{\includegraphics[width=1.4in,angle=0]{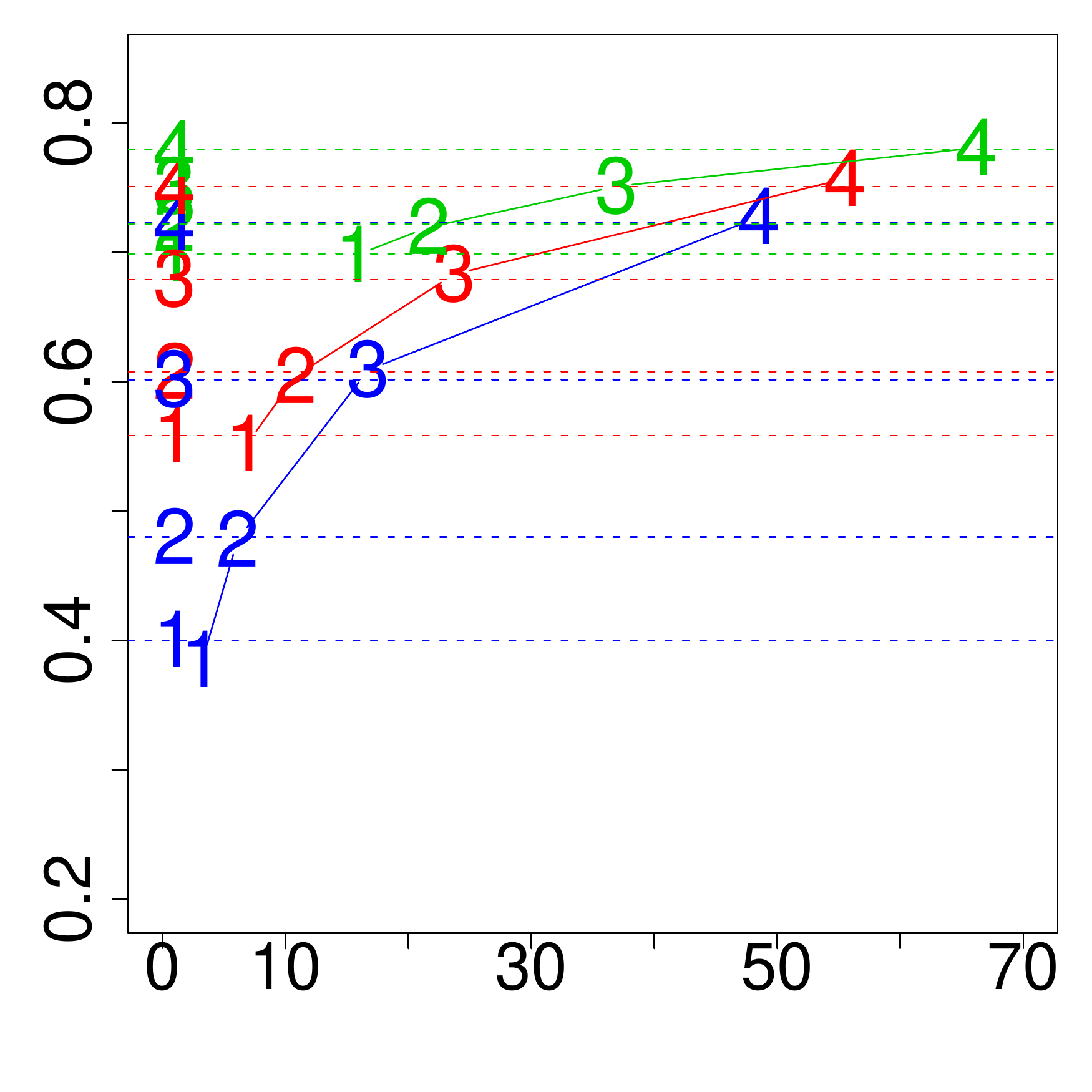}} &
\subfigure{\includegraphics[width=1.4in,angle=0]{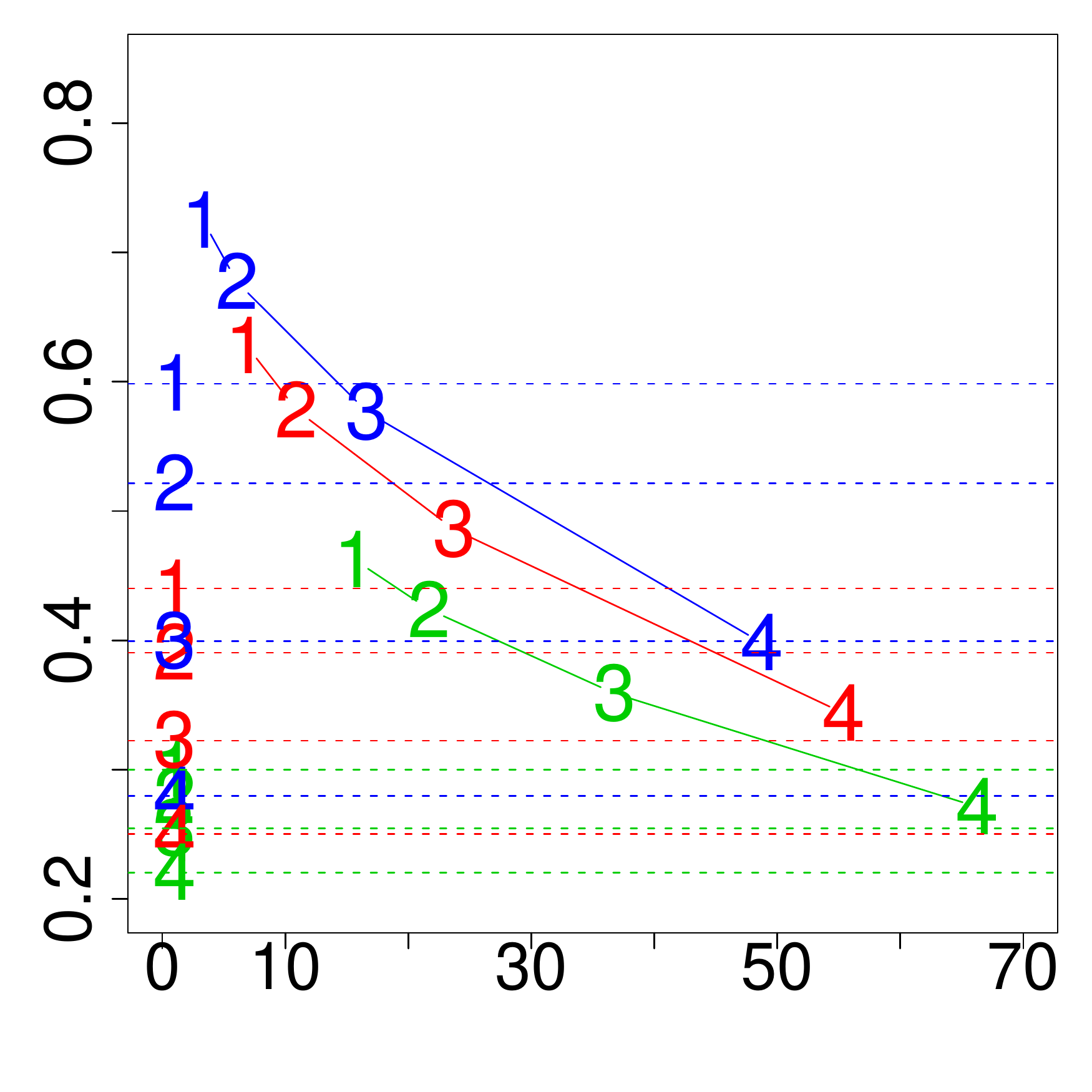}} &
\subfigure{\includegraphics[width=1.4in,angle=0]{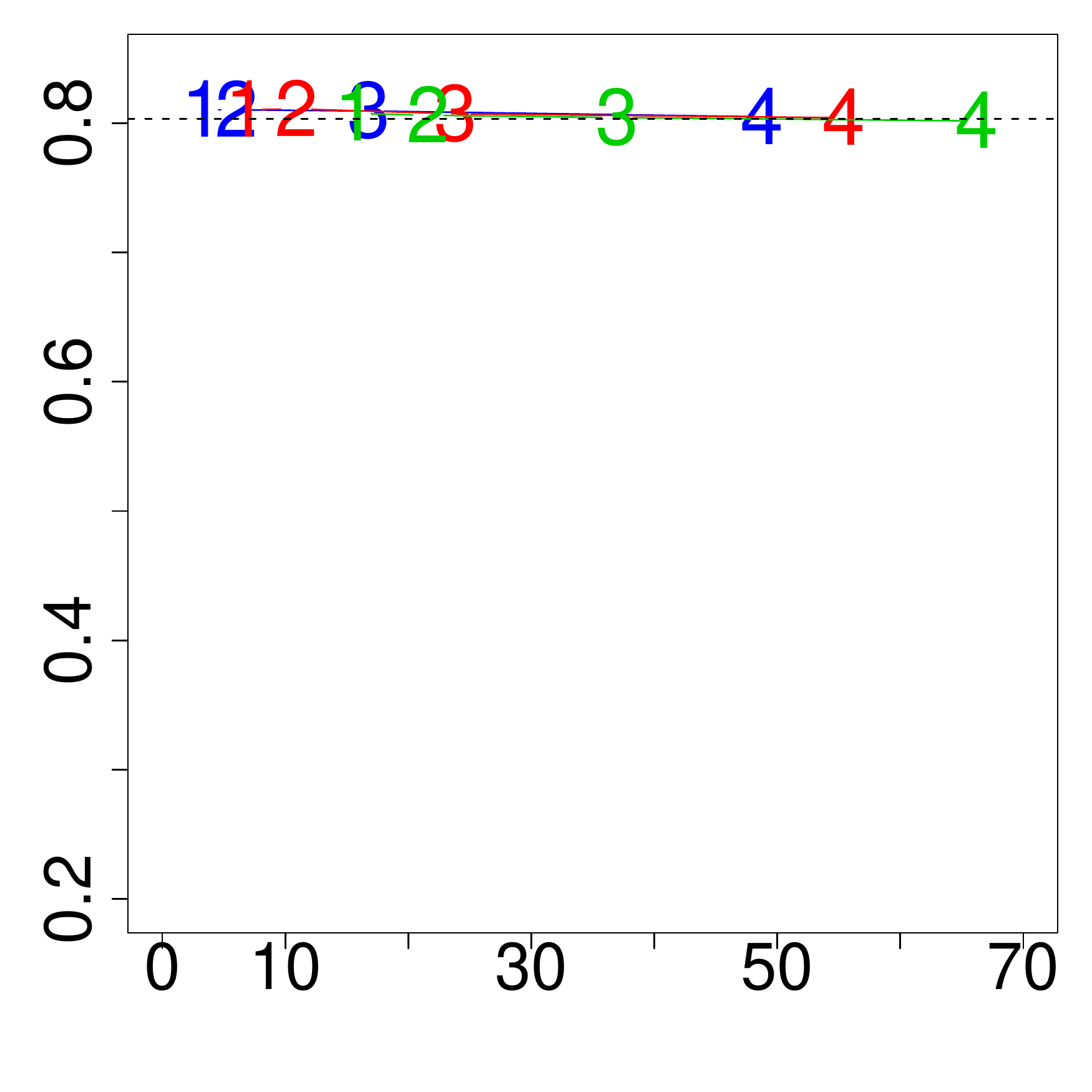}} \\
%%%
\begin{sideways} $\quad \quad \;\; \mbox{Fixed } \bar{\theta}_{D}, \bar{\theta}_{O} $ \end{sideways}& 
\subfigure{\includegraphics[width=1.4in,angle=0]{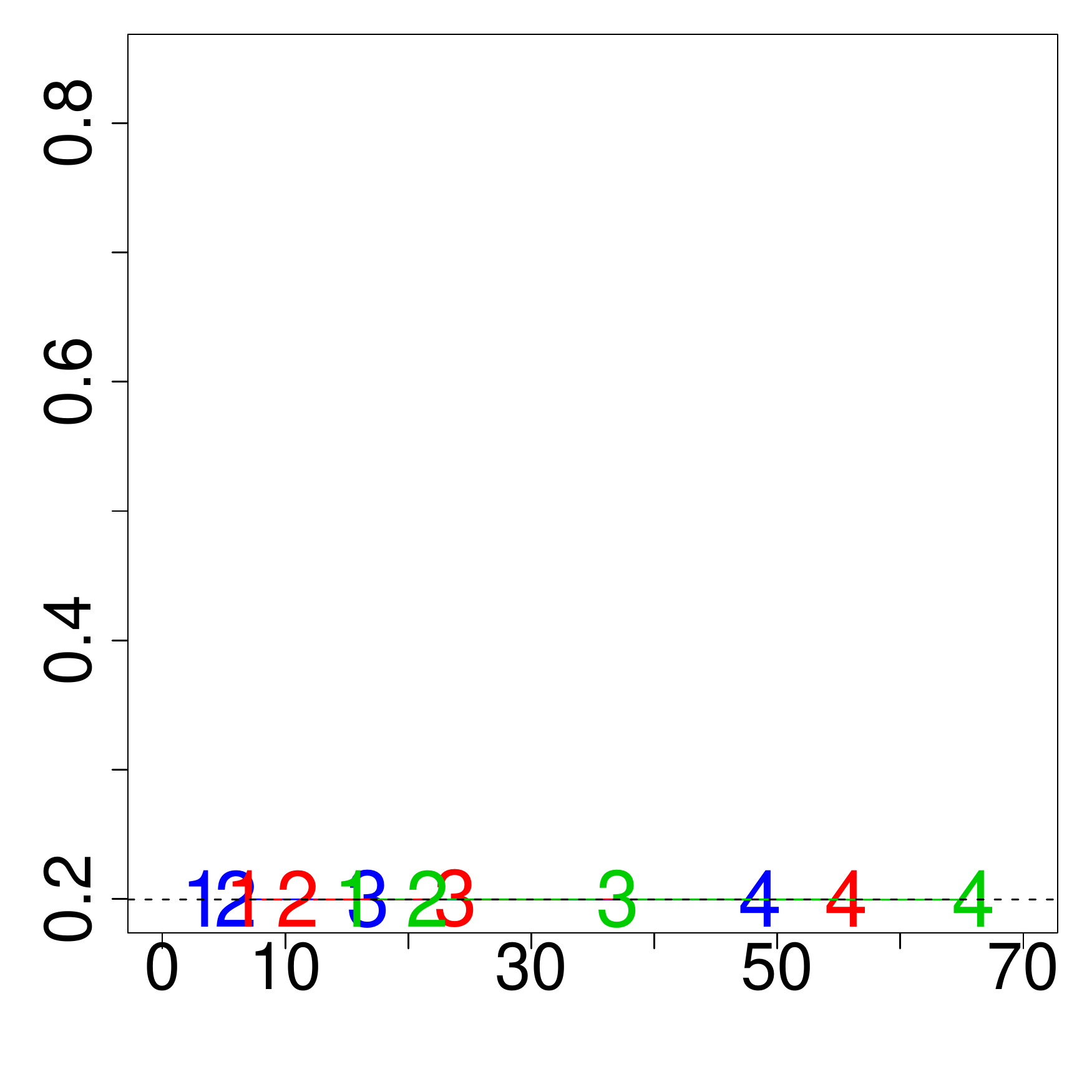}} &
\subfigure{\includegraphics[width=1.4in,angle=0]{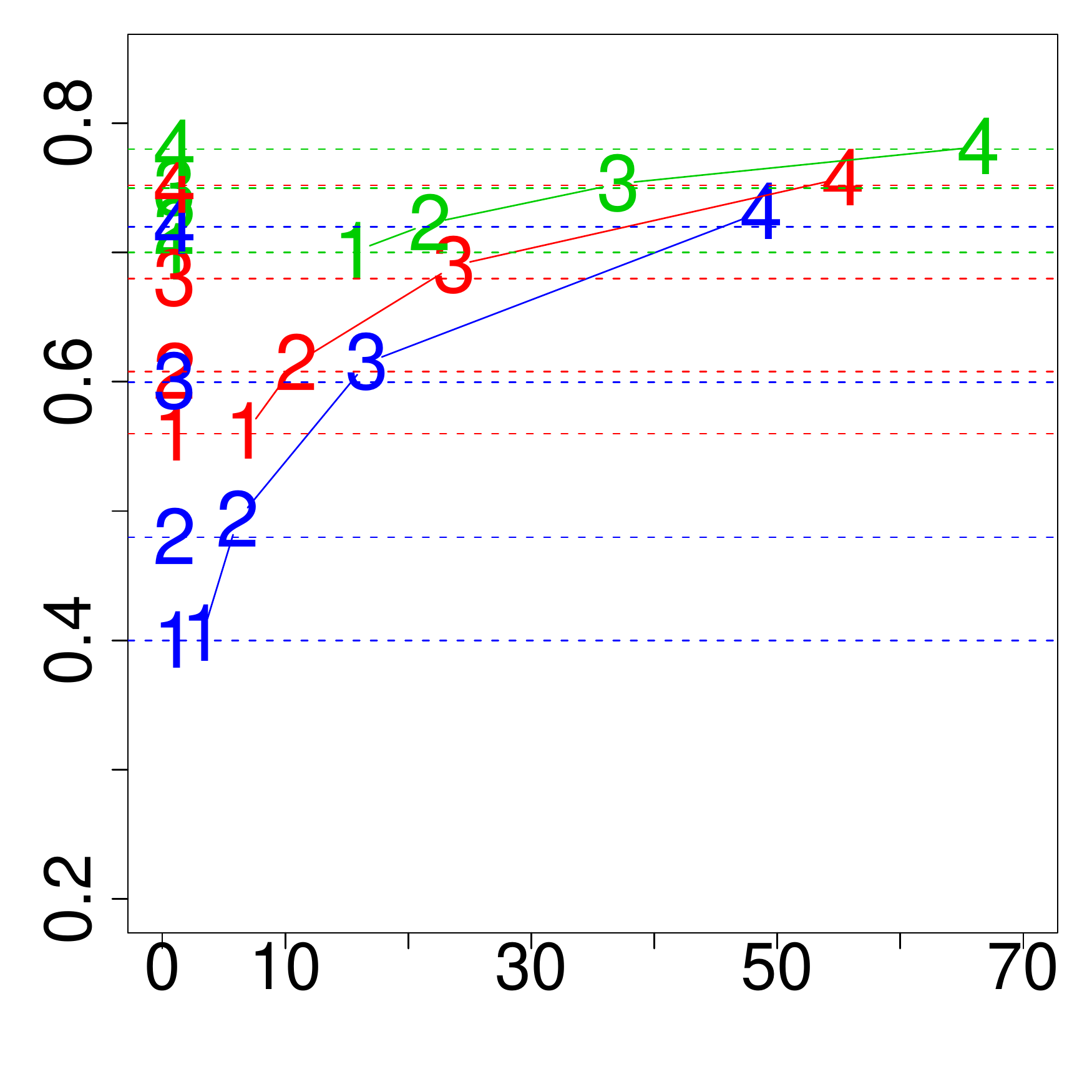}} &
\subfigure{\includegraphics[width=1.4in,angle=0]{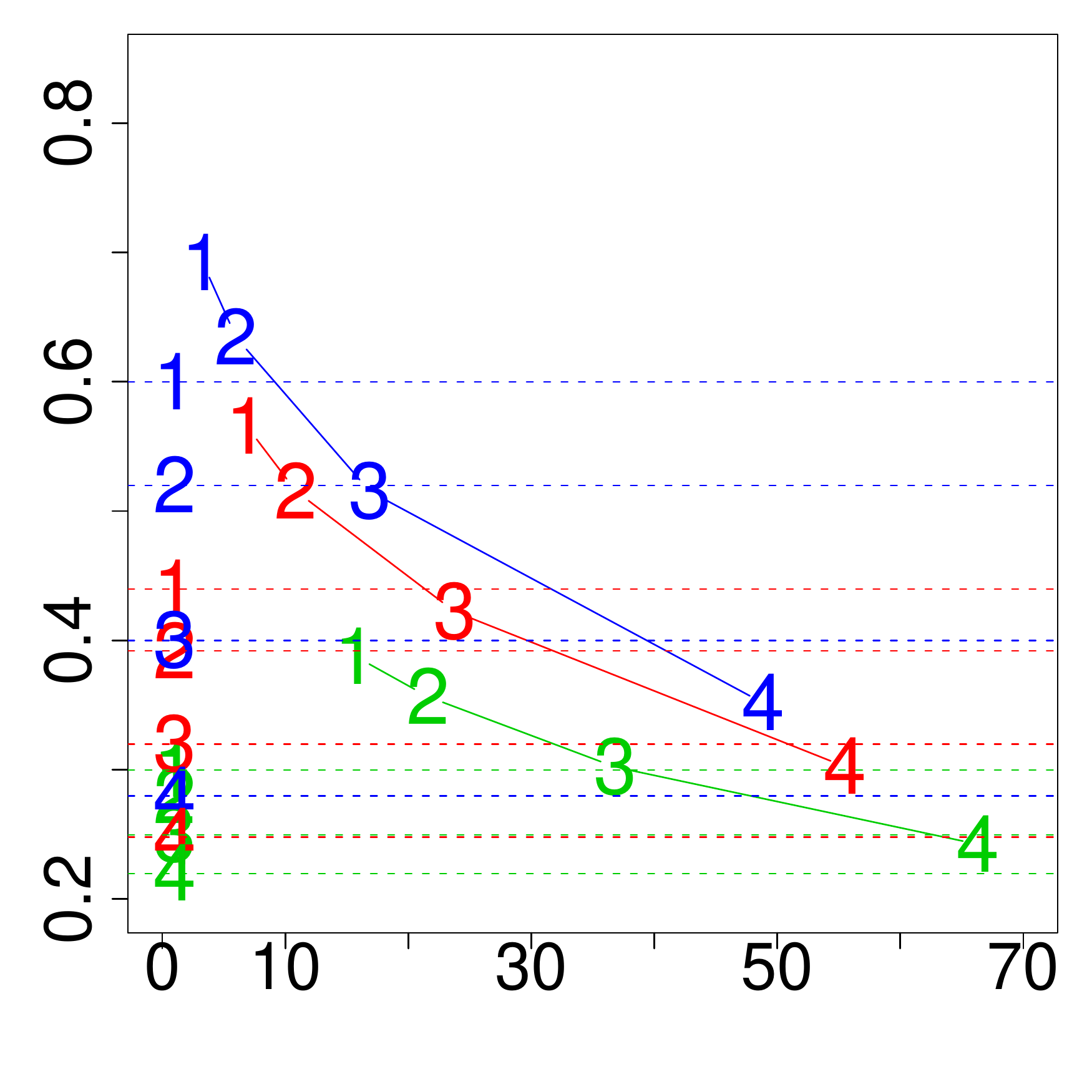}} &
\subfigure{\includegraphics[width=1.4in,angle=0]{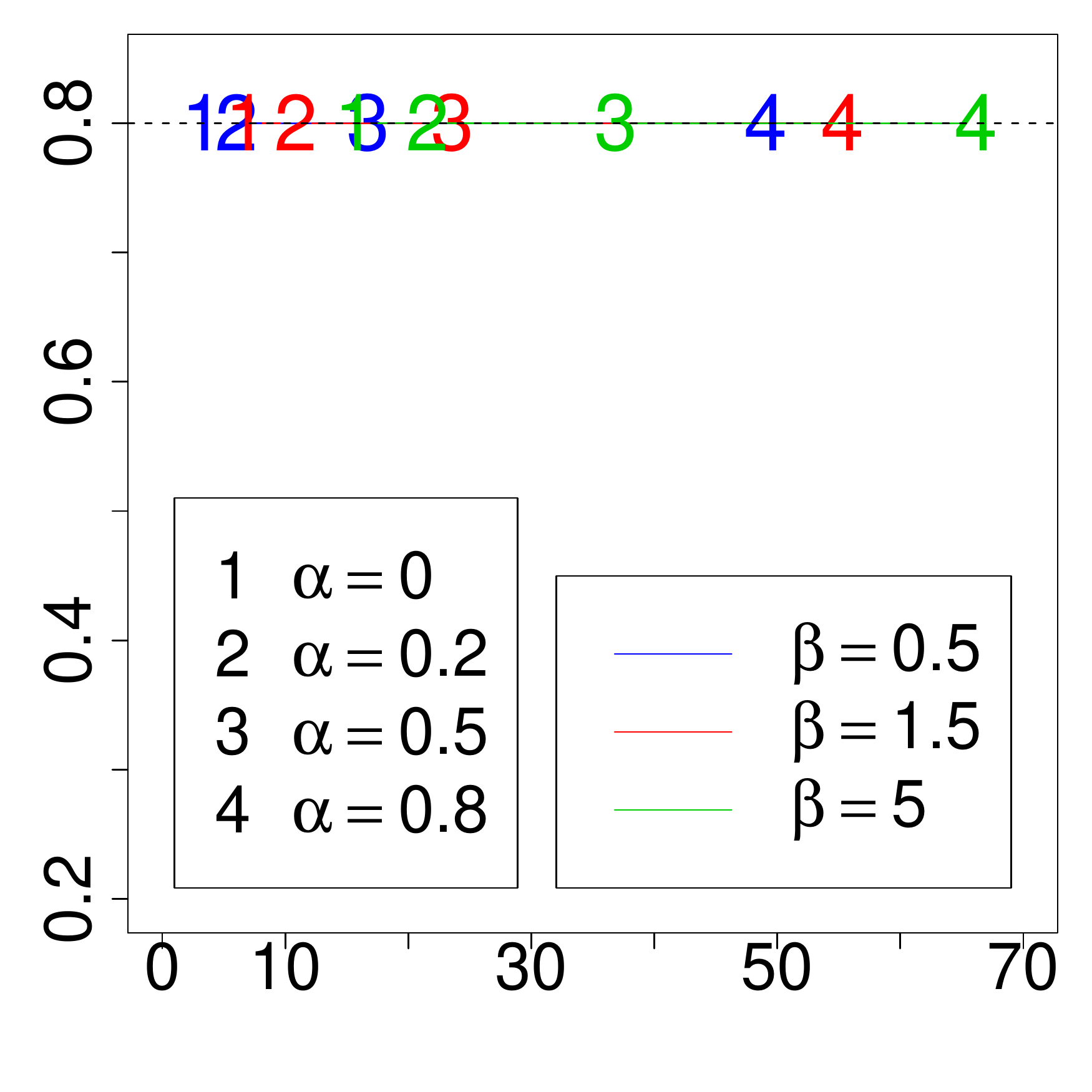}} \\
\end{tabular}
}
\caption{Transitivity Coefficient. Expected $C$ vs. expected number of clusters for different $\alpha$s and $\beta$s ({\small \tt continuous line}). Expected value of $C$ under a random graph or single component model with the same $\bar{\theta}$ ({\small \tt dotted line}). Numbers (1:4) represent increasing values of $\alpha=\{0, 0.2, 0.5, 0.8\}$, colors represent $\beta = 0.5$ ({\small \tt blue}), $1.5$ ({\small \tt red}), $5$ ({\small \tt green}).  } \label{fi:cluster}
\end{center}
\end{figure}

To better understand the effect of different parameters on the transitivity index we performed a second simulation study.  The transitivity index is empirically approximated using the so-called clustering coefficient  \citep{Ne03},
$$
C = \frac{ 3 \times \mbox{number of triangles}}{ \mbox{number of connected triples}}
$$
where a ``connected triple" means a single actor that interacts with two actors. The factor of three in the numerator guaranties that $0 \leq C \leq 1$ by accounting for the fact that each triangle contributes with three triples.  Using the same simulation setting than in the case of the degree distribution, 10,000 networks with $I=100$ actors each were generated to obtain the mean (expected) clustering coefficient $C$. For comparison, we also found the expected value $C$ of 10,000 networks with $I=100$ actors each, assuming a single component model holding everything else the same, i.e. all actors belong to the same group and have probability of connection $\theta$, where the prior $p(\theta)$ agrees with  
$p(\bfTheta | \bflambda)$. Figure \ref{fi:cluster} reports the mean clustering coefficient, on the $y$ axis we have the (expected) clustering coefficient $C$, on the $x$ axis the mean number of factions formed by the stick-breaking process \eqref{eq:weightprior} for the diverse values of $\alpha$ and $\beta$. Each row represents different levels of incertitude in the prior of $\bfTheta$. In each plot, horizontal dotted lines marked the expected value of $C$ under the single component model specification.

From the simulation results we observed that the clustering coefficient $C$ is highly associated with $\E(\bfTheta | \bflambda)$. In fact, as with the degree distribution, $C \approx \E(\bfTheta | \bflambda)$ when $\bar{\theta}_{D} =  \bar{\theta}_{O}$, regardless the specified $\alpha$ and $\beta$. However, $C$ is further from $\bar{\theta}$ and always greater for small values of $\alpha, \beta$ (fewer factions) and high $\V (\bfTheta | \bflambda)$. For the dissasortative model ($\bar{\theta}_{D}  < \bar{\theta}_{O} $), $C$ is again very close to $\E(\bfTheta | \bflambda)$, the most interesting results are obtained for the assortative model where the transitivity of the model is always larger than $\E(\bfTheta | \bflambda)$ and for small values of $\alpha, \beta$, $C$ could be greater than $\bar{\theta}$ by $5$ - $10$ percentage points.  This is not the case for the single component model, whose $C$ resulted equal to $\E(\theta) $ even for different levels of variability.  
 
\section{Hierarchical stochastic blockmodels for collections of networks}\label{sec:multiple}

Consider now a situation where multiple networks are observed {\it for the same set of actors}, so that the data corresponds to an exchangeable collection matrices $\bfY_1,\ldots,\bfY_J$, where $\bfY_j = [y_{i,i',j}]$, $i,i' = 1,\ldots,I$ and $j = 1,\ldots, J$.  We are interested in jointly modeling $\bfY_1,\ldots,\bfY_J$ in order to improve estimation of the community structure associated with each network.  Moreover, we are interested in identifying groups of networks (relationships) with similar community structure.  As we discussed in the introduction, communities are meaningful constructs (e.g., in the case of social networks they can be interpreted as social roles or social positions) that are often driven by unobservable (or unobserved) variables.  Hence, clusters of networks can provide important insights about the underlying processes generating the networks. Furthermore, by jointly modeling through the community structure the model can accommodate different types of networks (binary, count, directed, undirected, etc).

%We assume that the measured relationships are the result of interactions governed by one or more unobserved processes. Those processes are the ones that dictate the formation of factions, which in turn have an effect on a class of relationships. In this context, we will say that two networks are similar or equivalent if the individuals  group into the same communities.  Formally, we would classify each of the $J$ relationships into $R$ classes. This classification could be indicated by $\bfzeta = (\zeta_1, \ldots, \zeta_J)$, $\zeta_{j} = 1,\ldots,R$ and we will assume that $\zeta_j$'s follow a Poisson-Dirichlet process with parameters $\alpha, \beta$, assumption that will be denoted by $\zeta_j \sim CRP(\alpha, \beta)$.

%Inside each class of relationships we will further assume there exists a natural grouping of actors into communities or factions that rule the interaction between actors. As in the case of the relationships the groups of actors for the $r$-th class of relations  will be indicate by $\bfxi_r = (\xi_{r,1}, \ldots, \xi_{r,I})$, $r=1,\ldots,R$. For each $r$-th class $\xi_{r,i} \sim CRP(\alpha_r, \beta_r)$ this hierarchy cluster definition induces the following model for $y_{ii'j}$. 

From now on, dyads and networks are assumed to be conditionally independent, so that 
\begin{align}\label{eq:collect}
%  y_{i,i',j} & \sim p(y_{i,i',j} | \theta_{\xi_{\zeta_j, i}, \; \xi_{\zeta_j, i'}, \; j})
y_{i,i',j} & \sim_{iid} \psi_j(y_{i,i',j} | \theta_{\gamma_{j, i}, \gamma_{j, i'}, j}),
\end{align} 	
where
$\psi_j$ is a parametric distribution associated with network $j$, $\theta_{k,l,j}$ is the parameter that controls the rate of interaction among factions $k$ and $l$ in network $j$, and $\gamma_{j,i}$ is the faction membership indicator for actor $i$ in network $j$.

In order to identify groups of networks with similar community structures, we introduce a set of indicators $\zeta_1, \ldots, \zeta_J$ associated with each of the networks, such that $\zeta_j = \zeta_{j'}$ if and only if $\gamma_{j,i} = \bfgamma_{j',i}$ for all $i=1,\ldots,I$.  Hence, inferences on $\zeta_1,\ldots,\zeta_J$ allow us to asses how similar the community structures are across different networks.  A joint prior for $\bfgamma_1, \ldots, \bfgamma_J$ is then obtained by setting
\begin{align}\label{eq:prioronzeta}
\zeta_j | \bfvarpi &\sim_{iid} \sum_{k=1}^{\infty} \varpi_k \delta_k,  & \varpi_k &=v_k \prod_{s < k} (1 - v_s),  & v_k &\sim_{iid} \bet (1- \alpha_1, \beta_1 + \alpha_1 k)
\end{align}
and letting $\bfgamma_j = \bfxi_{\zeta_j}$, where vectors $\bfxi_1, \bfxi_2, \ldots$ (which encode the unique community structures associated with each group of networks) are independently sampled according to 
\begin{align}\label{eq:prioronxi}
\xi_{k,i} | \bfw &\sim_{iid} \sum_{l=1}^{\infty} w_{k,l} \delta_l,  & w_{k,l} &=u_{k,l} \prod_{s < l} (1 - u_{k,s}),  & u_{k,l} &\sim_{iid} \bet (1- \alpha_2, \beta_2 + \alpha_2 l).
\end{align}
The conditional independence assumptions implicit in \eqref{eq:prioronzeta} and \eqref{eq:prioronxi} ensure exchangeability across the networks, i.e., that the model is invariant to the order in which the networks are included in the model.
%In the above expressions $\bfxi_k = (\xi_{k,1}, \ldots, \xi_{k,J})$ represents the community structure associated with the $k$-th group of networks.  

Equation \eqref{eq:prioronxi} implies a joint distribution for $ (\xi_{k,1}, \ldots, \xi_{k,J})$ that can be described by a generalized Chinese restaurant process \citep{Pi95} that sets $\xi_{k,1} = 1$ and
\begin{align}\label{eq:polyaurn2}
\xi_{k,i} | \xi_{k,i-1}, \ldots, \xi_1 &\sim \sum_{l=1}^{L^{i-1}} \frac{n_{k,l}^{i-1} - \alpha_2}{\beta_2 + i - 1}\delta_{l} +  \frac{\beta_2 + \alpha_2 L^{i-1}}{\beta_2 + i - 1}\delta_{L^{i-1} + 1},   & 2 \le i &\le I,
\end{align}
where $n_{k,l}^{i-1} = \sum_{j=1}^{i-1} \mathbf{1}_{(\xi_{k,j} = l)}$ is the number of members of faction $l$ among the first $i-1$ subjects in the $k$-th cluster of networks, and $L^{i-1} = \max_{j < i} \{ \xi_{k,j} \}$ is the number of factions that are represented among the first $i-1$ subjects.  Note that taking $\alpha_2 = 0$  takes us back to the specification in \eqref{eq:DPstickbreaking}.  

%;  for example, $\Pr(\xi_{k,i} = \xi_{k,j}) = (\beta_2 + \alpha_2)/(\beta_2 + 1)$ for all $i$ and $j$
Similar comments apply to the implied joint distribution on the cluster indicators $\zeta_1, \ldots, \zeta_J$.  In particular, note that integrating out the prior weights $\bfw$ we obtain the prior probability that two networks have the same community structure, $\Pr(\zeta_{j} = \zeta_{j'} | \alpha_1, \beta_1, \alpha_2, \beta_2) = (1 - \alpha_1)/(\beta_1 + 1)$; taking $\beta_1 \to \infty$ implies that the networks are modeled independently and no information is borrowed, while taking $\beta_1 \to 0$ implies that all networks share the same community structure. In addition, note that 
\begin{multline*}
\Pr(\gamma_{j,i} = \gamma_{j,i'} | \gamma_{j',i} = \gamma_{j',i'}, \alpha_1, \beta_1, \alpha_2, \beta_2) = \frac{1 - \alpha_1}{\beta_1 + 1} + \frac{\beta_1 + \alpha_1}{\beta_1 + 1} \frac{1 - \alpha_2}{\beta_2 + 1} \\ \ge \frac{1 - \alpha_2}{\beta_2 + 1} = \Pr(\gamma_{j,i} = \gamma_{j,i'} | \alpha_2, \beta_2)
\end{multline*}
with the equality happening only if $\beta_1 \to \infty$. Hence the prior probability that two subjects belong to the same faction under the joint model is strictly larger than that implied by independently modeling each network.

The model is completed by specifying a prior distribution on the interaction matrices $\bfTheta_{1}, \ldots, \bfTheta_J$.  As in Section~\ref{se:modelproperties} we let
\begin{align}\label{eq:hierarchicalprior}
p_j(\bfTheta_{j} | \bflambda_{j} )= 
\begin{cases}
 \left\{ \prod_{k = 1}^{\infty} H_j^{\bflambda_{j,D}}(\theta_{k,k,j}) \right\} \left\{ \prod_{k = 1}^{\infty} \prod_{l = k+1}^{\infty} H_j^{\bflambda_{j,O}}(\theta_{k,l,j}) \delta_{\theta_{k,l,j}}(\theta_{l,k,j}) \right\}  & \mbox{$\bfY_j$ undirected }  \\
 \left\{ \prod_{k = 1}^{\infty} H_j^{\bflambda_{j,D}}(\theta_{k,k,j}) \right\} \left\{ \prod_{k = 1}^{\infty} \prod_{l = k+1}^{\infty} H_j^{\bflambda_{j,O}}(\theta_{k,l,j}) H_j^{\bflambda_{j,O}}(\theta_{l,k,j}) \right\}  & \mbox{$\bfY_j$ directed }  \\
 \end{cases},
\end{align}
where $\bflambda_j = (\bflambda_{j,D}, \bflambda_{j,O})$.  %Hence, unlike traditional IRMs, we allow for a different prior distribution on the diagonals and off-diagonal elements of $\bfTheta_j$.  As we discuss in the next Section, this simple extension greatly increases the flexibility and interpretability of the model. 

%A key feature of the model we just described is that networks are assumed to be exchangeable, i.e., the model is invariant to the order in which the networks are included in it.

%We are not aware of any work studying the properties of blockmodels as a prior on networks.
%In the next subsection we consider generalizations of the IRM where 1) the distribution for the prior on the diagonal elements of $\bfTheta$ are given a prior $H_D$, which is allowed to be different from the prior on the off-diagonal elements, $H_O$, and 2) the stick-breaking rations $v_1, v_2, \ldots$ are defined through a more general prior such as $v_i \sim \bet( 1 - \alpha, \beta + \alpha l )$ \citep{Pi95}.  As before, this prior distribution implies that the joint distribution can be written through a series of predictive distributions where $\xi_1 = 1$ and
%\begin{align}\label{eq:polyaurn2}
%\xi_{i} | \xi_{i-1}, \ldots, \xi_1 &\sim \sum_{k=1}^{K} \frac{m_k^{i-1} - \alpha}{\beta+ i - 1}\delta_{k} +  \frac{\beta + \alpha K}{\beta+ i - 1}\delta_{K + 1}   & 2 \le i &\le I
%\end{align}
%where $0 \le \alpha < 1$ and $\beta > -\alpha$ are two constants.  If we take $\alpha = 0$, we recover Equation \eqref{eq:polyaurn1}; values of $\alpha$ greater than 0 increase the relative probability of creating new clusters.  As a consequence, the model allows the expected number of clusters to increase at a rate $I^{-\alpha}$
%, which is strongly related to the Poisson-Dirichlet process,

\subsection{Markov chain Monte Carlo inference}\label{sec:MCMC}

%{\color{red} We need to be careful in the notation of the MCMC and use $\psi_j(y | \theta)$ rather than $p(y | \theta)$, $H_j(\theta_{k,l})$ instead of $p(\theta_{k,l})$ and to check consistency of the notation overall}

The posterior distribution implied by the model described in Section \ref{sec:multiple} is
\begin{multline}\label{eq:posterior}
p(\{ \bfTheta_j \}, \{ \bfxi_k \} , \bfzeta , \{ \bflambda_j \}, \alpha_1, \beta_1, \bfalpha_2, \bfbeta_2 | \{ \bfY_j \}) \propto \left\{ \prod_{j=1}^{J} \prod_{i=1}^{I} \prod_{i'=1, i\ne i'}^{I}  \psi_j( y_{i,i',j} | \theta_{\xi_{\zeta_j, i},\xi_{\zeta_j,i'},j})  \right\} \\
\left\{ \prod_{j=1}^{J} p_j(\bfTheta_j | \bflambda_j) \right\}  \left\{ \prod_{k=1}^{\max \{ \zeta_j \}} p( \bfxi_k | \alpha_{2,k}, \beta_{2,k}) \right\}   p(\bfzeta | \alpha_1, \beta_1) \left\{ \prod_{j=1}^{J} p(\bflambda_j) \right\} \\ 
p(\alpha_1, \beta_1) \left\{ \prod_{k=1}^{\max\{ \zeta_j \}}p(\alpha_{2,k}, \beta_{2,k}) \right\}
\end{multline}
where
\begin{align*}
p(\bfzeta | \alpha_1, \beta_1) &= \frac{\Gamma(\beta_1 + 1)}{(\beta_1 + K \alpha_1) \Gamma(\beta_1 + J) } \prod_{k=1}^{K} (\beta_1 + k \alpha_1) \frac{\Gamma(n_k - \alpha_1)}{\Gamma(1 - \alpha_1)}, & K &= \max_{j \le J} \{ \zeta_j \},  &n_k &=\sum_{j=1}^{J} \mathbf{1}_{(\zeta_j = k)}
\end{align*}
and
\begin{align}\label{eq:XIdist}
p(\bfxi_k | \alpha_{2,k}, \beta_{2,k}) & =  \frac{\Gamma(\beta_{2,k} + 1)}{  (\beta_{2,k} + L_k \alpha_{2,k}) \Gamma(\beta_{2,k} + I) } \prod_{l=1}^{L_k}  (\beta_{2,k} + l \alpha_{2,k}) \frac{\Gamma(m_{k,l} - \alpha_{2,k})}{\Gamma(1 - \alpha_{2,k})}
\end{align}
with $ L_k = \max_{i \le I} \{ \xi_{k,i} \}$,  $m_{k,l} =\sum_{i=1}^{I} \mathbf{1}_{(\xi_{k,i} = l)}$. 
This posterior distribution is computationally intractable, even if the baseline measure $H_j^{\bflambda}$ is chosen to be conjugate to the kernel $\psi_j$.  Indeed, the number of possible groups of networks and the number of factions within each group of networks grows exponentially fast with $J$ and $I$, making it impossible to explicitly enumerate all possible models.

To overcome this difficulty we develop a Markov chain Monte Carlo (MCMC) sampler to jointly explore the posterior distribution in \eqref{eq:posterior}.  We exploit the conjugacy of $\psi_j$ and $H_j$ and factorize the posterior distribution as
\begin{multline}\label{eq:posterior2}
p(\{ \bfTheta_j \}, \{ \bfxi_k \} , \bfzeta , \{ \bflambda_j \}, \alpha_1, \beta_1, \bfalpha_2, \bfbeta_2 | \{ \bfY_j \}) = \\
p(\{ \bfTheta_j \} | \{ \bfxi_k \} , \bfzeta , \{ \bflambda_j \} , \{ \bfY_j \} ) 
p(\{ \bfxi_k \} , \bfzeta , \{ \bflambda_j \}, \alpha_1, \beta_1, \bfalpha_2, \bfbeta_2 | \{ \bfY_j \}). 
\end{multline}
Sampling from $p(\{ \bfTheta_j \} | \{ \bfxi_k \} , \bfzeta , \{ \bflambda_j \}, \alpha_1, \beta_1, \bfalpha_2, \bfbeta_2, \{ \bfY_j \})$ is straightforward under conjugacy.  To sample from the marginal posterior $p(\{ \bfxi_k \} , \bfzeta , \{ \bflambda_j \}, \alpha_1, \beta_1, \bfalpha_2, \bfbeta_2 | \{ \bfY_j \})$ we iteratively sample from the following six sets of full conditional distributions 
\begin{enumerate}
\item $p(\bfzeta | \{ \bfxi_k \} , \{ \bflambda_j \} , \alpha_1, \beta_1, \bfalpha_2, \bfbeta_2, \{ \bfY_j \})$.

\item $p(\{ \bfxi_k \} , \bfzeta | \{ \bflambda_j \} , \alpha_1, \beta_1, \bfalpha_2, \bfbeta_2, \{ \bfY_j \})$.

\item $p(\bfxi_k | \bfzeta , \{ \bflambda_j \} , \alpha_1, \beta_1, \bfalpha_2, \bfbeta_2, \{ \bfY_j \})$ for $k = 1,\ldots, \max_{j \le J} \{ \zeta_j \}$.

\item $p( \{ \bflambda_j \}  | \{ \bfxi_k \} , \bfzeta, \{ \bfY_j \})$ for $j=1,\ldots, J$.

\item $p( \alpha_1, \beta_1  |  \bfzeta )$.

\item $p( \{ \alpha_{2,k}, \beta_{2,k} \}  |  \bfxi_k)$ for $k = 1,\ldots, \max_{j \le J} \{ \zeta_j \}$.
\end{enumerate}

To sample from $p(\{ \bfxi_k \} , \bfzeta | \{ \bflambda_j \} , \alpha_1, \beta_1, \bfalpha_2, \bfbeta_2, \{ \bfY_j \})$ we develop a split-merge algorithm that combines ideas from \cite{Da03} and \cite{JaNe04}.  More specifically, at each iteration of the MCMC we randomly select two networks.  If they currently belong to two separate clusters we propose to merge them into a single cluster.  On the other hand, if they belong to the same cluster we propose to split it into two different clusters.  The faction structure within each cluster of networks is proposed by sequentially allocating actors to factions, in the spirit \cite{Da03}.  A similar approach is employed to sample from $p(\bfxi_k | \bfzeta , \{ \bflambda_j \} , \alpha_1, \beta_1, \bfalpha_2, \bfbeta_2, \{ \bfY_j \})$.  These long-range moves are combined with more traditional short-range moves that individually update each component of $\bfzeta$ and $\bfxi_{k}$ given the rest of the components.  Details on the algorithm can be seen in Appendix \ref{se:MCMCdetails}.
%\begin{align*}
%p(\{ \bfxi_k \} , \bfzeta | \{ \bflambda_j \} , \alpha_1, \beta_1, \alpha_2, \beta_2 \{ \bfY_j \}) \\
%\end{align*}
%
%\begin{align*}
%p(\{ \{ \bflambda_j \} | \bfxi_k \} , \bfzeta , \alpha_1, \beta_1, \alpha_2, \beta_2 \{ \bfY_j \}) \\
%\end{align*}
%
%\begin{align*}
%p(\{ \alpha_1, \beta_1  |  \bfzeta )
%\end{align*}
%and
%\begin{align*}
%p(\{ \{ \alpha_{2,k}, \beta_{2,k} \}  |  \bfxi_k)
%\end{align*}

The posterior distributions associated with $\bfzeta$ and $\{ \bfxi_k \}$ can be summarized through posterior pairwise incidence matrices.  For example, the pairwise incidence matrix associated with $\bfzeta$, $\mathbf{D}_{\zeta}$ is an $J \times J$ matrix such that $[\mathbf{D}_{\zeta}]_{j,j'} = \Pr(\zeta_j = \zeta_{j'} | \bfY_1, \ldots, \bfY_J)$.  To obtain point estimates of the partition structure we follow \cite{LaGr06} and take a decision theoretic approach.  For example, a point estimator $\tilde{\bfzeta}$ for $\bfzeta$ is obtained by minimizing the expected loss function
\begin{align}\label{eq:expecutil}
L(\tilde{\bfzeta}) &=  \E \left\{ \sum_{j=1}^{J} \sum_{j'=j+1}^{J} \left[ a \mathbf{1}_{(\zeta_j = \bfzeta_{j'}, \tilde{\zeta}_j \ne \tilde{\bfzeta}_{j'})} + b \mathbf{1}_{(\zeta_j \ne \bfzeta_{j'}, \tilde{\zeta}_j = \tilde{\bfzeta}_{j'})} \right] \mid \bfY_1, \ldots, \bfY_J \right\}
\end{align}
Minimizing \eqref{eq:expecutil} is equivalent to maximizing 
$$
U(\tilde{\bfzeta}) =  \sum_{j=1}^{J} \sum_{j'=j+1}^{J} \mathbf{1}_{(\tilde{\zeta}_j = \tilde{\bfzeta}_{j'})} \left\{ \Pr(\zeta_j = \zeta_{j'} | \bfY_1, \ldots, \bfY_J) - \frac{b}{a+b} \right\}
$$
The constants $a$ and $b$ represent the costs of misclassification errors; setting $b=0$ leads to a point estimate that includes all networks into a single partition, while $a=0$ leads to a point estimate the places each network into an individual partition.

\section{Illustrations}

\subsection{Simulation study}
            \begin{figure}[htbp]
	\begin{center}
          \begin{tabular}{c c | cc }
	\multicolumn{2}{c |}{\Large Individual Network Model} & \multicolumn{2}{ c}{\Large Multiple Network Model }\\
	&& &\\
	\multicolumn{2}{c |}{Class 1} & \multicolumn{2}{c}{Class 1} \\
	 { \small Network 1 } & {\small Network 6 } &  { \small Network 1} & {\small Network 6 } \\
	  \multicolumn{2}{c |}{  \subfigure{\includegraphics[height=3.8cm,angle=0]{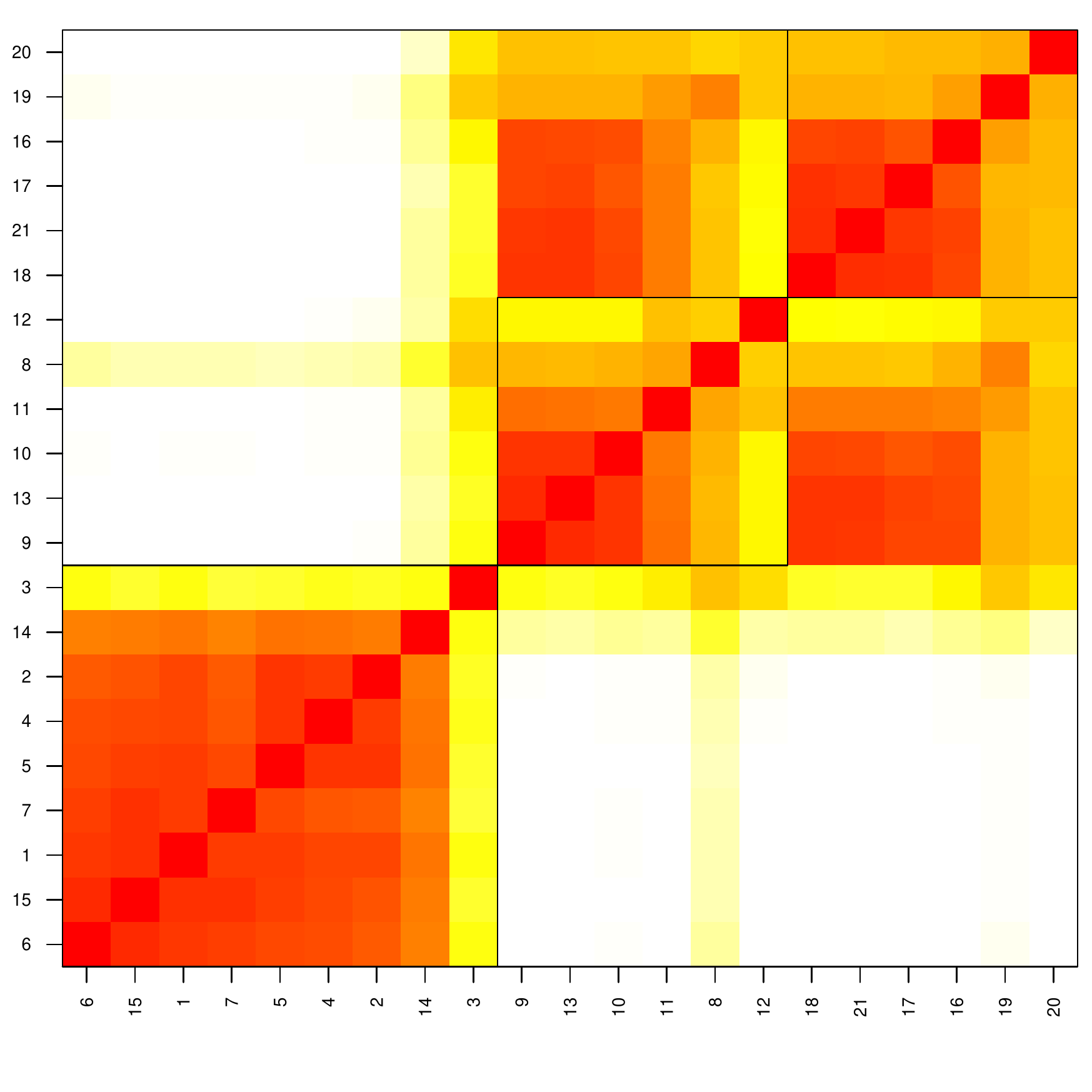}}
          \subfigure{\includegraphics[height=3.8cm,angle=0]{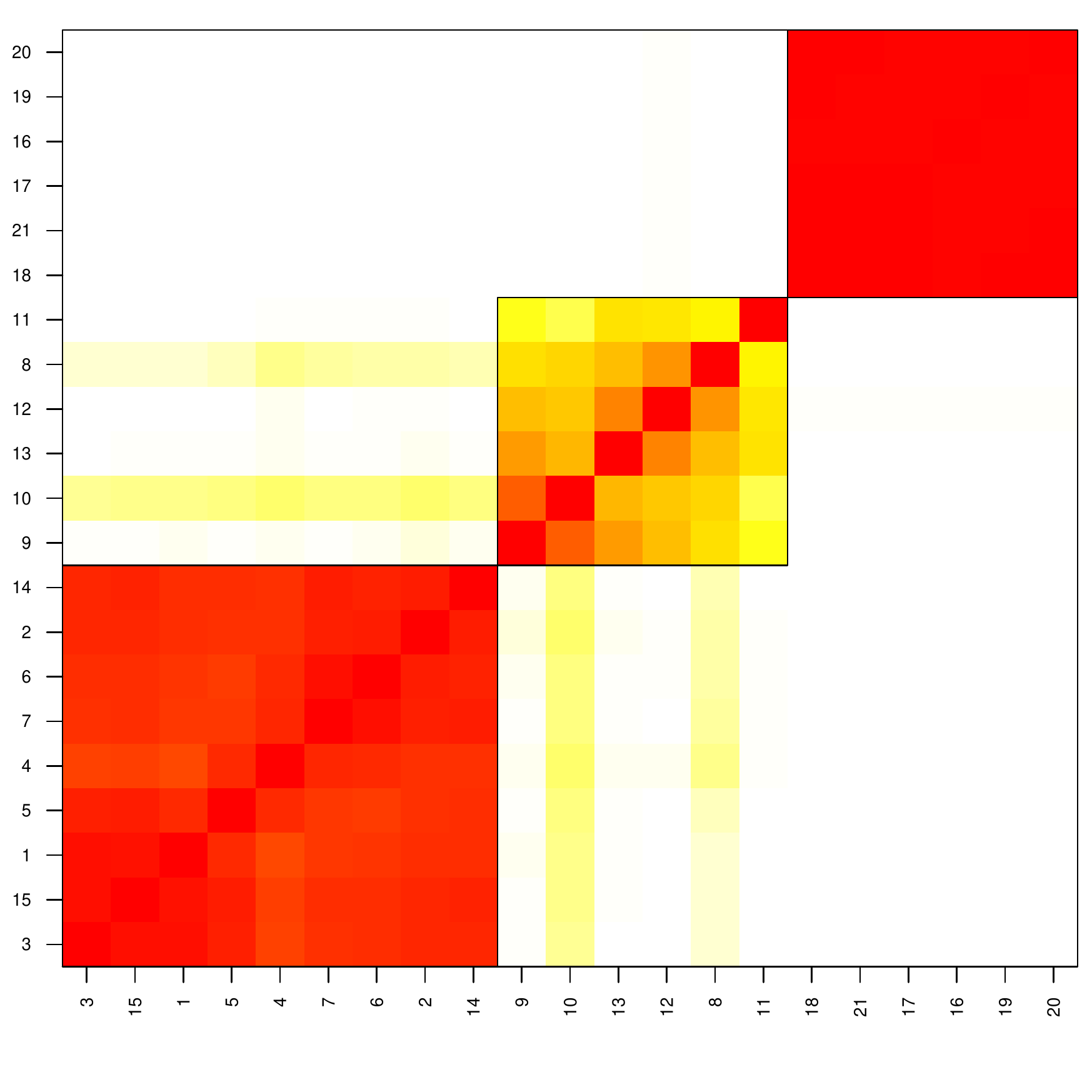}} 
          }& 
          \multicolumn{2}{c}{ \subfigure{\includegraphics[height=3.8cm,angle=0]{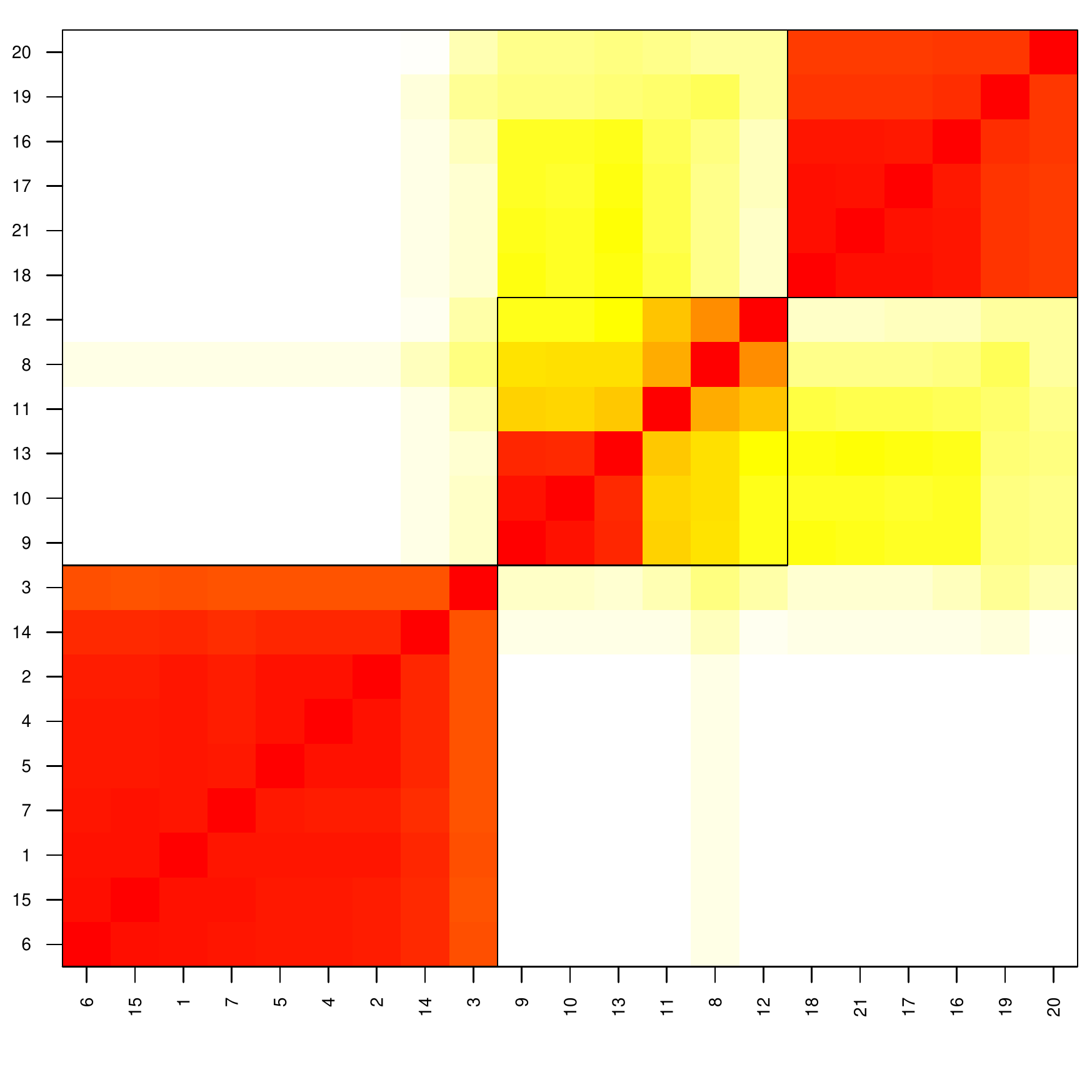}}
          \subfigure{\includegraphics[height=3.8cm,angle=0]{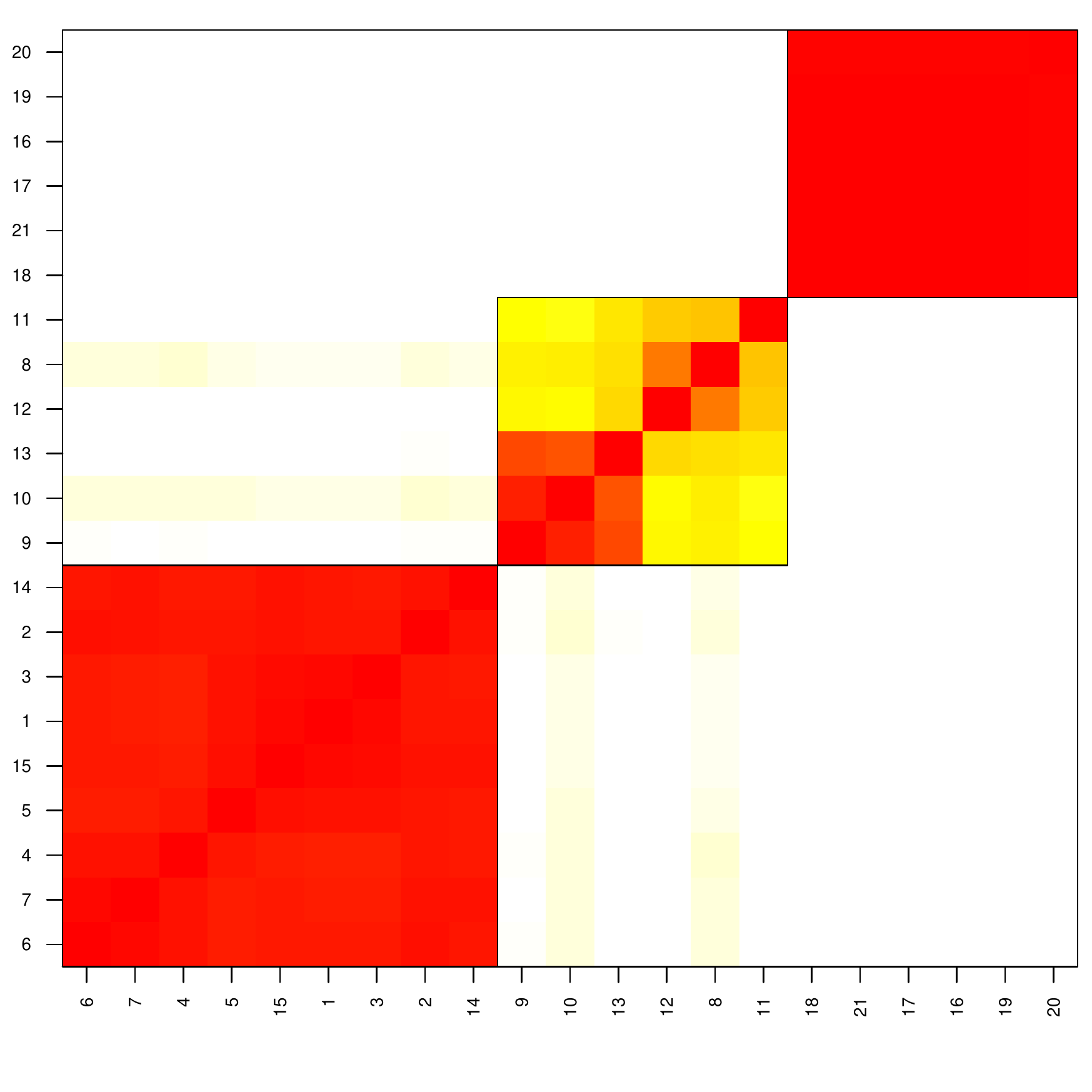}}
          }\\
	\multicolumn{2}{c |}{Class 2} & \multicolumn{2}{c }{Class 2} \\
             { \small Network 2} & { \small Network 5 } &  { \small Network 2} &{ \small Network 5 } \\
            \multicolumn{2}{c |}{ \subfigure{\includegraphics[height=3.8cm,angle=0]{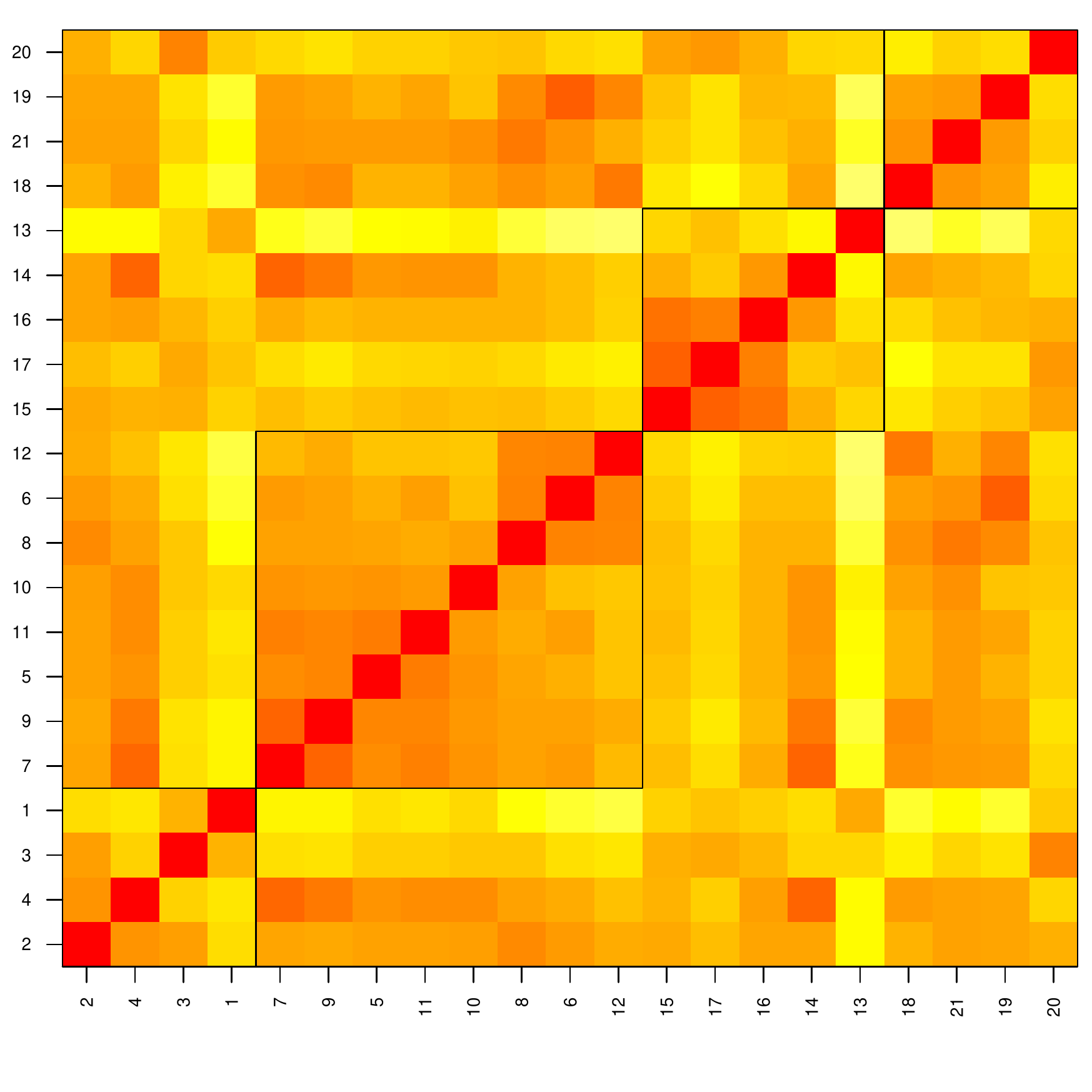}}
          \subfigure{\includegraphics[height=3.8cm,angle=0]{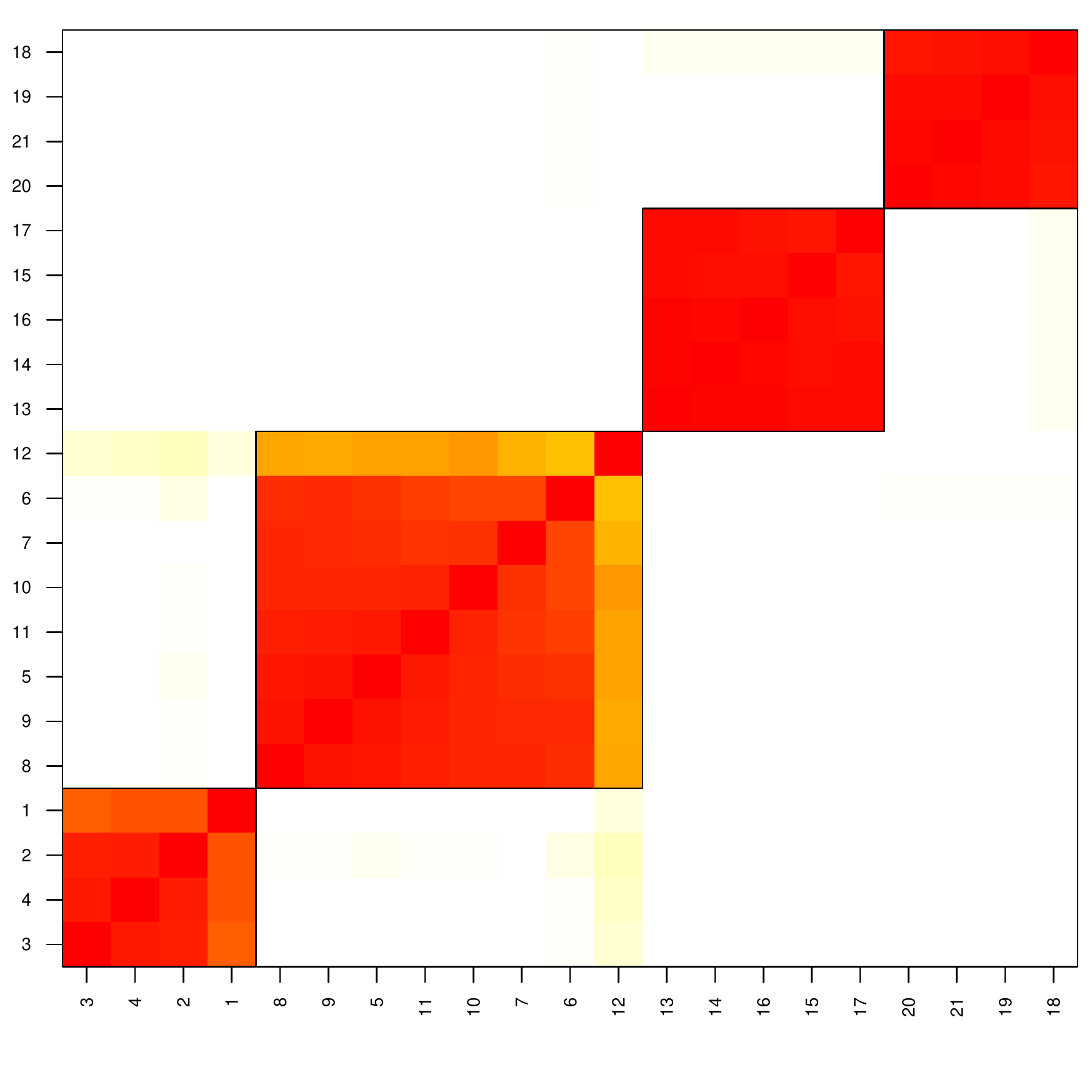}}
	} &            
          \multicolumn{2}{c }{ \subfigure{\includegraphics[height=3.8cm,angle=0]{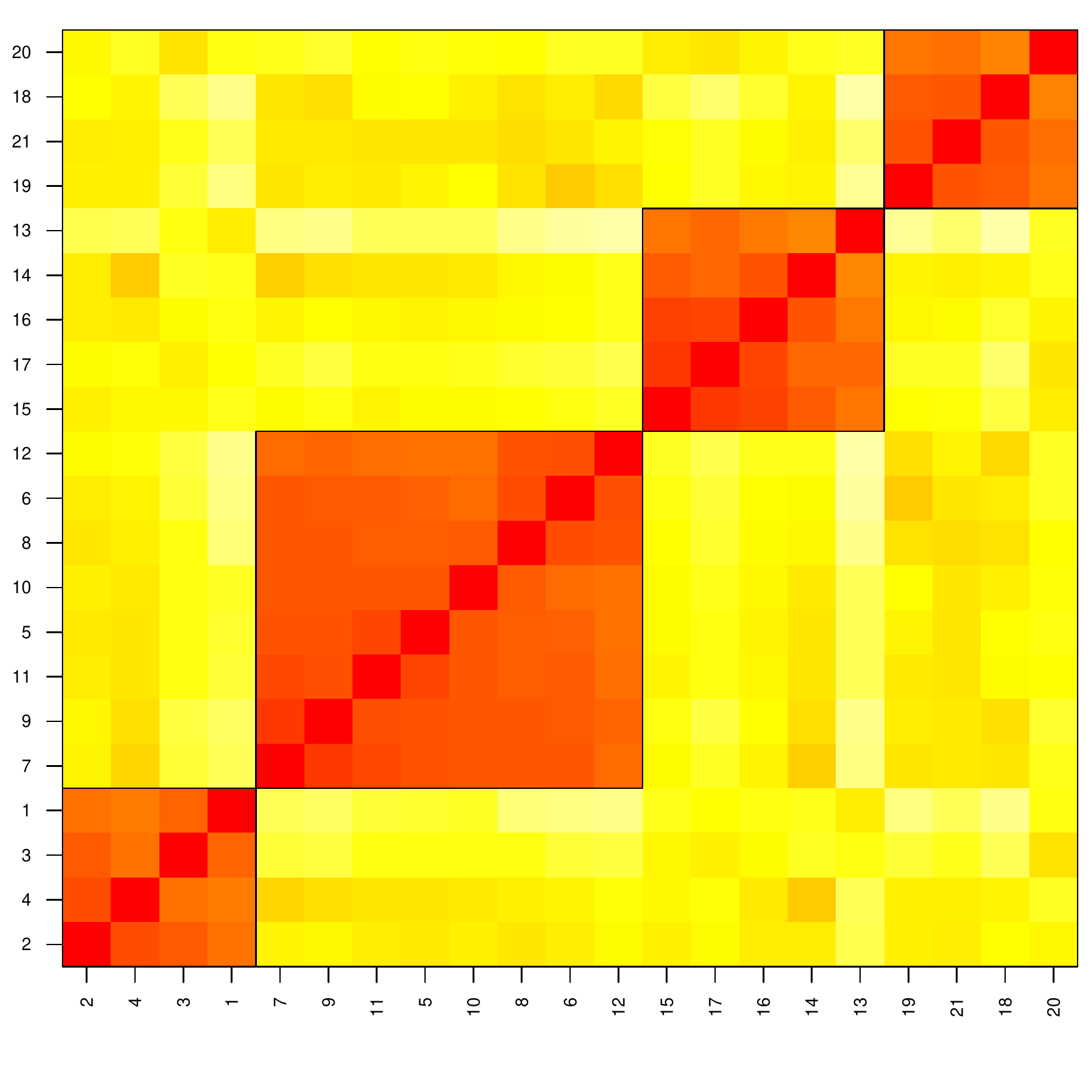}}
          \subfigure{\includegraphics[height=3.8cm,angle=0]{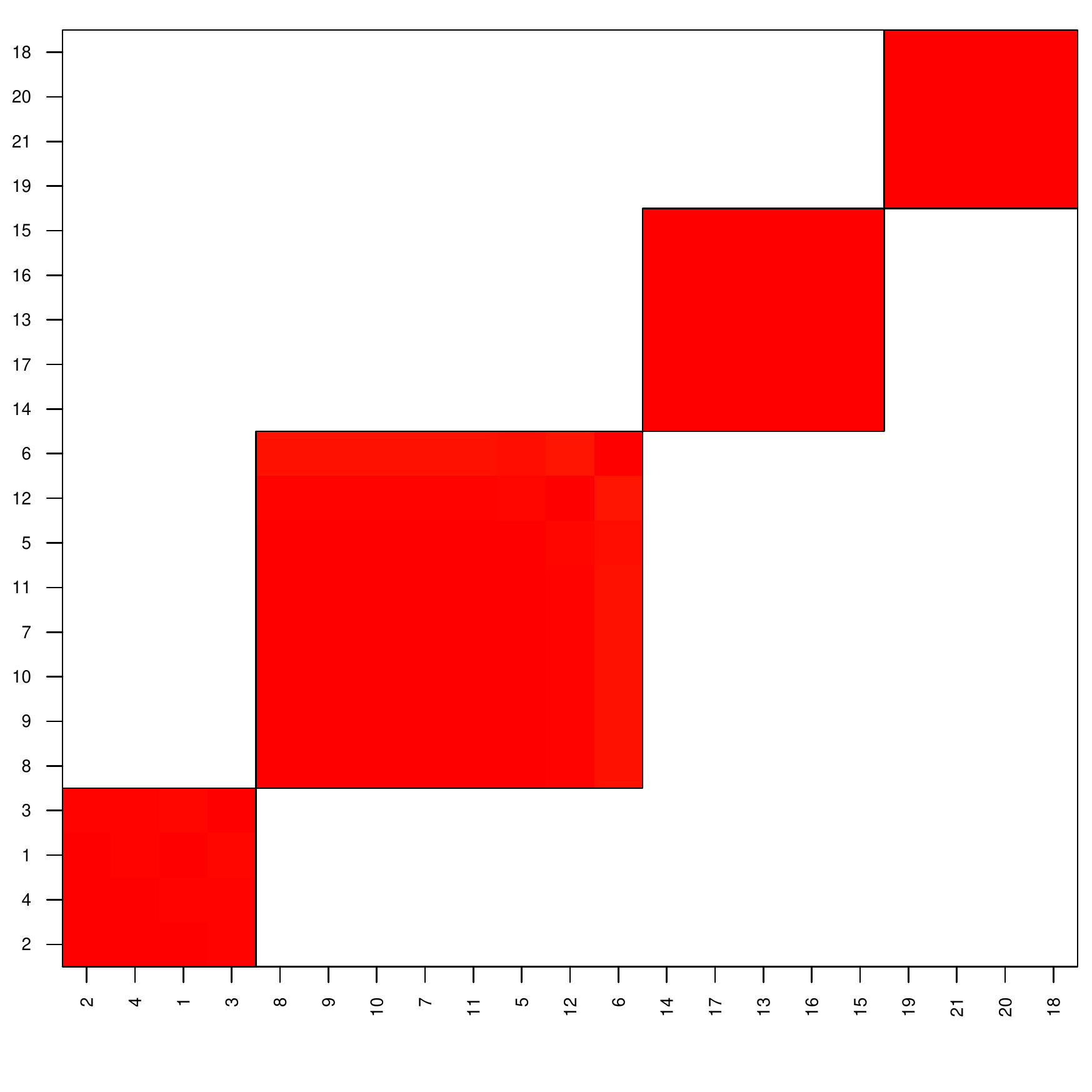}}
          } \\
            \multicolumn{2}{c |}{\small Network 7} &   \multicolumn{2}{c }{\small Network 7} \\
           \multicolumn{2}{c |}{\subfigure{\includegraphics[height=3.8cm,angle=0]{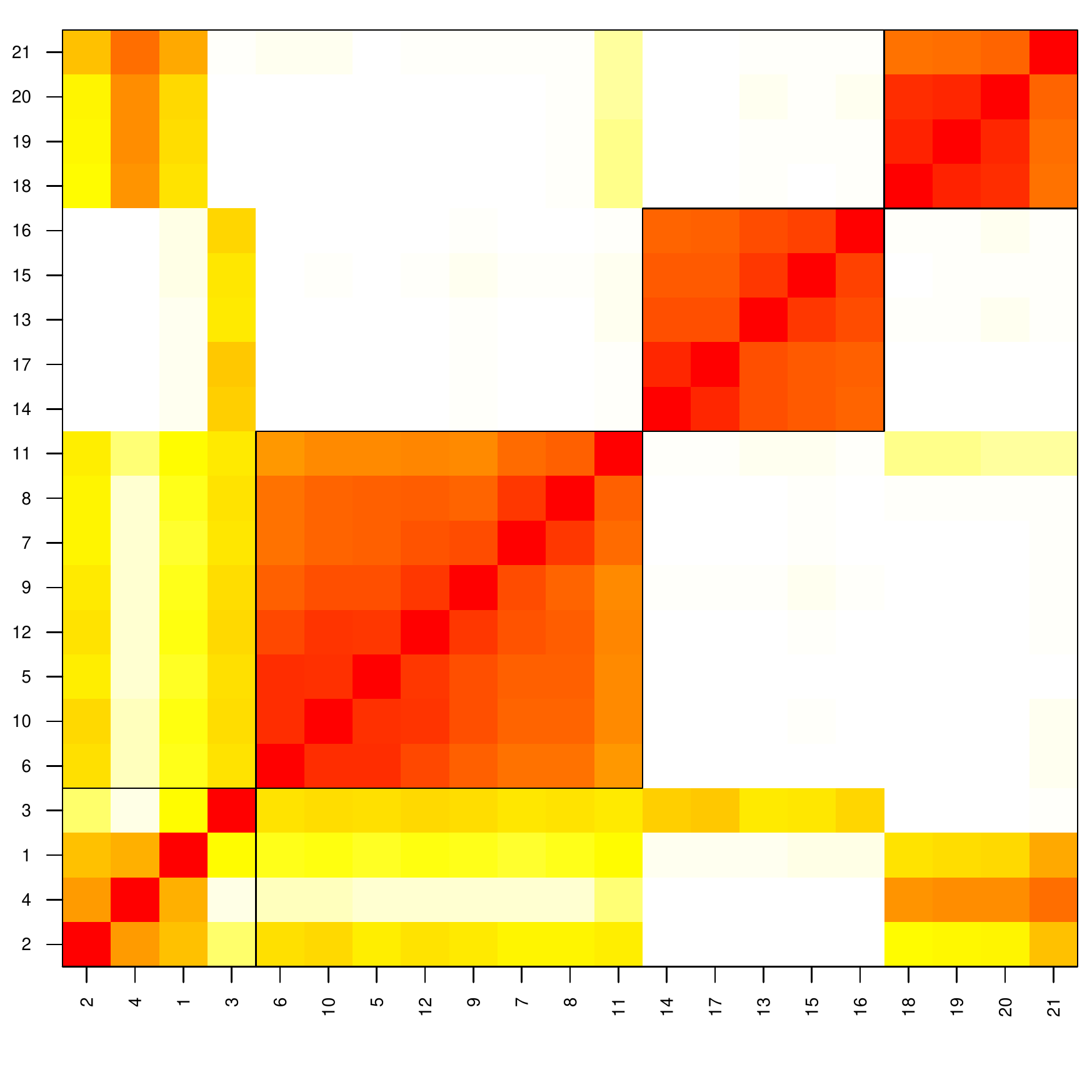}} 
           } &
           \multicolumn{2}{c }{\subfigure{\includegraphics[height=3.8cm,angle=0]{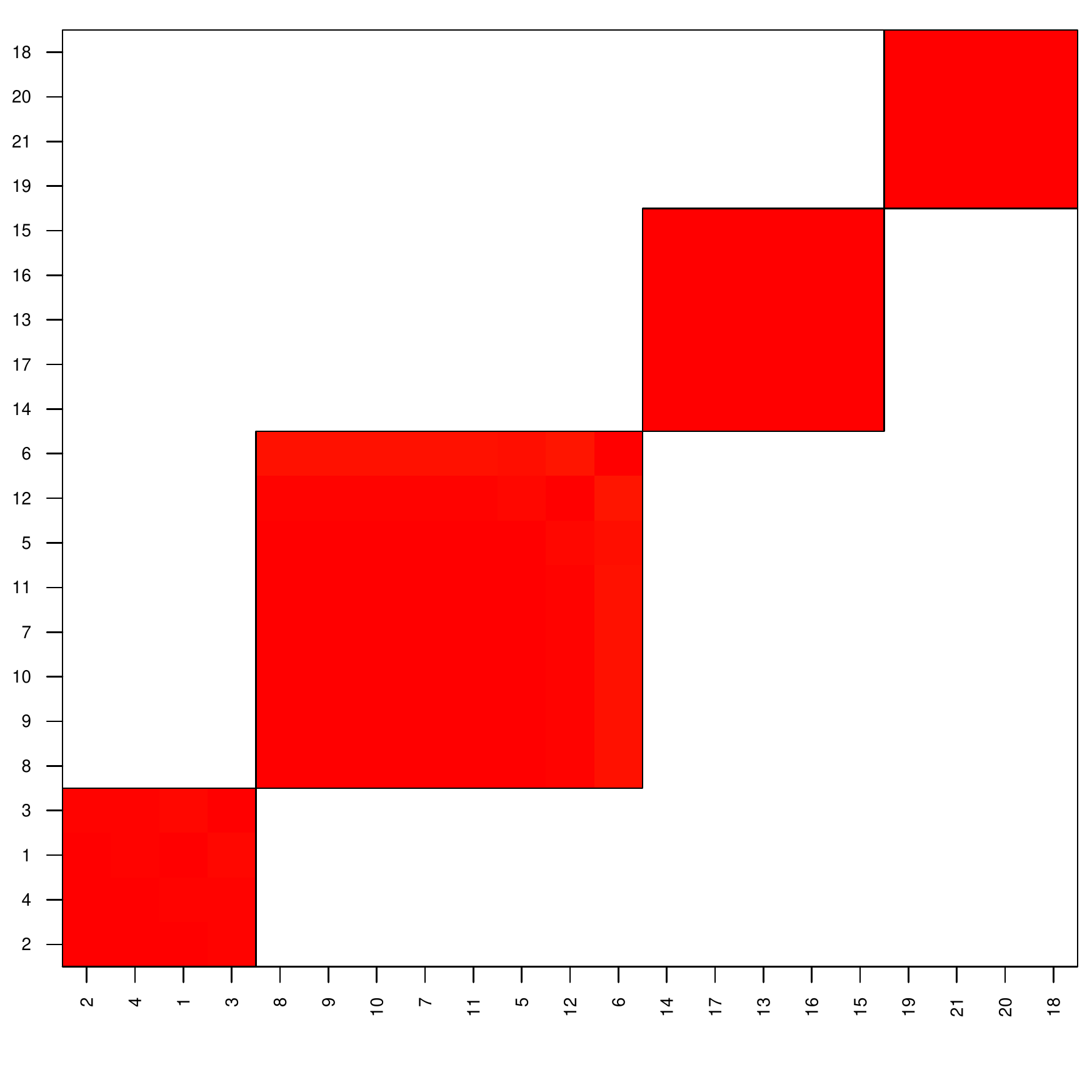}} 
           }\\
	   Class 3 & Class 4  &  	   Class 3 & Class 4  \\
	&& &\\
	 { \small Network 3 } & { \small Network 4 } &  { \small Network 3} & { \small Network 4 } \\
      \subfigure{\includegraphics[height=3.8cm,angle=0]{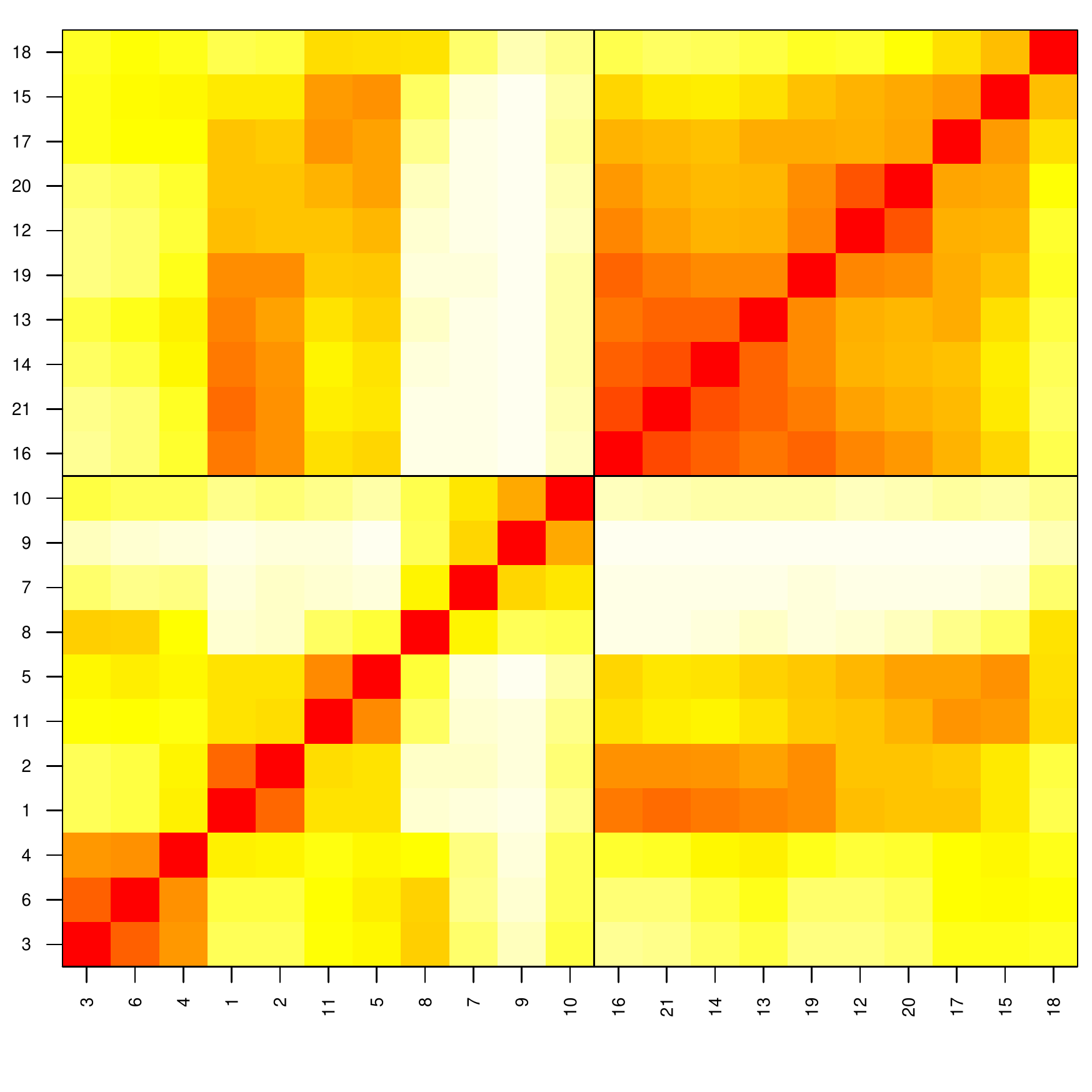}} &
          \subfigure{\includegraphics[height=3.8cm,angle=0]{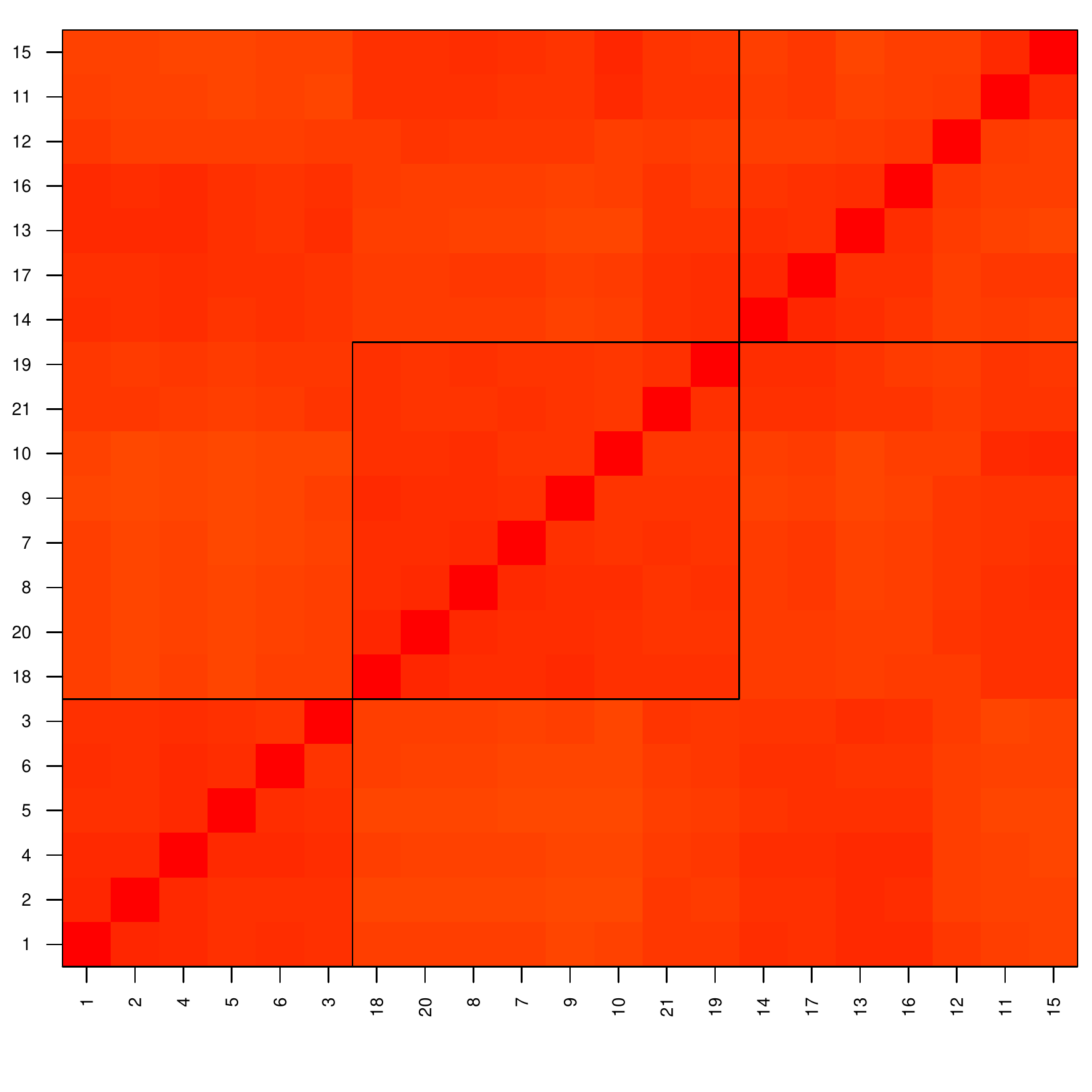}}
          &  
                \subfigure{\includegraphics[height=3.8cm,angle=0]{figures/Test_Gibbs_XI3.pdf}}&
          \subfigure{\includegraphics[height=3.8cm,angle=0]{figures/Test_Gibbs_XI4.pdf}}\\

	\end{tabular}
         \includegraphics[height=0.9cm,angle=0]{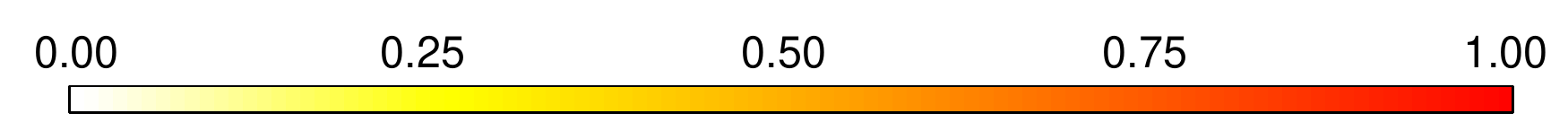}\\
         \caption{Simulation Results. Estimated posterior probability that two actors belong to the same group for each network. The squares mark the real groups used to create the data.}\label{fi:simul}
          \end{center}
          \end{figure}
         
          A simulation study was conducted to compare the results between modeling a collection of networks individually applying model (\ref{eq:blockmodel}) or using the model proposed in Section~\ref{sec:multiple}. Seven networks with 21 actors each were simulated. The first four are binary and undirected, the fifth is Poisson undirected, and the last two are binary and directed (this setting is similar to the real data we consider in Section \ref{se:wiring}). The simulated networks group as follow (1,6), (2,5,7), (3), (4).           

          Figure~\ref{fi:simul} presents the results in terms of the estimated posterior probability that two actors belong to the same group. Comparing in particular the first networks of classes 1 and 2, we can observe that those networks borrow strength from the other networks on the group to get a better estimation of the factions when the multiple network model is used. Moreover, by looking to the estimation on left-hand side of figure \ref{fi:simul} alone, it would have been difficult to identify the similitude of the networks 1 and 2 to the rest in classes 1 and 2, respectively.  For classes 4 and 5, there is not sufficient information on the data to detect the true factions, notice that in this case both modeling techniques are equivalent, and considering MCMC error, they lead to similar results as expected.
          
          This experiment provides evidence that in the presence of similar factions for two or more networks, the estimation benefits from modeling multiple networks simultaneously. Moreover, the model does not force artificial groupings of networks and eventually leads to equivalent results when networks need to be modeled independently. 
 
\subsection{Modeling cognitive social structures:  The Krackhardt dataset}\label{se:wiring}

\begin{table}[htb]
\caption{Krackhardt data. Manager's attributes by class}\label{tb:attributes}
\begin{center}
\begin{tabular}{rr rr cc}
\hline \\
Class & Actor & Age & Tenure & Department & Level \\
\\ 
\hline \\
1 & 4 & 33 & 7.5 & 4 & 3 \\
 & 18 & 33 & 9.1 & 3 & 2 \\
 & 21 & 36 & 12.5 & 1 & 2 \\
 & 16 & 27 & 4.7 & 4 & 3 \\
 & 2 & 42 & 19.6 & 4 & 2 \\
 & 8 & 34 & 11.3 & 1 & 3 \\
 & 12 & 34 & 8.9 & 1 & 3 \\
 & 1 & 33 & 9.3 & 4 & 3 \\
 & 15 & 40 & 8.4 & 2 & 3 \\
\\ 
\hline \\
2 & 3 & 40 & 12.8 & 2 & 3 \\
 & 9 & 62 & 5.4 & 2 & 3 \\
\\ 
\hline \\
3 & 5 & 32 & 3.3 & 2 & 3 \\
 & 19 & 32 & 4.8 & 2 & 3 \\
 & 14 & 43 & 10.4 & 2 & 2 \\
\\ 
\hline \\
4 & 6 & 59 & 28.0 & 1 & 3 \\
5 & 7 & 55 & 30.0 & 0 & 1 \\
6 & 10 & 37 & 9.3 & 3 & 3 \\
7 & 11 & 46 & 27.0 & 3 & 3 \\
8 & 13 & 48 & 0.3 & 2 & 3 \\
9 & 17 & 30 & 12.4 & 1 & 3 \\
10 & 20 & 38 & 11.7 & 2 & 3 \\
\\ 
\hline 
\end{tabular}
\end{center}
\end{table}%

\cite{Kr87} presents information about the relationship among management personnel of a small manufacturing company producing high-tech machinery
in the west coast of the U.S. At the time, it had about one hundred employees, including 21 managers (CEO, 4 Vice-presidents and 16 managers) that are the set of actors of this example.  We focus on friendship relationships; each manager was asked not only about their connections (``Who are your friends?''), but also about their perception of other manager's connections (``Who is a friend of ... ?''), resulting in 21 networks, one for each actor's perception of the friendship network.  %Here we concentrate in friendship to illustrate our technique. Krackhardt was particularly interested in the perceptions that actor had of the entire network. Therefore, e

The multiple networks model presented in Section~\ref{sec:multiple} groups networks accordingly with how actors form factions in each network. Three classes were identified with 9, 2 and 3 networks, while the other 7 networks were left ungrouped.   Many of these singleton clusters arise because the individuals see themselves as friends of everybody else, a vision that is not shared by the other members of the network.  Table~\ref{tb:attributes} presents additional information on the managers listed accordingly with the classes obtained by the proposed technique. The four attributes for the managers obtained from table B.4 in \cite{WaFa94} are age, tenure (time employed by the company, in years), department and level in the hierarchy (1: CEO, 2: vice-presidents and 3: managers).  Note that all the managers in department 4 belong to class one, while most of department 2 subjects are in classes two and three. Class 3 is formed by department 2's manager and its two newest employees. In general, newer employees appear to share the perception of others, either with more tenure or with more authority. The only exception, subject 13 may be too new in the company to have been influenced.  

\begin{figure}[htbp]
	\begin{center}
          \begin{tabular}{c@{}c@{}c @{ }c}
          Factions & Network & $P(y_{i,i',j} = 1)$ &
        \multirow{8}{*}{ \includegraphics[height=19cm,width=1cm,angle=0]{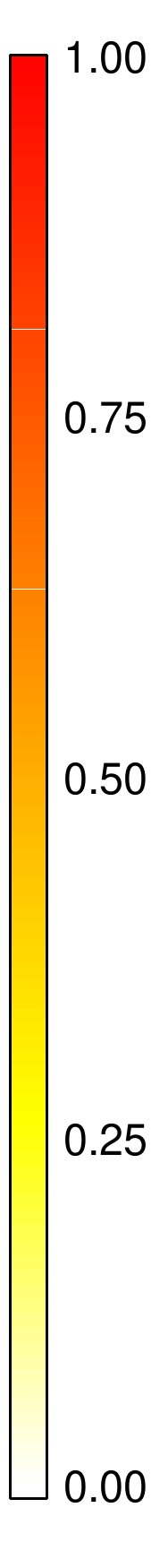} } \\
	\begin{sideways}  \qquad \qquad \; Class 1  \end{sideways} 
	\begin{sideways}  \qquad \qquad Subject 4  \end{sideways} 
	\subfigure{\includegraphics[width=0.27\textwidth,angle=0]{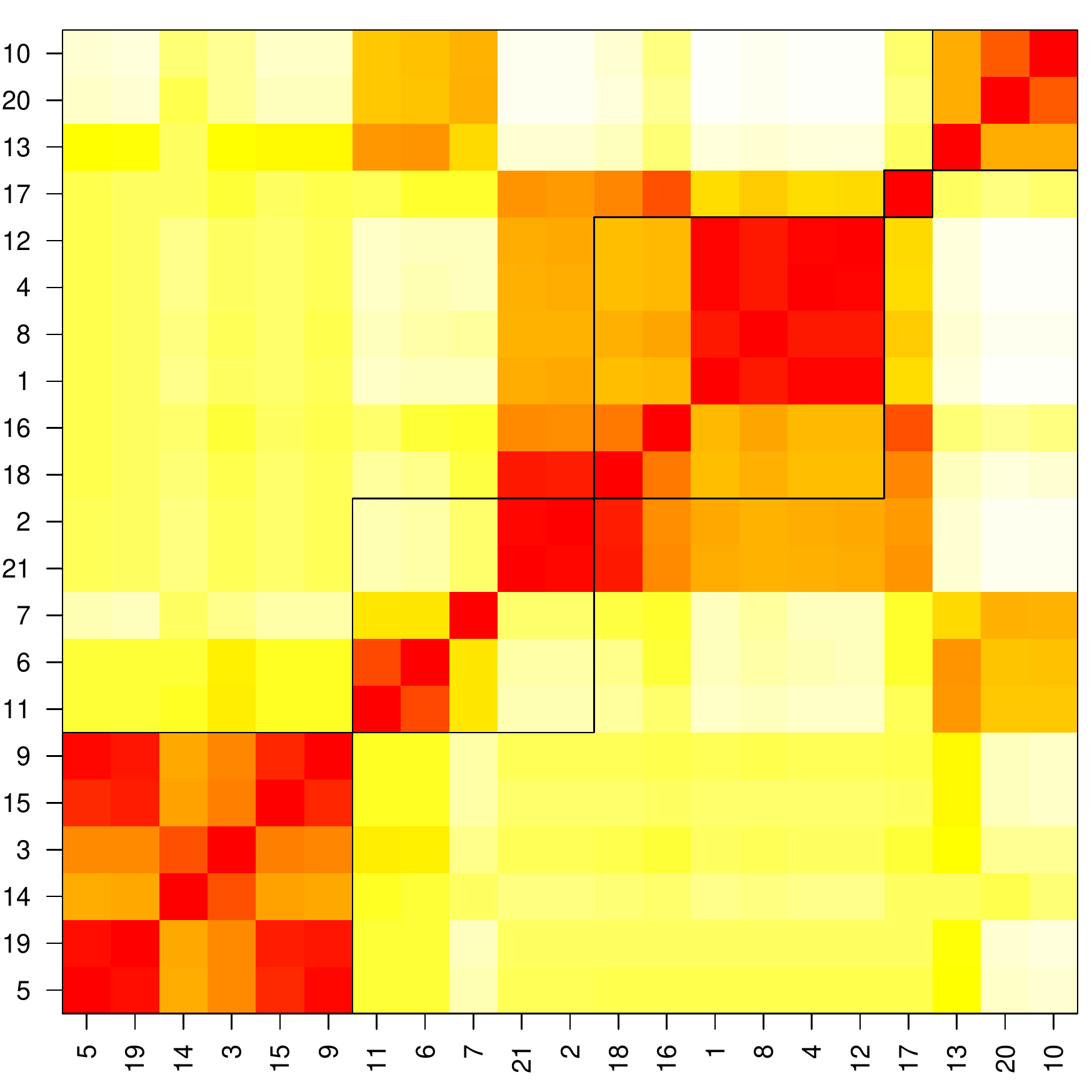} }& 
	\subfigure{\includegraphics[width=0.27\textwidth,angle=0]{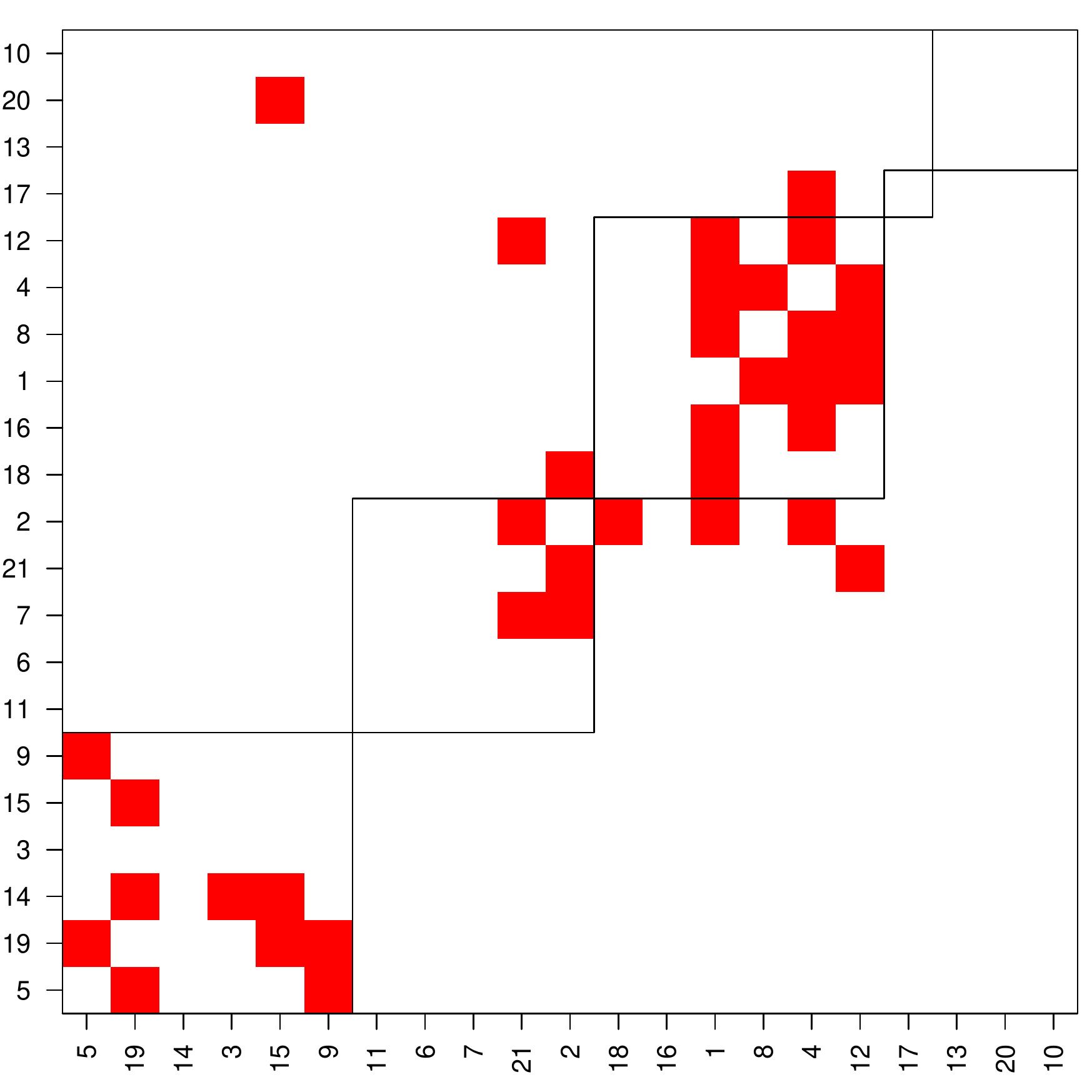} }& 
	\subfigure{\includegraphics[width=0.27\textwidth,angle=0]{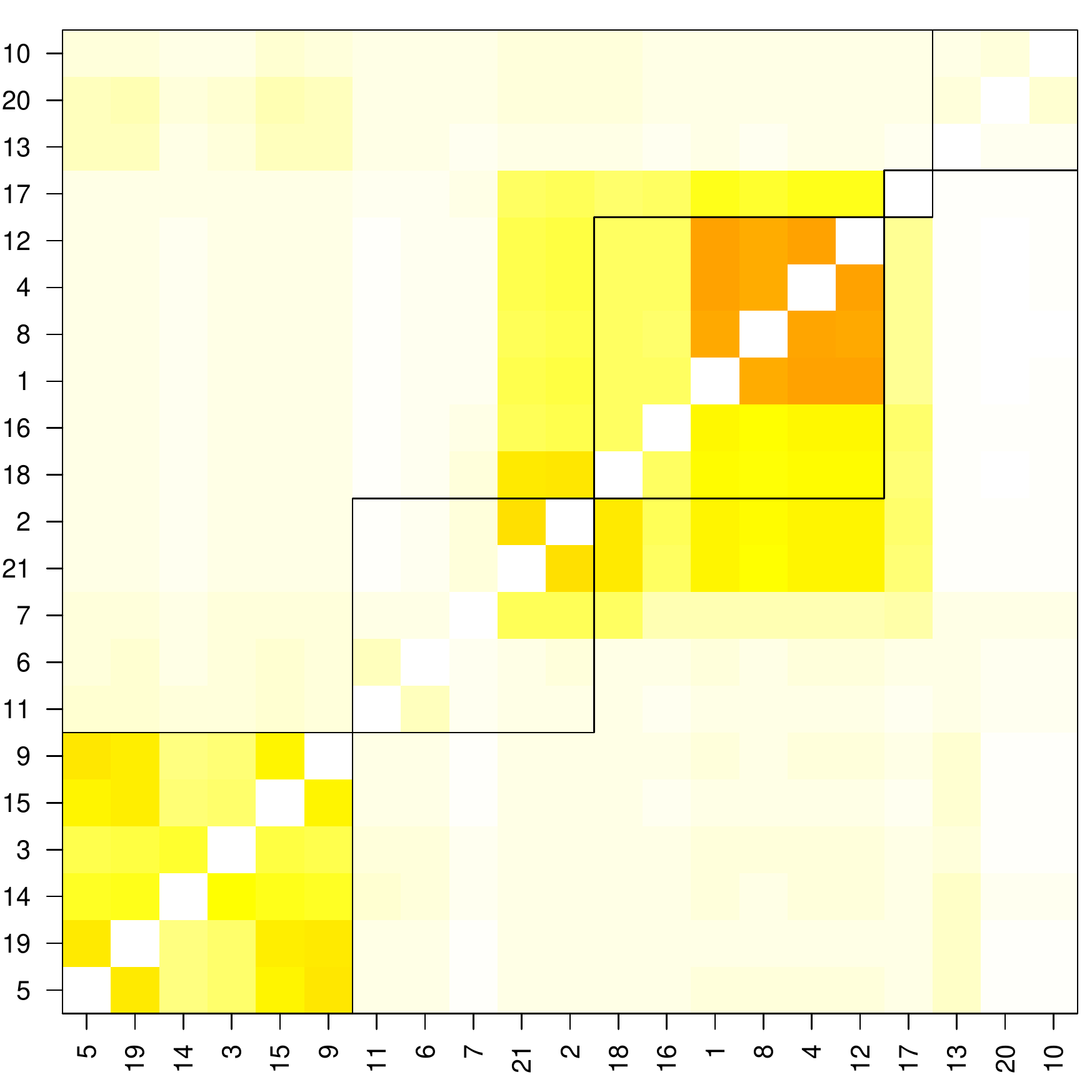} } \\
	%%%
	\begin{sideways}  \qquad \qquad \; Class 2  \end{sideways} 
	\begin{sideways}  \qquad \qquad Subject 3  \end{sideways} 
	\subfigure{\includegraphics[width=0.27\textwidth,angle=0]{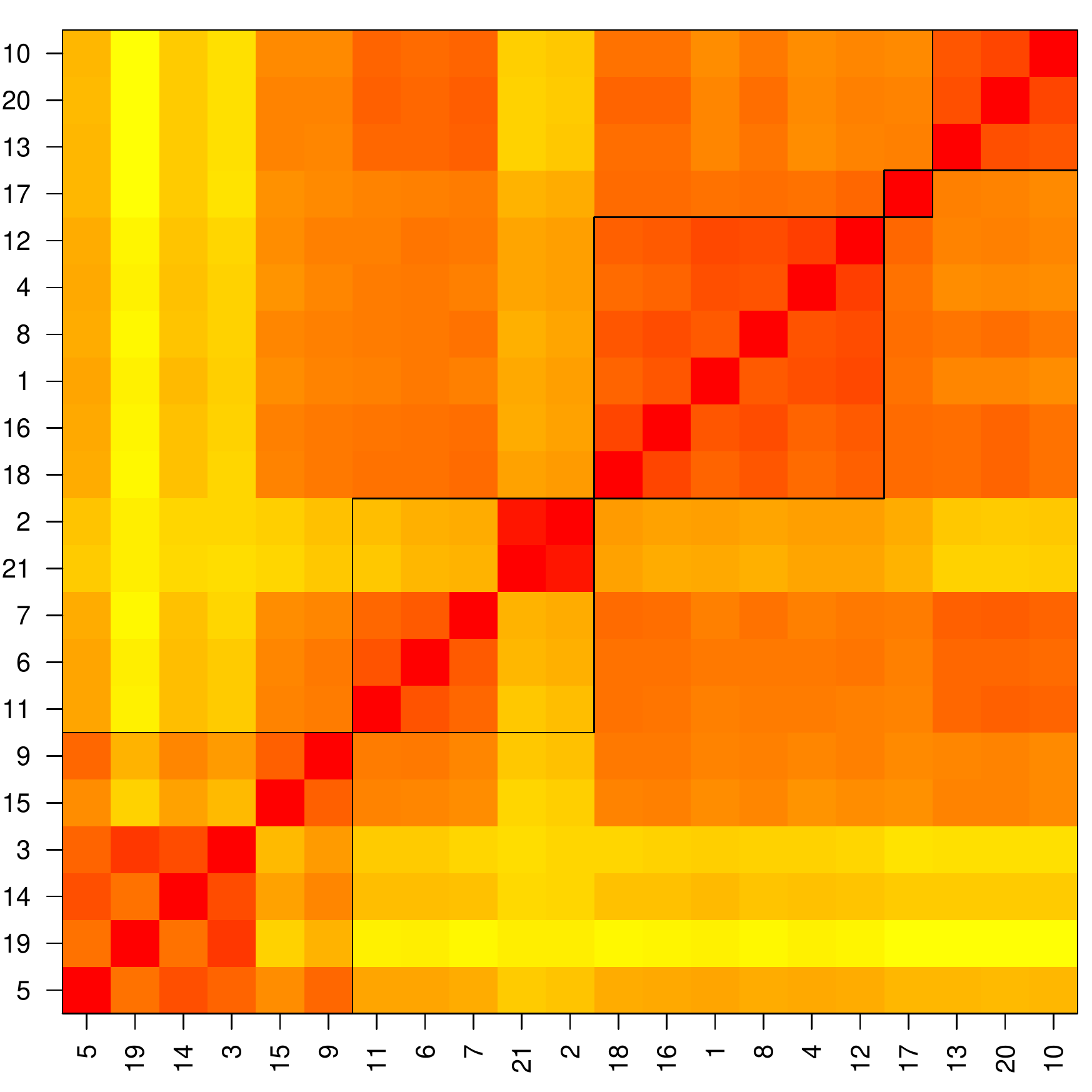} }& 
	\subfigure{\includegraphics[width=0.27\textwidth,angle=0]{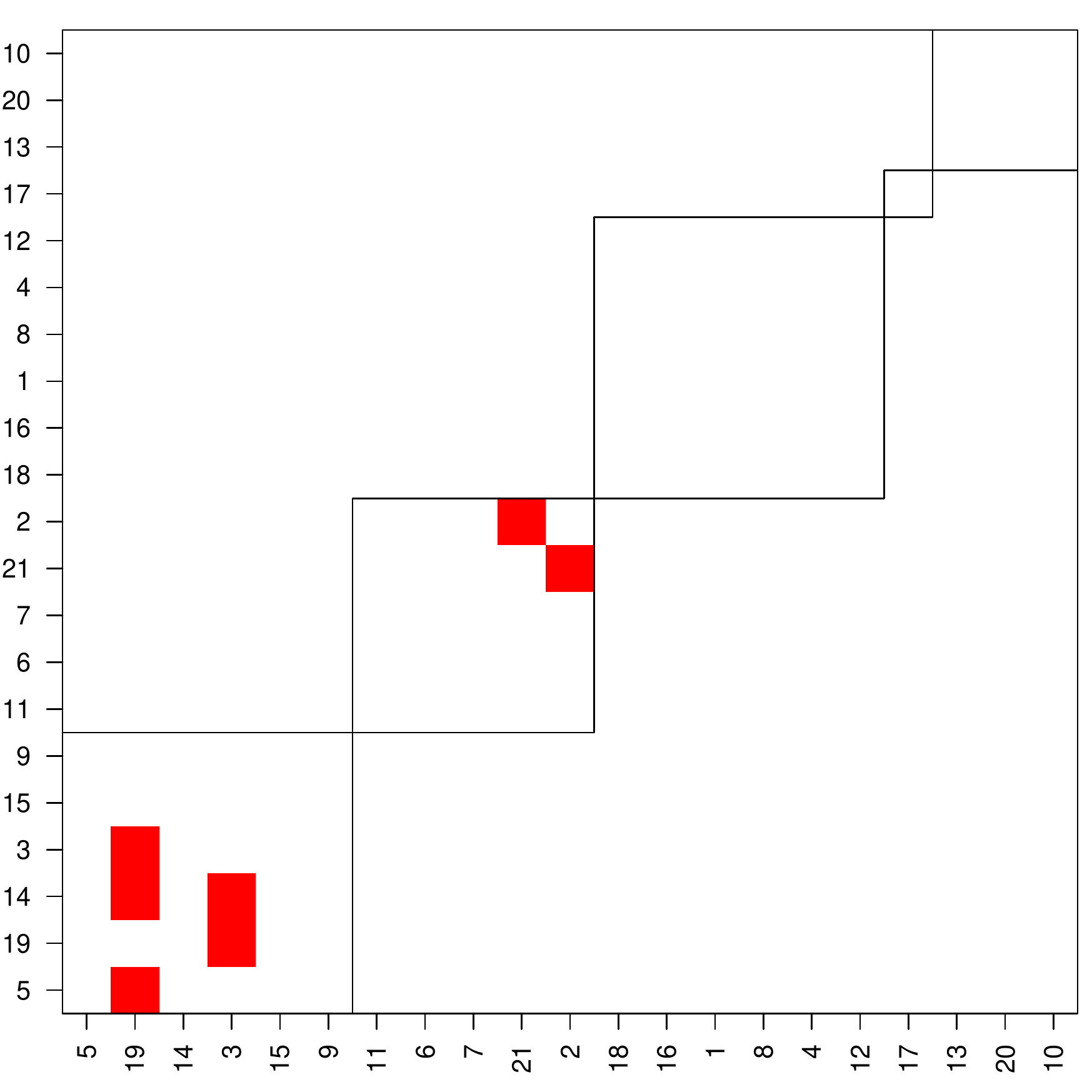} }& 
	\subfigure{\includegraphics[width=0.27\textwidth,angle=0]{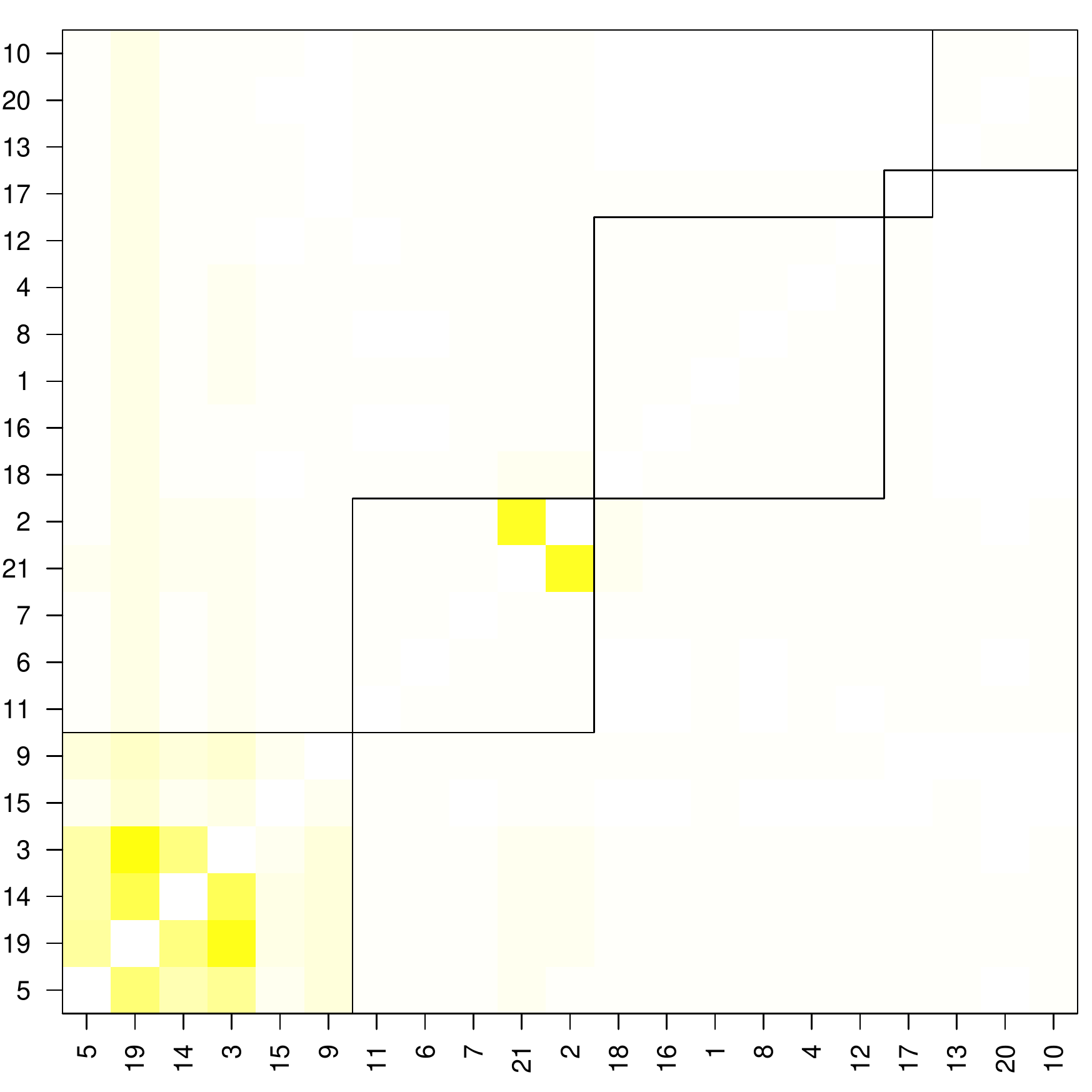} } \\
	%%%
	\begin{sideways}  \qquad \qquad \; Class 3  \end{sideways} 
	\begin{sideways}  \qquad \qquad Subject 5  \end{sideways} 
	\subfigure{\includegraphics[width=0.27\textwidth,angle=0]{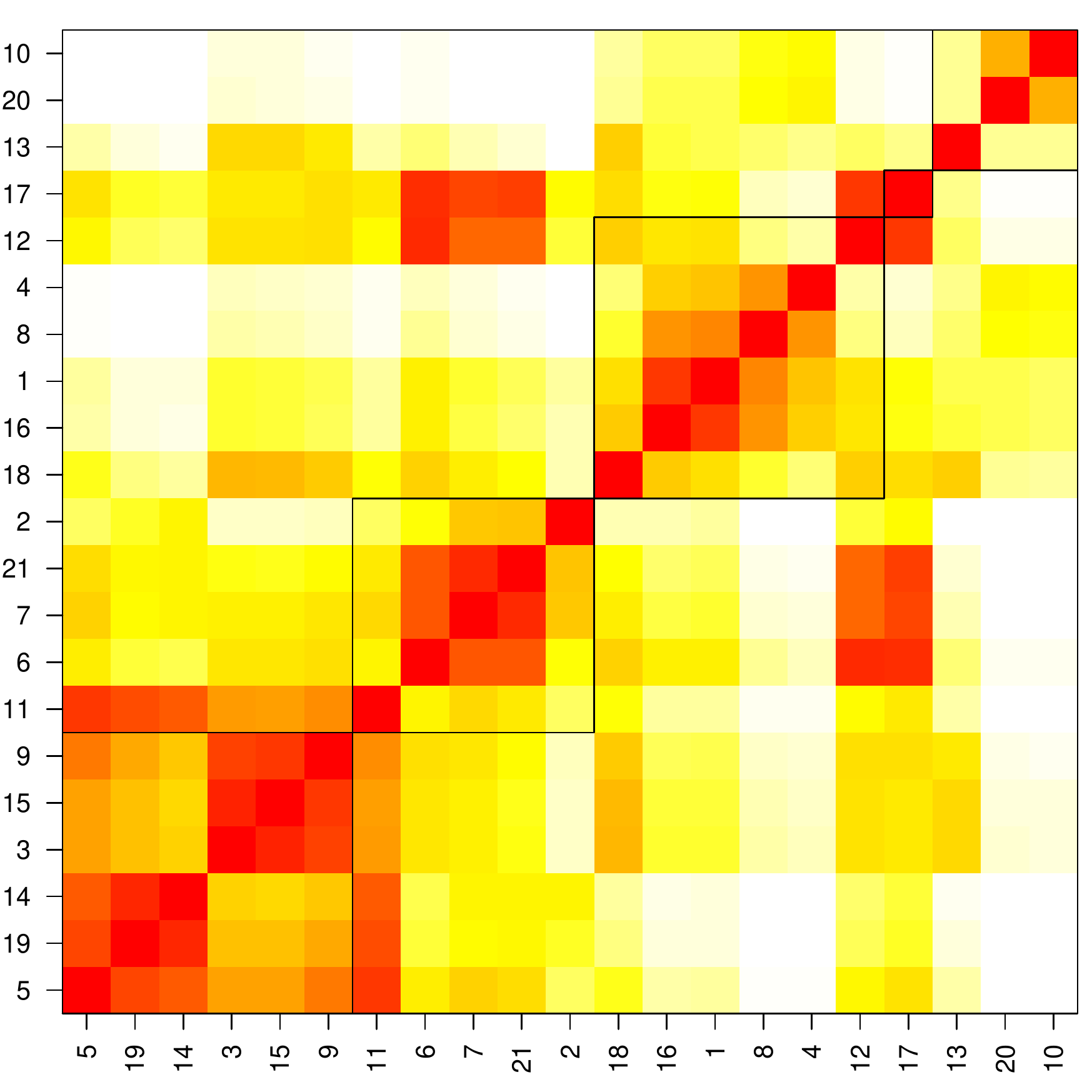} }& 
	\subfigure{\includegraphics[width=0.27\textwidth,angle=0]{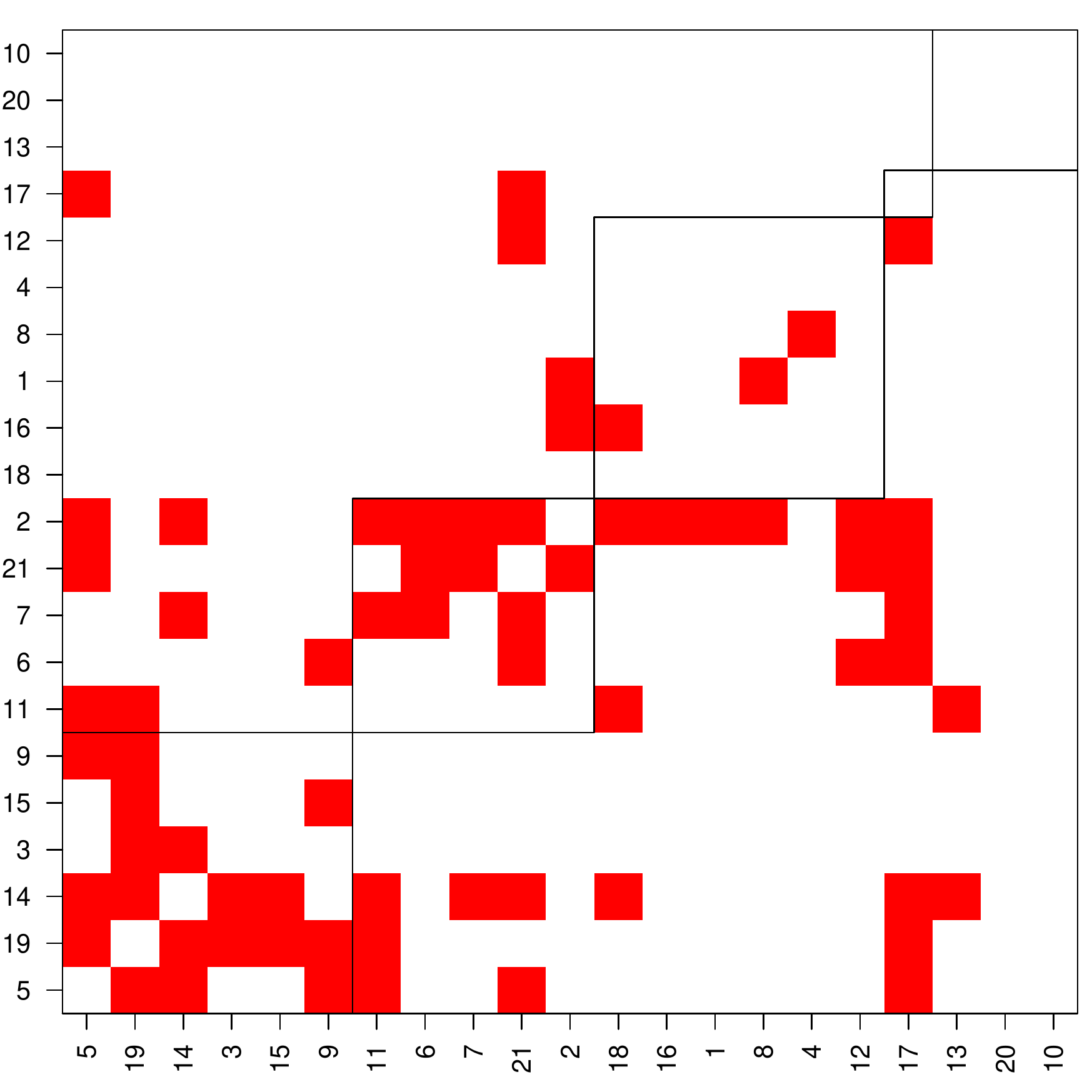} }& 
	\subfigure{\includegraphics[width=0.27\textwidth,angle=0]{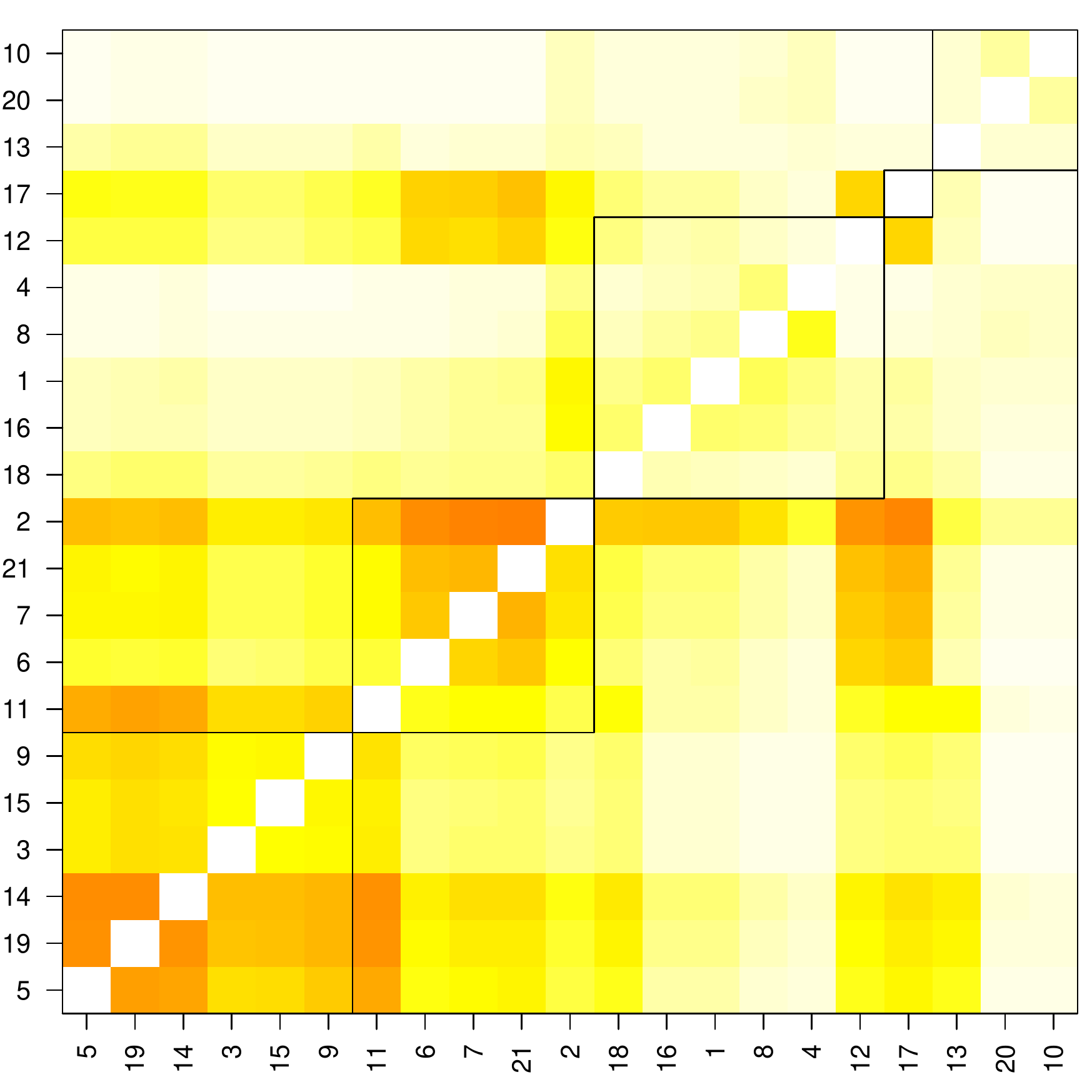} } \\
	%%%
	\begin{sideways}  \qquad \qquad Individual  \end{sideways} 
	\begin{sideways}  \qquad \qquad Subject 17  \end{sideways} 
	\subfigure{\includegraphics[width=0.27\textwidth,angle=0]{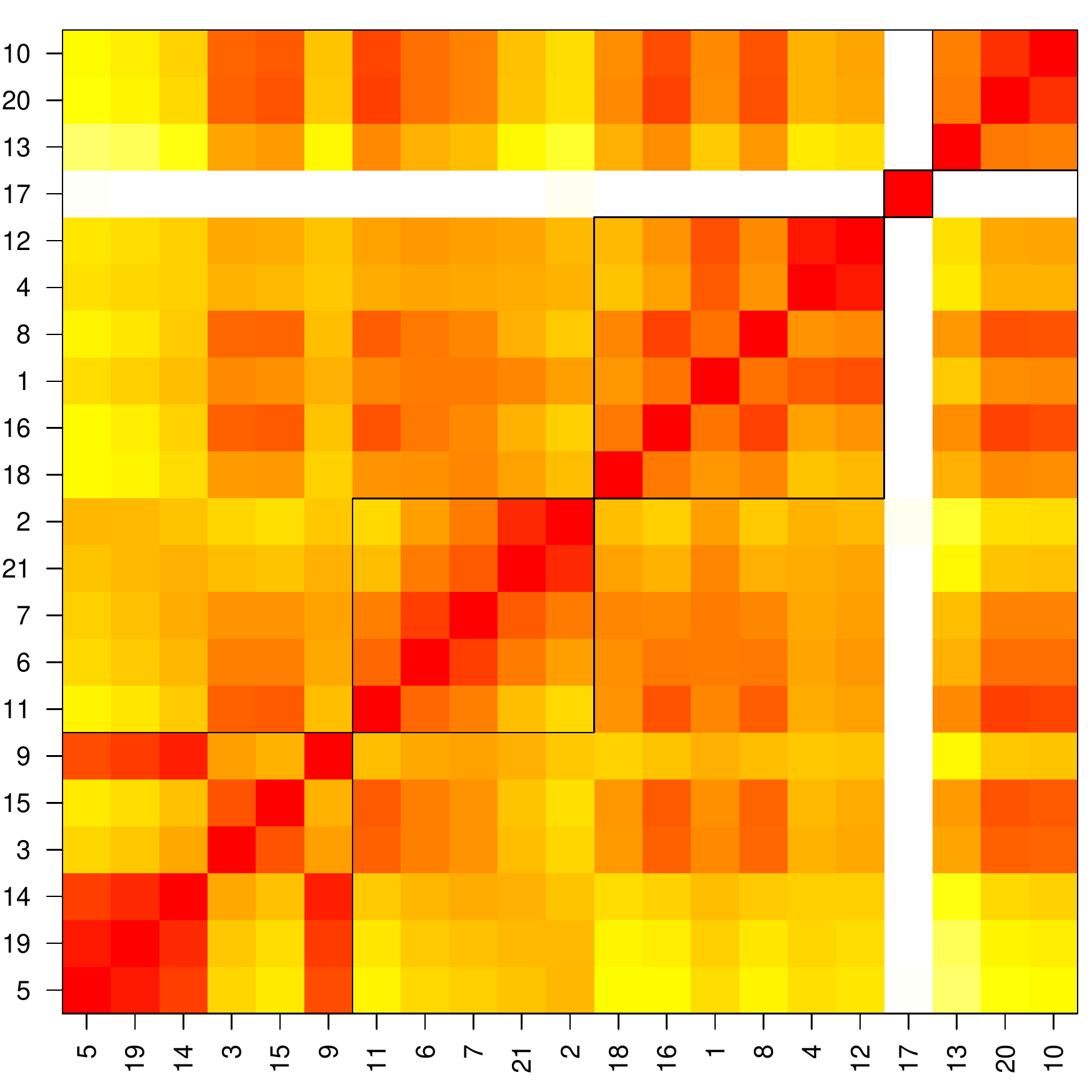} }& 
	\subfigure{\includegraphics[width=0.27\textwidth,angle=0]{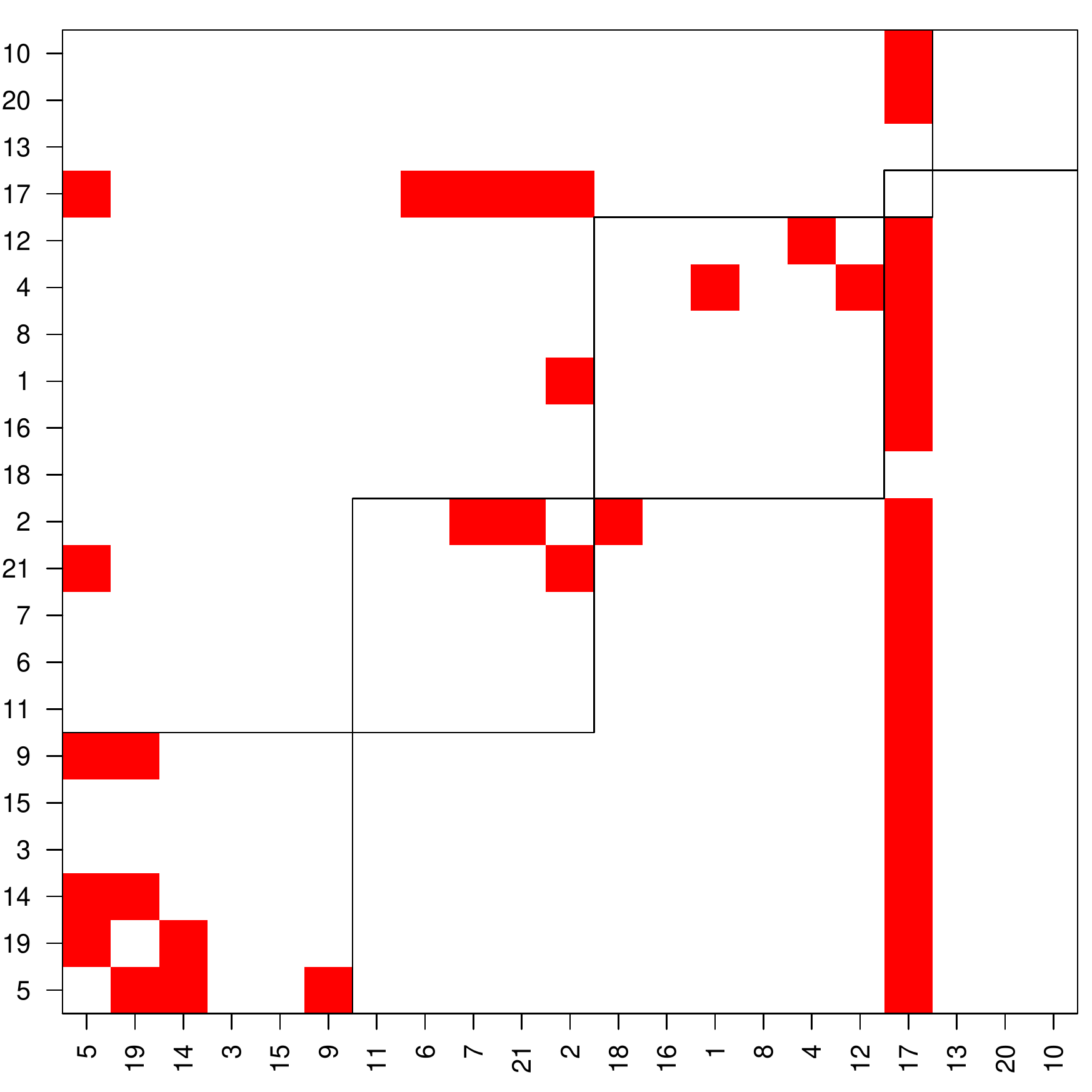} }& 
	\subfigure{\includegraphics[width=0.27\textwidth,angle=0]{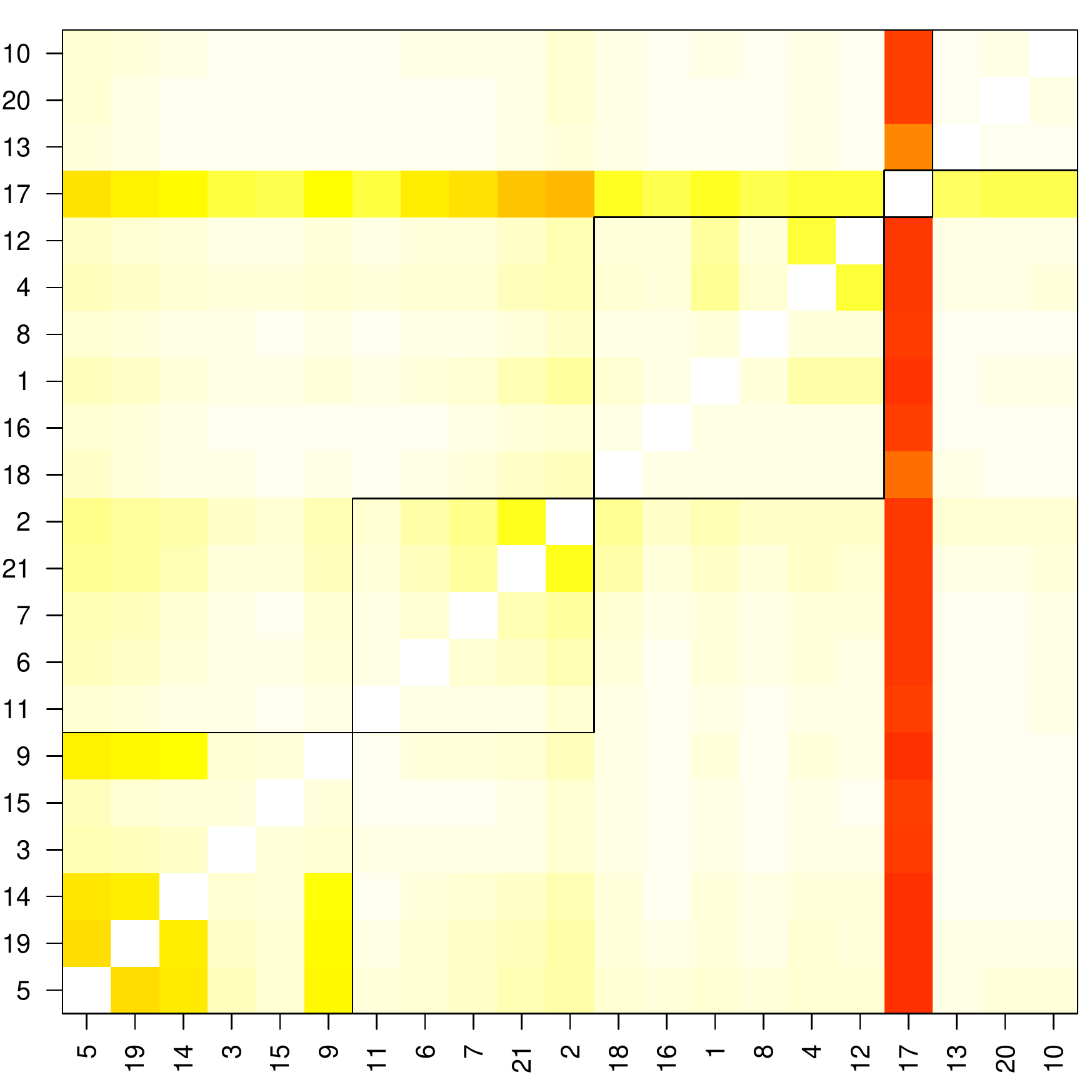} } \\
	%%%
          \end{tabular}
   \end{center}
 	   \caption{Krackhardt data. Individual community structures (estimated probability of being in the same community), friendship network, and estimated probability of connection between actors.}\label{fi:KrackRes1}
           \end{figure}

Figure~\ref{fi:KrackRes1} shows the resulting community structures for one representative actor from each class. Although all plots have the same ordering and inner squares have been added to help with comparisons, neither of them are the result of a formal estimation procedure. Class one actors described the network having several connections, the actors tend to be grouped in factions of friendship groups (more connected within groups than between). Class two subjects indicate very few friendship connections in their networks. Class three shows more connections in their networks and a tendency of forming groups according to friendship patterns. The ungrouped managers showed diverse oddities that made their networks different than the rest, subject 17 is an example of one of the most common of this differences. Note that, as we highlighted before, he marked himself as being friend of almost all of the managers. 
%REFERENCE
%
%Krackhardt D. (1987). Cognitive social structures. Social Networks, 9, 104-134.
%Wasserman, S. and Faust, K. (1999). Social Network Analysis. Cambridge University Press. New York.

%Results
%	\begin{figure}[htbp]
%	  \begin{center}     	
%            \includegraphics[width=\textwidth,angle=0]{figures/Jul26/NewFR5_zeta_All.pdf} \\
%%       	\vspace{-.7cm}
%          \includegraphics[height=1cm,angle=0]{figures/scale_hor.pdf}
%	   \caption{Krackhardt data. Comparing actors perceptions, probability of two actors of having the same perception of the network grouping.}\label{fi:Krackzeta}
% 	  \end{center}
%        \end{figure}

            \begin{figure}[htbp]
            \begin{center}
          \includegraphics[height=7in, width=7in,angle=0]{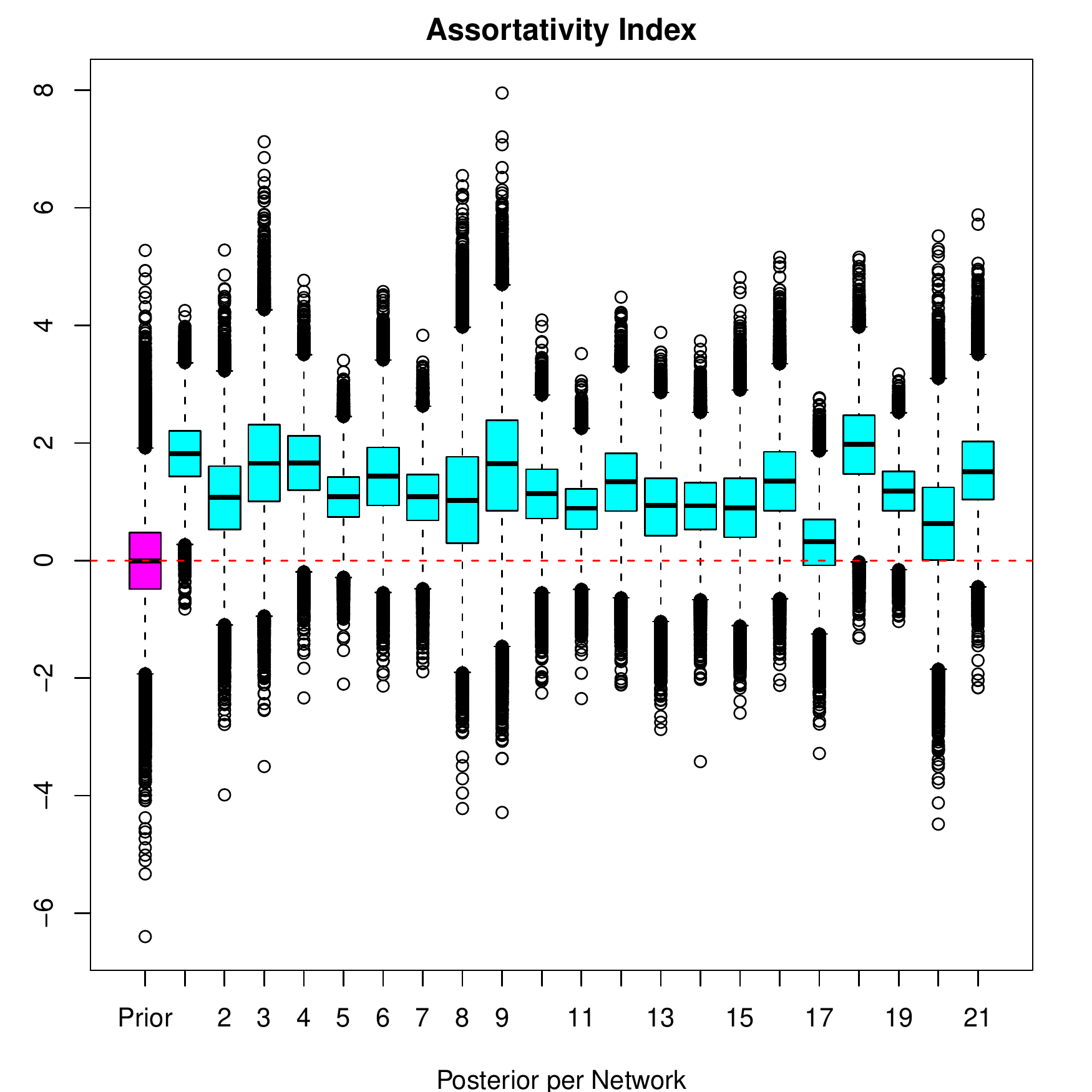}
          \caption{Krackhardt data. Prior ({\tt left}) and Posterior distribution of the assortativity index per network} \label{fi:assortKrack}
          \end{center}
            \end{figure}

	   In addition to comparing actors' perceptions of community structures, we studied the assortativity for each of the 21 networks. As we discussed in Section \ref{se:modelproperties}, a simple summary of this property is the assortativity index $\Upsilon$ defined in section~\ref{sec:assortative}.  Figure~\label{fi:assortKrack} shows boxplots of the posterior distributions of $\Upsilon$ for each of the 21 networks, along with a sample from the prior distribution (which was identical for all networks). Evidently all actors describe the network as assortative, i.e. interactions among actors in the same faction tend to be more common than interactions among actors in different factions.  This is exactly what we expected to see for this type of relationship.

\subsection{The Wiring dataset}\label{se:wiring}

 %The data are better known through a scrutiny made of the interactions in Homans (1950), and the CONCOR analyses presented in Breiger et al (1975).

%Breiger R., Boorman S. and Arabie P. (1975). An algorithm for clustering relational data with applications to social network analysis and comparison with multidimensional scaling. Journal of Mathematical Psychology, 12, 328-383.
%Homans G. (1950). The human group. New York: Harcourt-Brace.
%Roethlisberger F. and Dickson W. (1939). Management and the worker. Cambridge: Cambridge University Press.

%The employees worked in a single room and include two inspectors (I1 and I3), three solderers (S1, S2 and S3), and nine wiremen or assemblers (W1 to W9). The interaction categories include: RDGAM, participation in horseplay; RDCON, participation in arguments about open windows; RDPOS, friendship; RDNEG, antagonistic (negative) behavior; RDHLP, helping others with work; and RDJOB, the number of times workers traded job assignments.

   \begin{figure}[htbp]
	\begin{center}
          \begin{tabular}{r ccc l}
          \hline \\
        \multirow{4}{*}{ 	\begin{sideways} {\large Individual Networks Model  }  \qquad   \qquad   \qquad  \end{sideways} } &
            Friendship & Horseplay & Help  &
        \multirow{8}{*}{ \includegraphics[height=19cm,width=1cm,angle=0]{figures/scale_ver.pdf} } \\
           &
        \includegraphics[width=0.26\textwidth,angle=0]{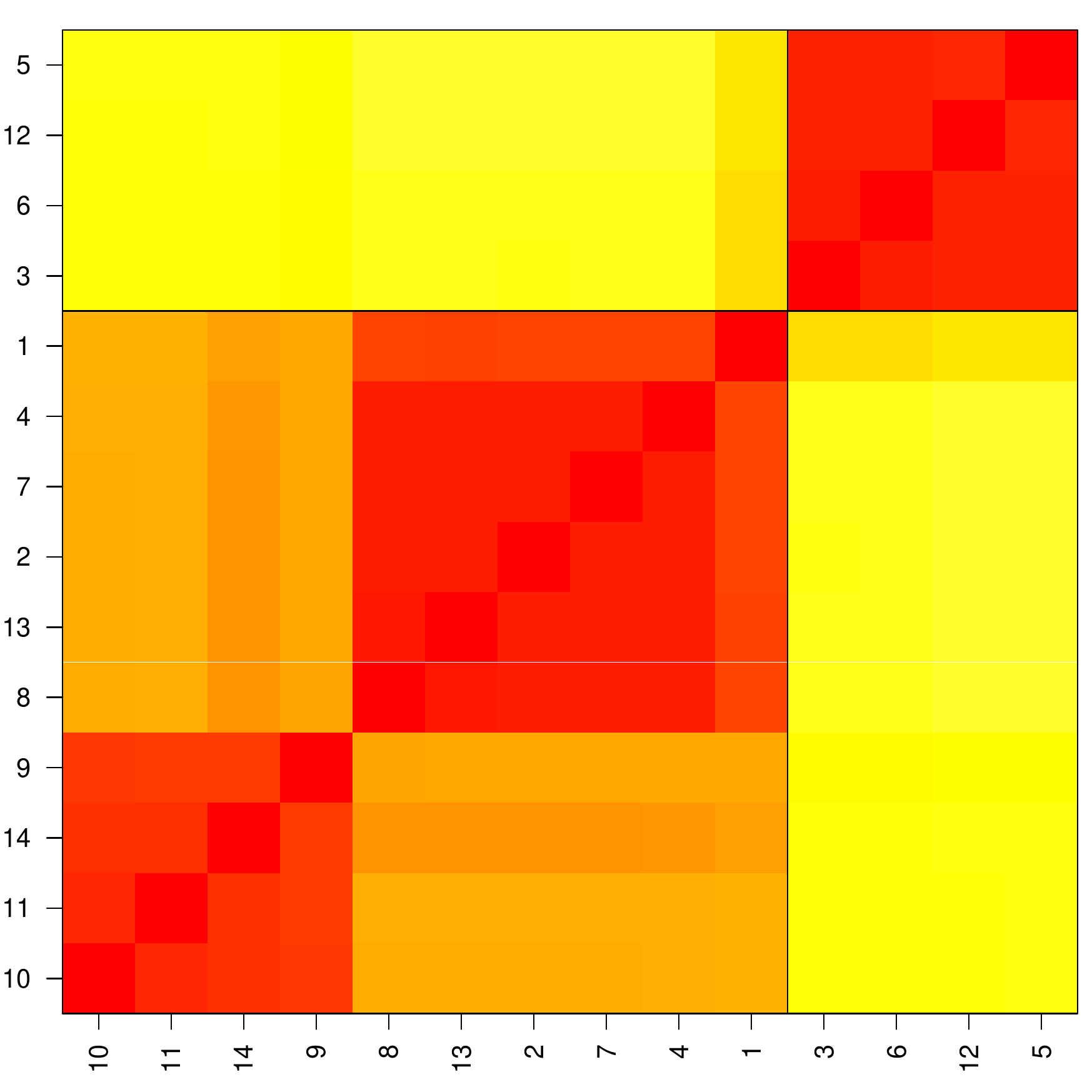} &
        \includegraphics[width=0.26\textwidth,angle=0]{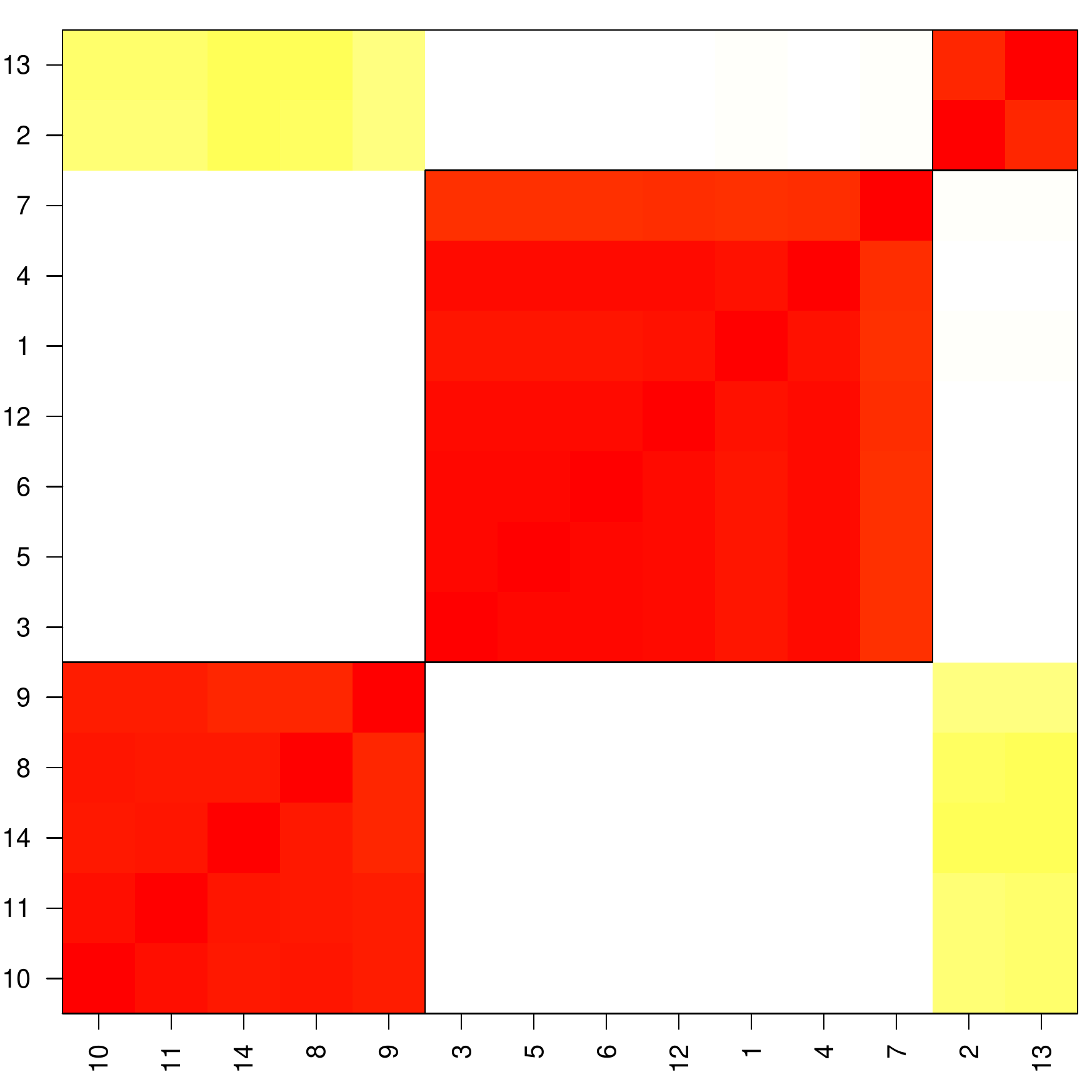} &       
        	\includegraphics[width=0.26\textwidth,angle=0]{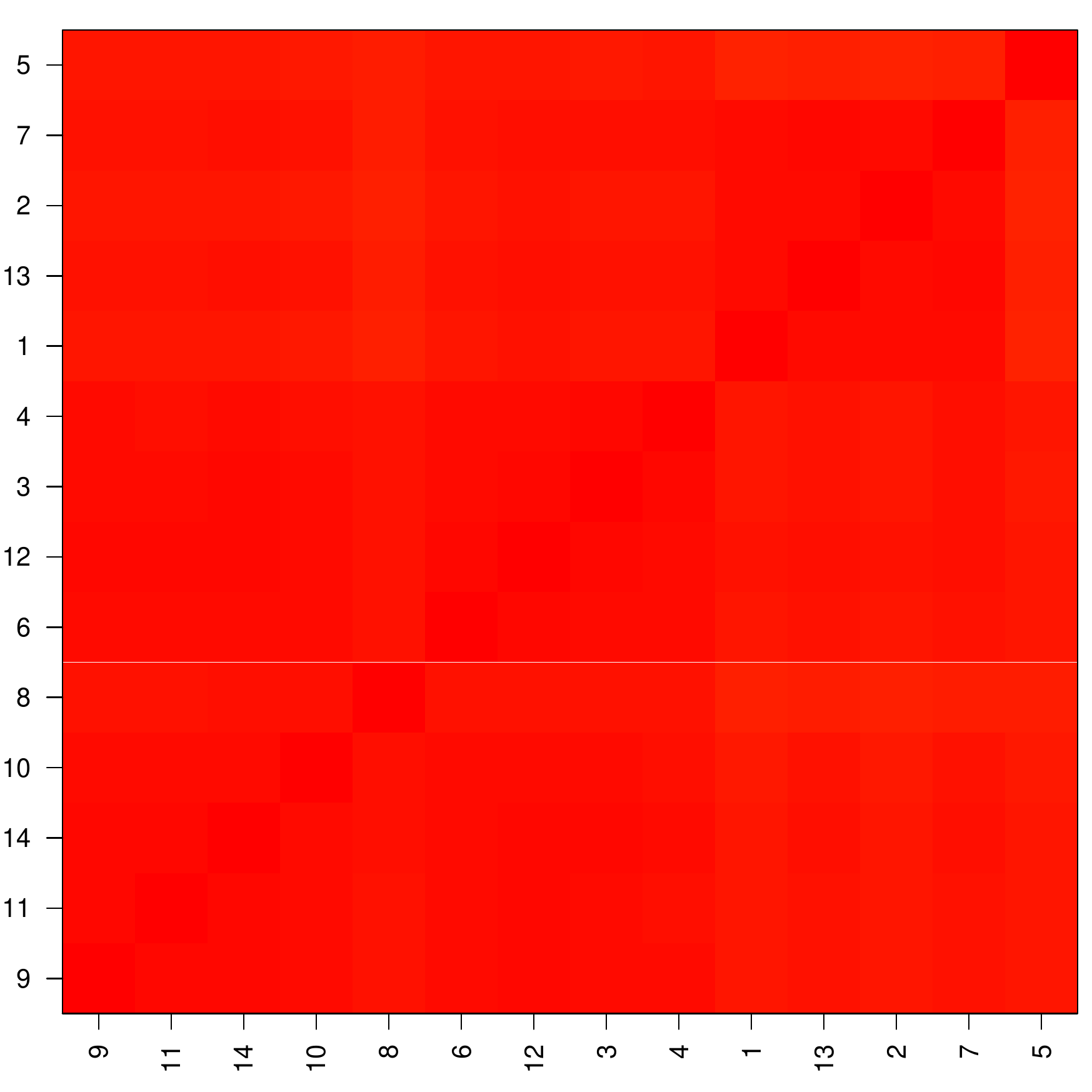} \\
       	&
 	    Antagonist & Open Window &	    Trade Job\\
       	&
    \includegraphics[width=0.26\textwidth,angle=0]{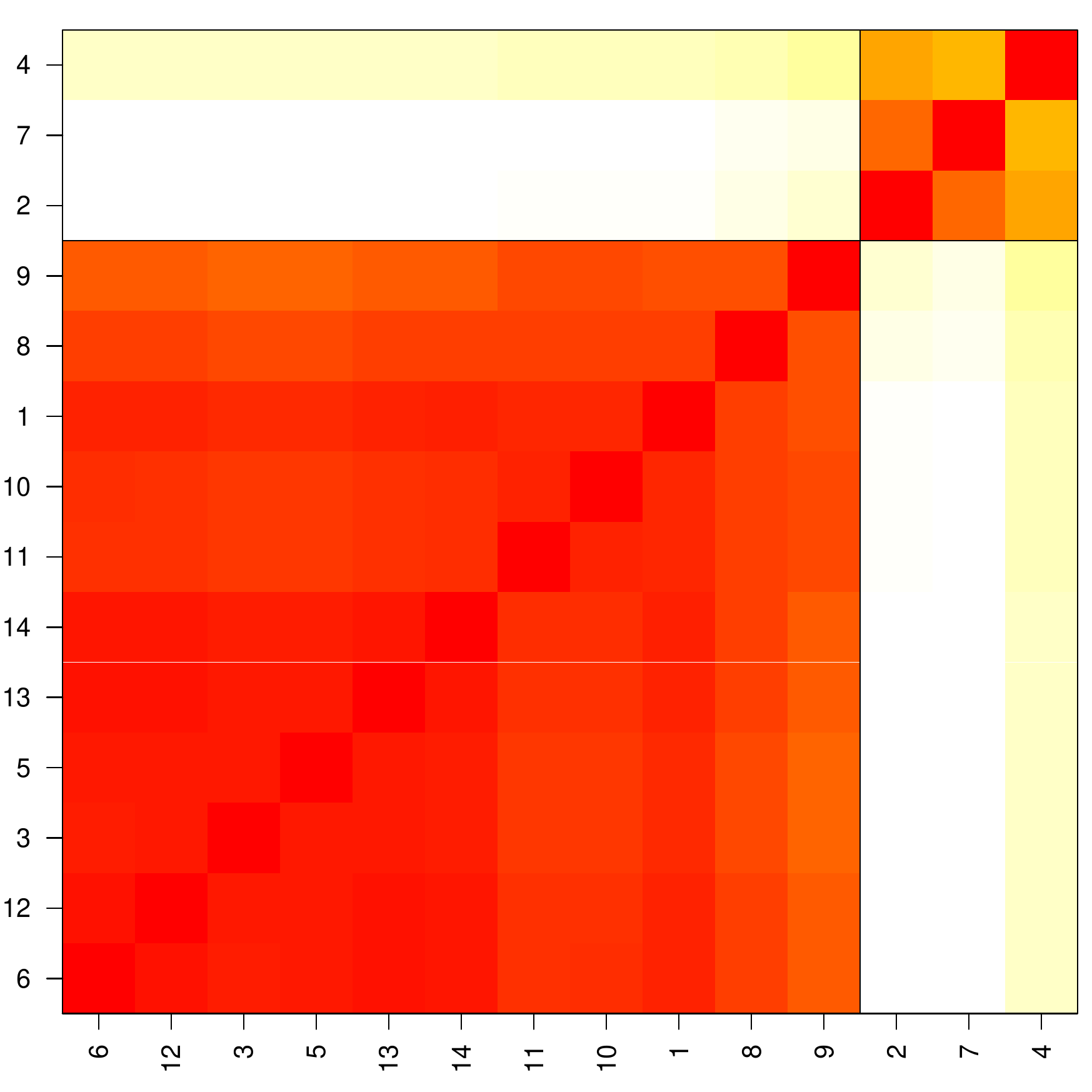} &
    \includegraphics[width=0.26\textwidth,angle=0]{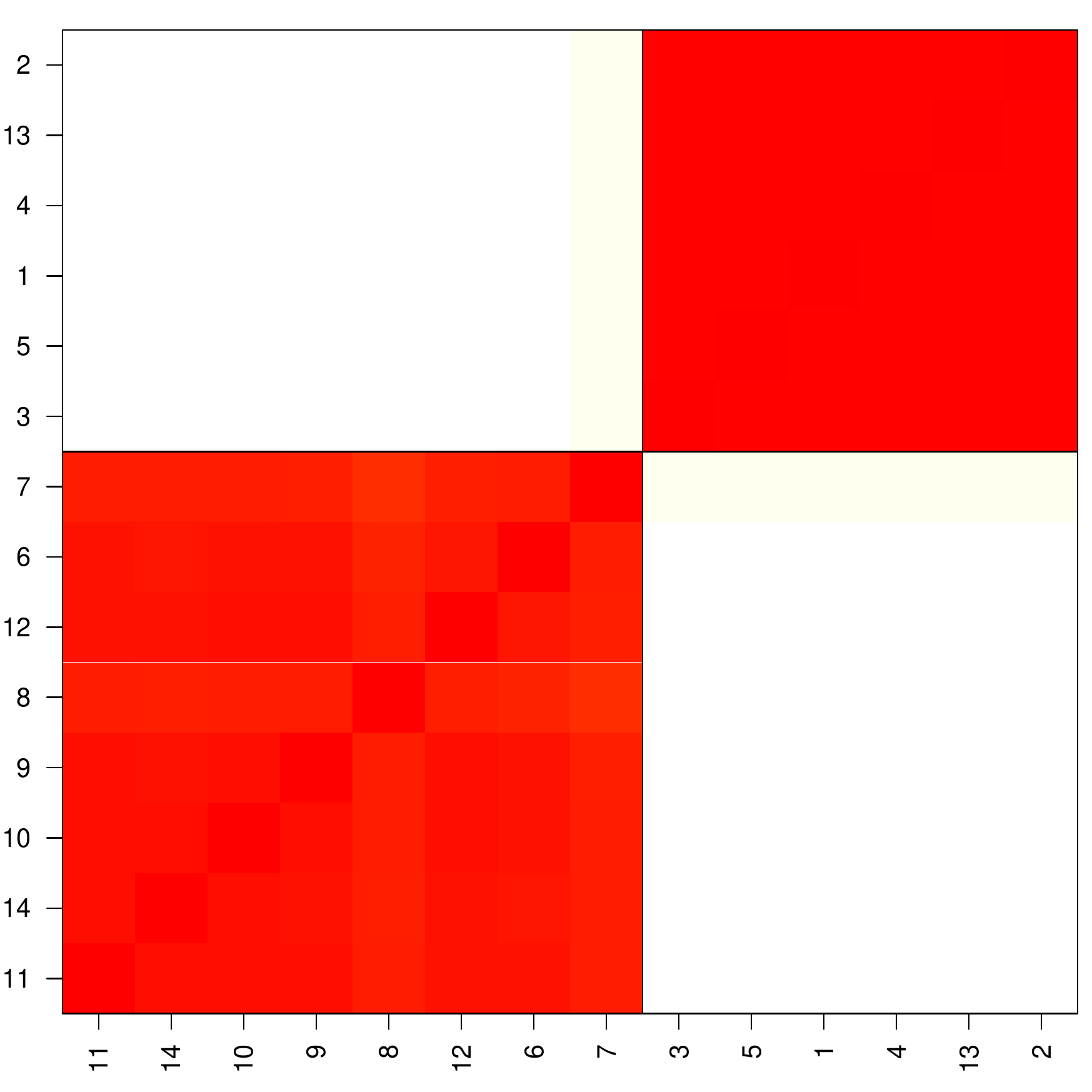} &
    \includegraphics[width=0.26\textwidth,angle=0]{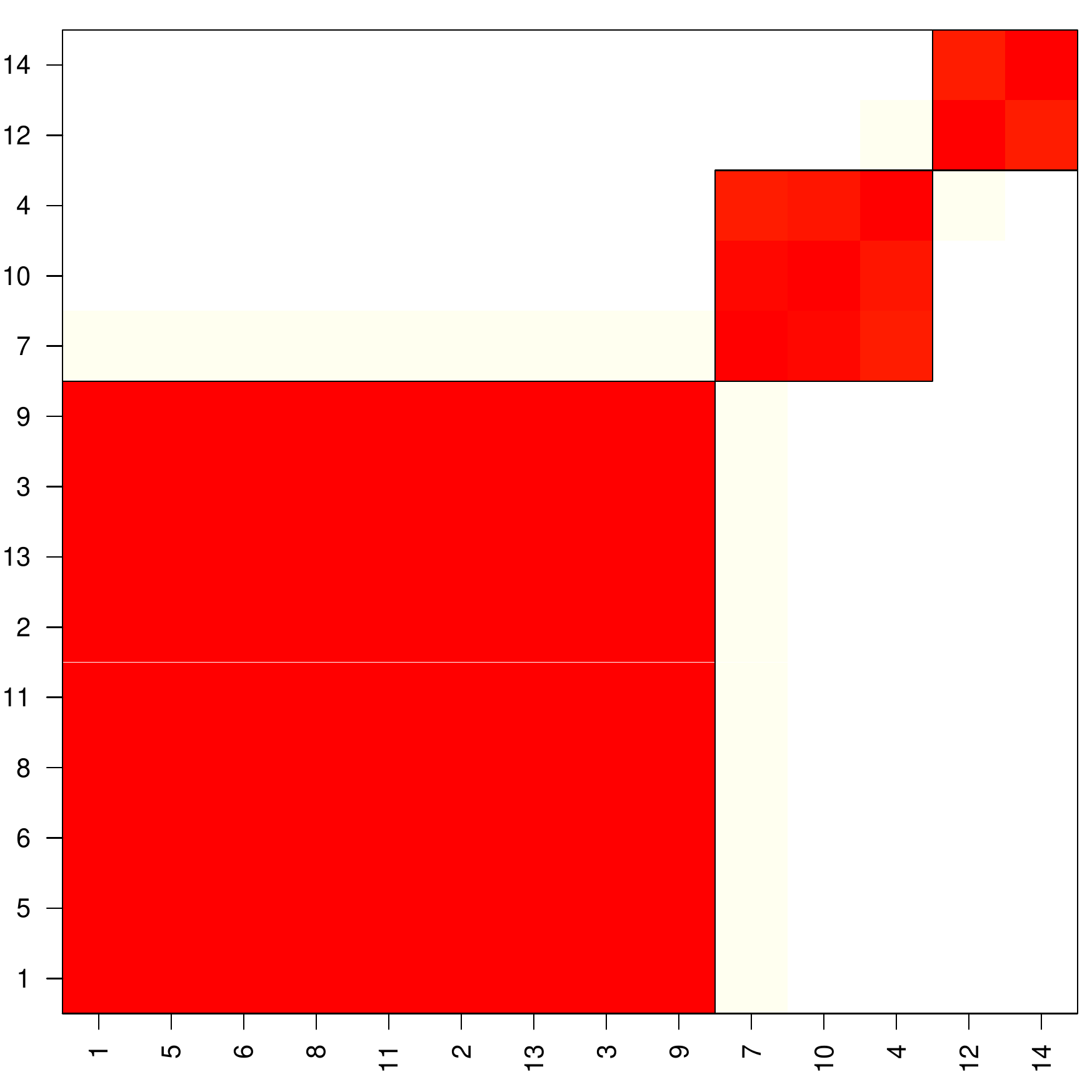} \\
          \cline{1-4} \\
    
        \multirow{4}{*}{ 	\begin{sideways} {\large Multiple Networks Model  }  \qquad   \qquad   \qquad  \end{sideways} } &
            Friendship & Horseplay & Help  \\
           &
        \includegraphics[width=0.26\textwidth,angle=0]{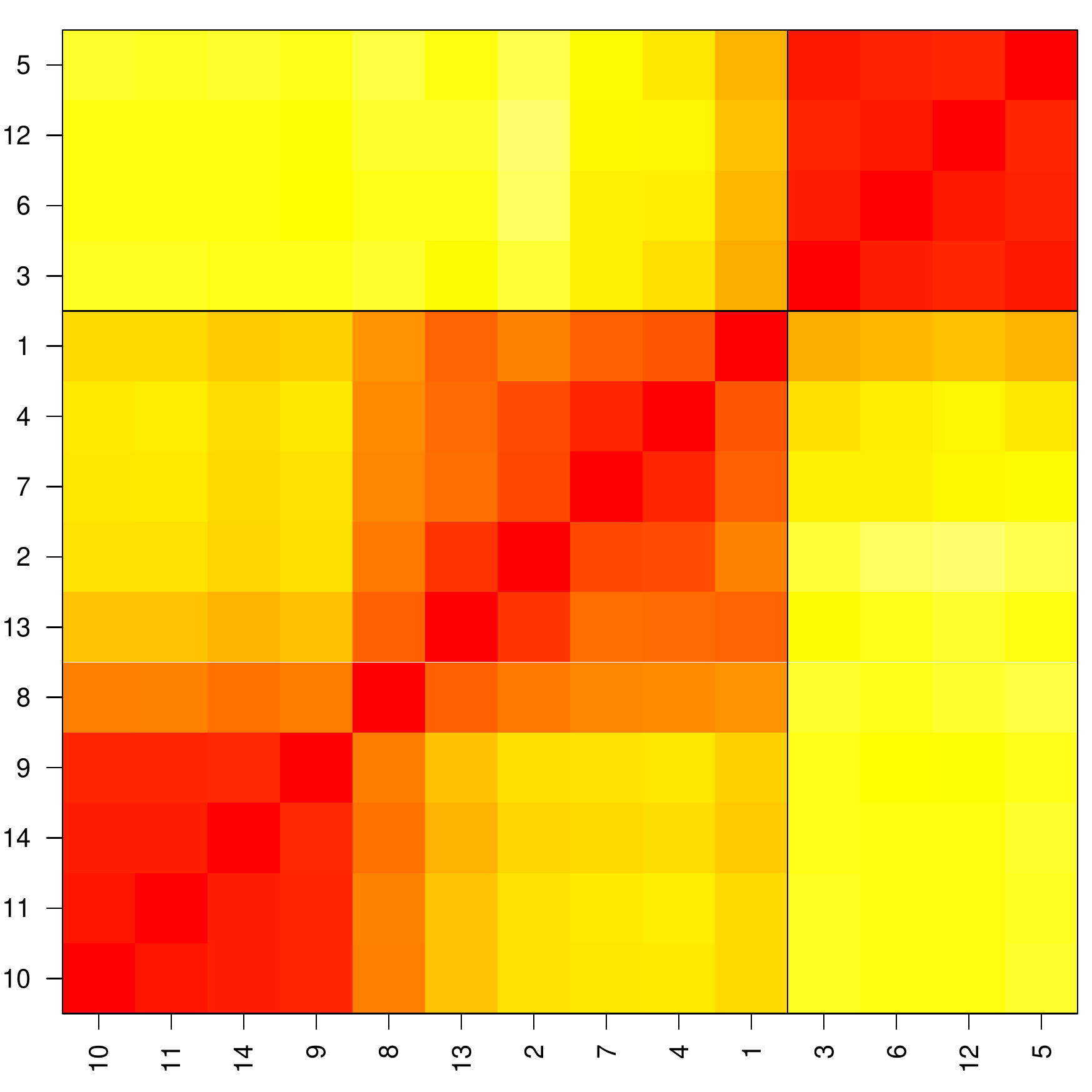} &
        \includegraphics[width=0.26\textwidth,angle=0]{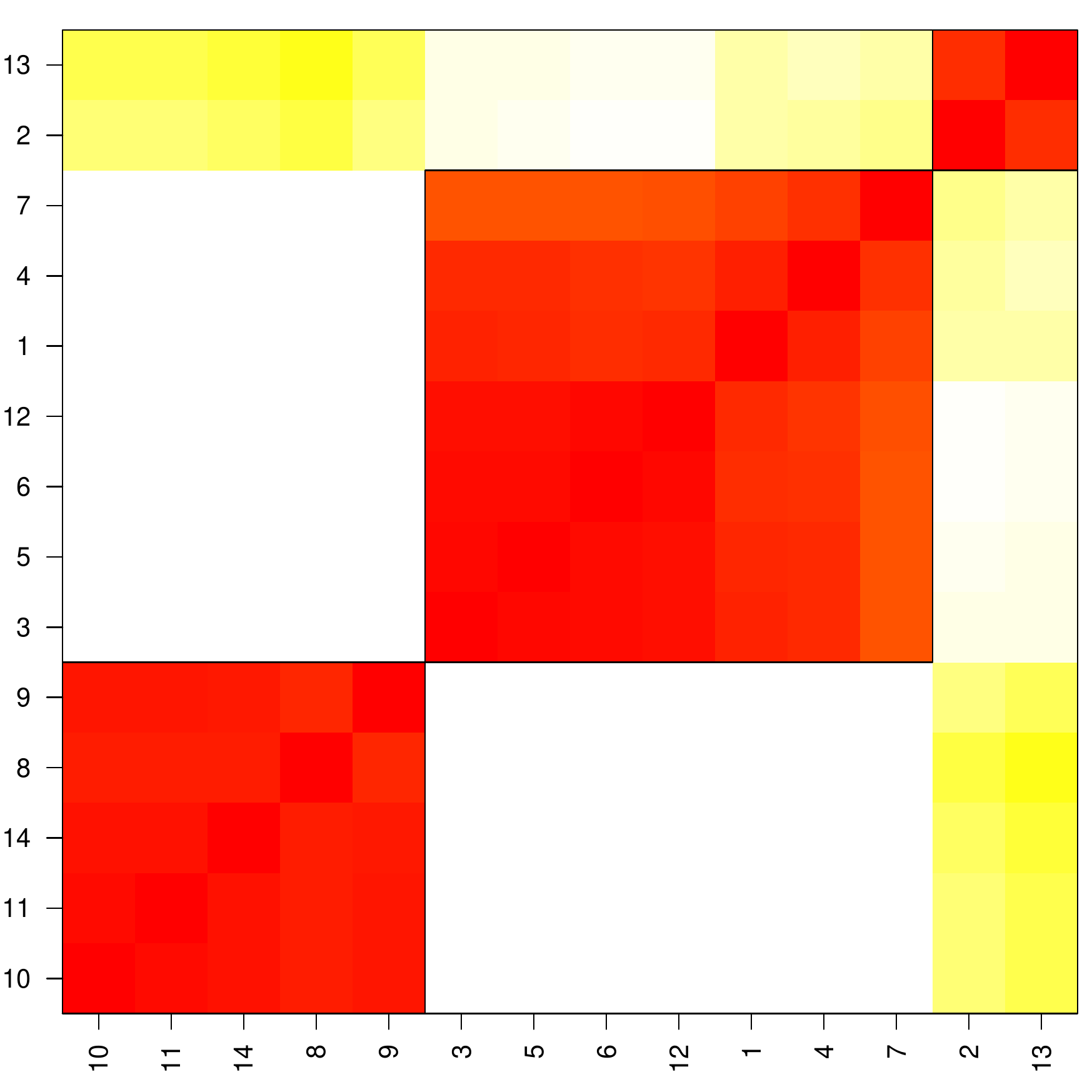} &       
        	\includegraphics[width=0.26\textwidth,angle=0]{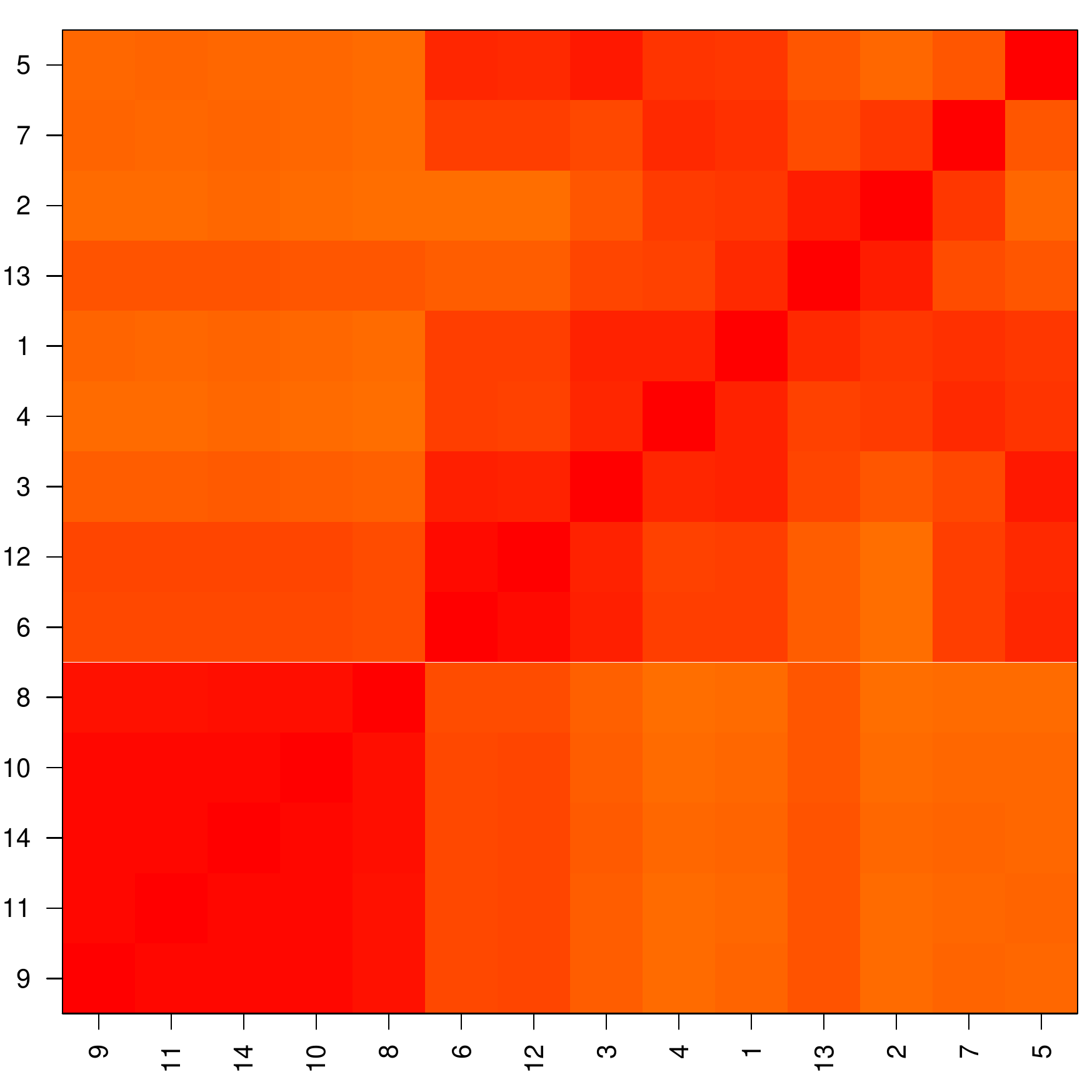} \\
       	&
 	    Antagonist & Open Window &	    Trade Job\\
       	&
    \includegraphics[width=0.26\textwidth,angle=0]{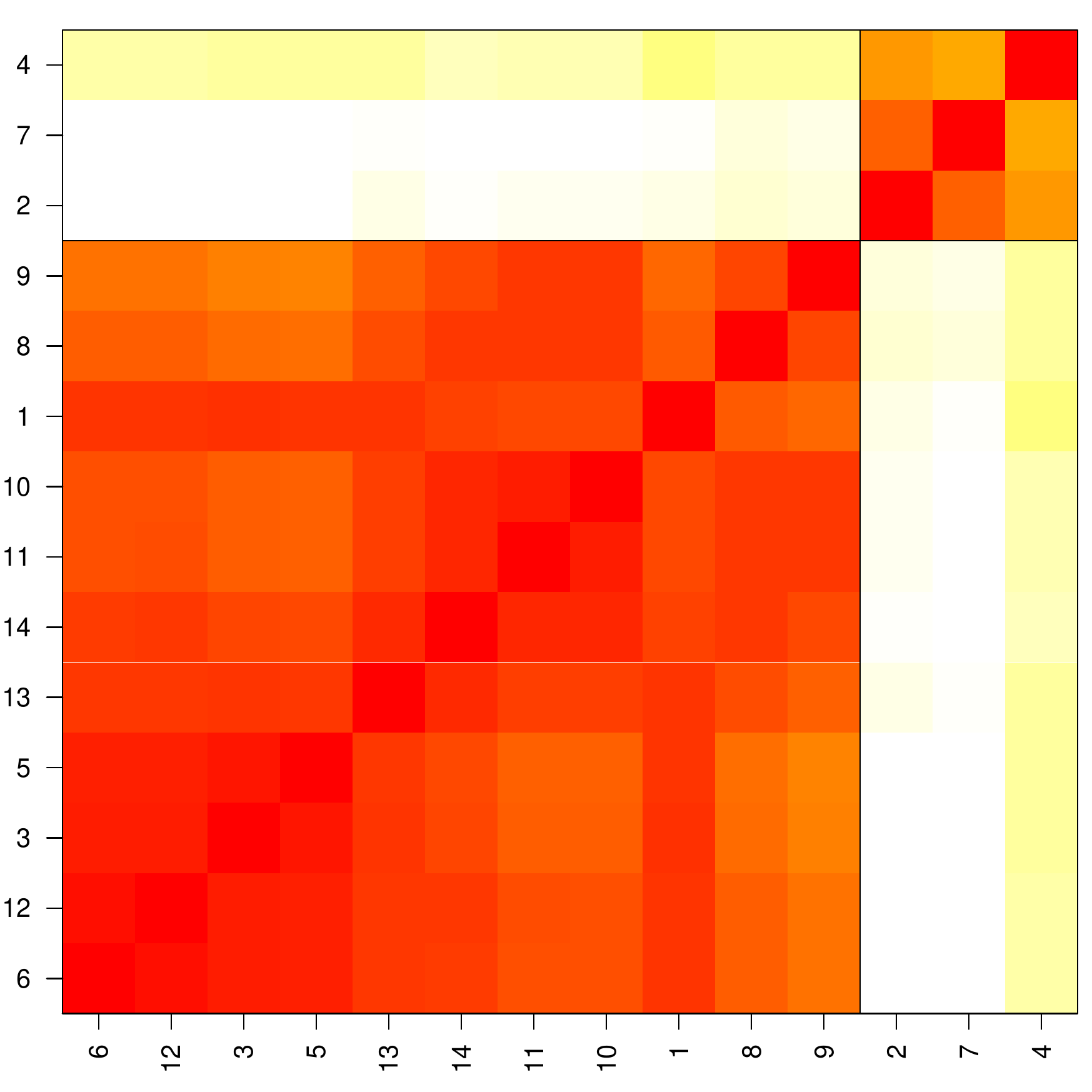} &
    \includegraphics[width=0.26\textwidth,angle=0]{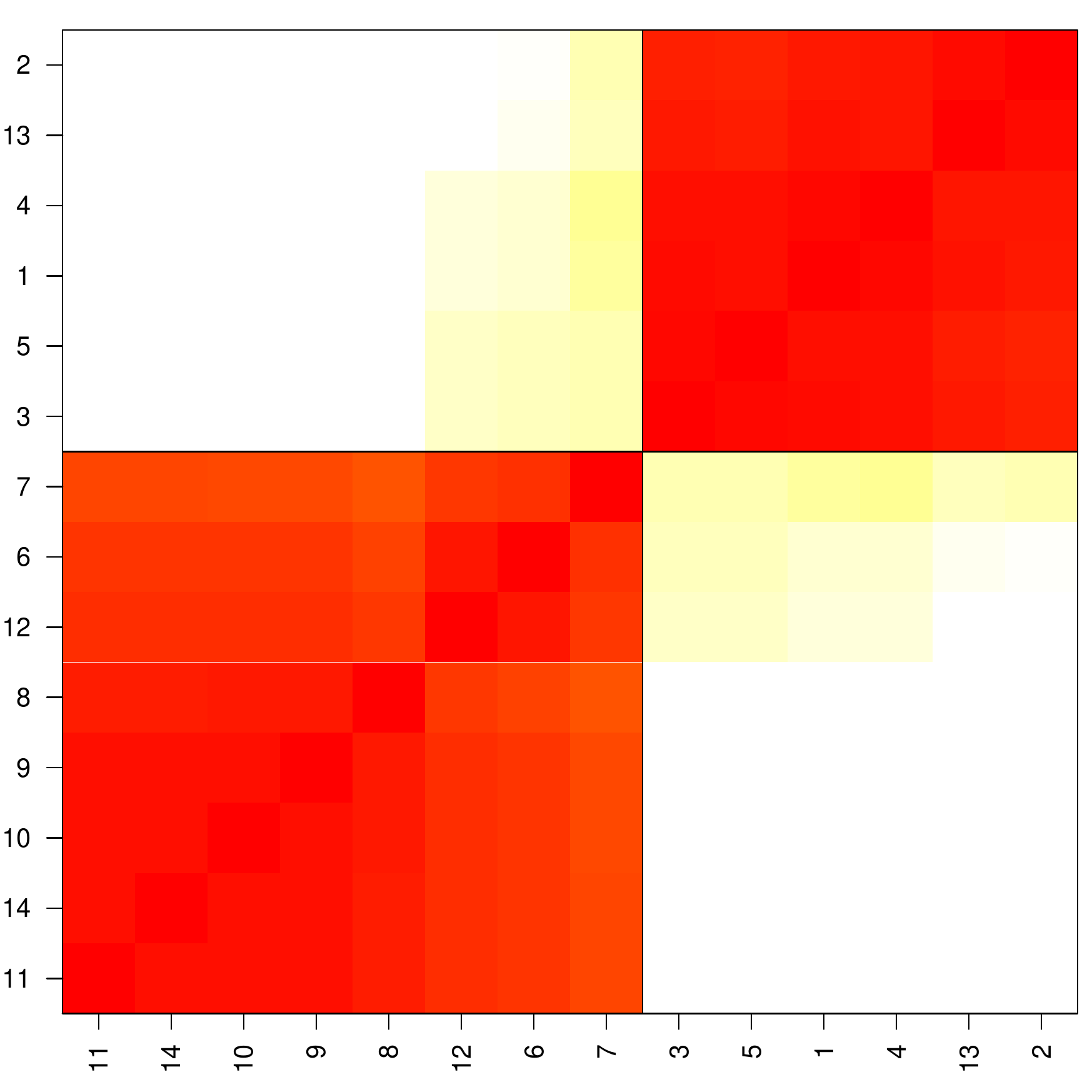} &
    \includegraphics[width=0.26\textwidth,angle=0]{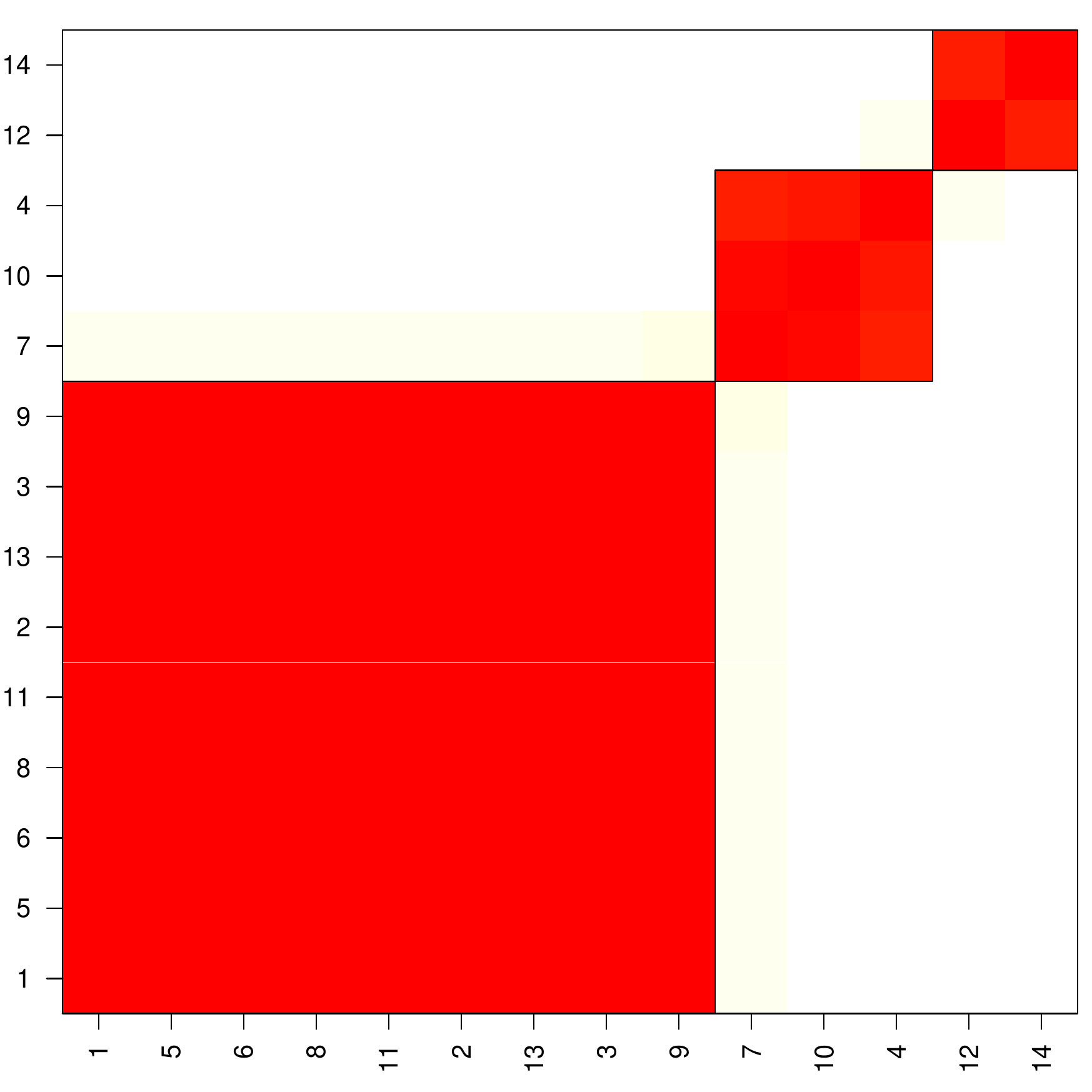} \\
          \hline
	\end{tabular}
         \caption{Wiring dataset. Estimated posterior probability that two actors belong to the same group for each network. }\label{fi:wiring1}
          \end{center}
          \end{figure}

Finally, we applied our model to a dataset consisting of the interactions among 14 Western Electric (Hawthorne Plant) employees working on the bank wiring room \citep{RoDi39}.  The employees worked in a single room and six relationships were recorded:  friendship (binary, undirected),  participation in horseplay (binary, undirected), helping others with work (binary, directed), antagonistic behavior (binary, undirected), arguments about open windows (binary, directed), and number of times workers traded job assignments (integer valued, directed).

The posterior distribution on $\bfzeta$ strongly suggests that the community structure associated with each network is different from the rest. Indeed, the highest posterior probability that two relationships have the same structure is 0.23 (for friendship and participation in horseplay).  Figure \ref{fi:wiring1} shows the estimated mean posterior probabilities/intensities and tentative groupings for each network. Most relationships form clear factions with the exception of Help, where the groups are not as delimited. Note that by clustering networks according to their underlying community structures we are able to simultaneously compare directed, undirected, binary and  valued networks. 

\section{Conclusions}

This manuscript makes two contributions to the literature on network analysis.  Firstly, we have described a general class of priors for blockmodels that allow us to incorporate available prior information about the structure of the network and construct model-based summary statistics for properties such as assortativity.  Secondly, we have constructed a joint model for multiple networks that allows us to compare their underlying social structures, even across networks of different types.

With regard to the first contribution, it is worthwhile noting that further generalizations of the model are possible.  For example, by constructing a prior for $\bfTheta$ where the entries for $\theta_{k,l}$ adn $\theta_{l,k}$ are identically distributed but dependent, the model can accommodate reflexivity as well more flexible specifications for the correlation among the in-degree and out-degree of directed networks.  With regard to the second contribution, the results from our analyses suggests that, for some applications, assuming that the community structure associated with various networks is the same might be too strong.  A model with softer constrains is presented in \cite{RoReVu11}, however, such model is not helpful when dealing with networks of different types.  In this regard, an interesting open problem is constructing joint distributions for multiple partitions of a given set of items; such models can in turn be used to create more flexible models for collections of networks.

\appendix

\section{Proof of lemma \ref{le:exchangeability}}\label{ap:exchangeability}

We focus on the case of directed networks, the proof for undirected networks follows along the same lines.  For any fixed $K$, consider the $K \times K$ matrix $\bfTheta = [\theta_{k,k}]$ with joint distribution 
\begin{align*}
p(\bfTheta) = \left\{ \prod_{k = 1}^{K} H^{\bflambda_D}(\theta_{k,k}) \right\} \left\{ \prod_{k = 1}^{K} \prod_{l = k+1}^{K} H^{\bflambda_O}(\theta_{k,l})H^{\bflambda_O}(\theta_{l,k}) \right\}.
\end{align*}
To show that $p(\bfTheta)$ is invariant to any permutation, it is enough to show that it is invariant to permutations that only exchange any two indexes $i$ and $j$, i.e., it is invariant to the permutation $\sigma(i) = j$, $\sigma(j) = i$, and $\sigma(k) = k$ for any other $k$ different from $i$ and $j$.  This is because any general permutation can be written as a composition of these simple permutations.  Let $\bfTheta^{*}$ be the permuted version of $\bfTheta$.  Now
\begin{align*}
p(\bfTheta^{*}) = \left\{ \prod_{k = 1}^{K} H^{\bflambda_D}(\theta^{*}_{k,k}) \right\} \left\{ \prod_{k = 1}^{K} \prod_{l = k+1}^{K} H^{\bflambda_O}(\theta^{*}_{k,l}) H^{\bflambda_O}(\theta^{*}_{l,k}) \right\}.
\end{align*}
Note that $H^{\bflambda_D}(\theta^{*}_{i,i}) = H^{\bflambda_D}(\theta_{\sigma(i),\sigma(i)}) = H^{\bflambda_D}(\theta_{j,j})$ and $H^{\bflambda_D}(\theta^{*}_{j,j}) = H^{\bflambda_D}(\theta_{\sigma(j),\sigma(j)}) = H^{\bflambda_D}(\theta_{ii})$, while 
\begin{align*}
H^{\bflambda_D}(\theta^{*}_{k,k}) &= H^{\bflambda_D}(\theta_{\sigma(k),\sigma(k)}) = H^{\bflambda_D}(\theta_{k,k})
\end{align*}
for any $k\ne i, j$.  Hence, 
$$
\prod_{k = 1}^{K} H^{\bflambda_D}(\theta^{*}_{k,k})  =   \prod_{k = 1}^{K} H^{\bflambda_D}(\theta_{k,k}).
$$

A similar argument can be made for the off-diagonal elements.  In this case, the only terms that are not obviously identical before and after the permutation are those associated with $(\theta_{i,k}, \theta_{k,i})$ for $k \ne i,j$, those associated with $(\theta_{j,k}, \theta_{k,j})$ for $k \ne i,j$, and the term associated with $(\theta_{i,j}, \theta_{i,j})$.  Since the pairs are independent, it is clear that 
\begin{align*}
& \prod_{k \ne i,j}  H^{\bflambda_O}(\theta^{*}_{i,k})H^{\bflambda_O}(\theta^{*}_{k,i})  \prod_{k \ne i,j}  H^{\bflambda_O}(\theta^{*}_{j,k})H^{\bflambda_O}(\theta^{*}_{k,j}) = \\
& \;\;\;\;\;\; \;\;\;\;\;\; \;\;\;\;\;\;  \prod_{k \ne i,j}  H^{\bflambda_O}(\theta_{\sigma(i),k})H^{\bflambda_O}(\theta_{k,\sigma(i)}) \prod_{k \ne i,j}  H^{\bflambda_O}(\theta_{\sigma(j), k}) H^{\bflambda_O}(\theta_{k, \sigma(j)}) = \\
& \;\;\;\;\;\; \;\;\;\;\;\; \;\;\;\;\;\; \;\;\;\;\;\; \;\;\;\;\;\; \;\;\;\;\;\; \;\;\;\;\;\; \prod_{k \ne i,j}  H^{\bflambda_O}(\theta_{j,k})H^{\bflambda_O}(\theta_{k,j}) \prod_{k \ne i,j}  H^{\bflambda_O}(\theta_{i,k})H^{\bflambda_O}(\theta_{k,i}).
\end{align*}
Finally, $H^{\bflambda_O}(\theta^{*}_{i,j})H^{\bflambda_O}(\theta^{*}_{j,i}) = H^{\bflambda_O}(\theta_{\sigma(i),\sigma(j)})H^{\bflambda_O}( \theta_{\sigma(j),\sigma(i)}) = H^{\bflambda_O}(\theta_{i,j})H^{\bflambda_O}( \theta_{j,i})$, which completes the proof.

\section{Proof of lemma \ref{le:problink}}\label{ap:problink}

A simple conditioning argument implies that
\begin{align*}
\bar{\theta} = \E \left\{ \E \left( y_{ij} | \xi_i, \xi_j, \bfTheta \right)  \right\} &= \E_{H^{\bflambda_D}} \{ \theta_{k,k} \} \Pr(\xi_i = \xi_j) + \E_{H^{\bflambda_O}} \{ \theta_{k,l} \} \Pr(\xi_i \ne \xi_j)
%\E \left\{  \sum_{k=1}^{\infty} \sum_{l=1}^{\infty} w_l w_k \theta_{k,l}  \right\} =   \sum_{k=1}^{\infty} \sum_{l=1}^{\infty} \E \left\{  w_l w_k \right\}  \E \left\{ \theta_{k,l}  \right\} \\
%&=  \bar{\theta}_D \left( \sum_{k=1}^{\infty} \E \left\{  w_k^2  \right\}  \right) +  \bar{\theta}_O \left( \sum_{k=1}^{\infty} \sum_{l=1, k \ne l}^{\infty} \E \left\{  w_l w_k \right\} \right) \\
%&=  \bar{\theta}_D \left( \sum_{k=1}^{\infty} \E \left\{  w_k^2  \right\}  \right) +  \bar{\theta}_O \left( 1 - \sum_{k=1}^{\infty} \E \left\{  w_k^2  \right\} \right)
\end{align*}
Now, because of the exchangeability of the observations we have $\Pr(\xi_i = \xi_j) = \Pr(\xi_1 = \xi_2) = \sum_{k=1}^{\infty} \E \{  \omega^2_k \} = (1-\alpha)/(\beta + 1)$.
%Now, due to the exchangeability of the indicators, $\sum_{k=1}^{\infty} w_k^2 = \Pr(\xi_1 = \xi_2)$, which in the case of of \eqref{eq:DPstickbreaking} euals $1/(1 + \beta)$, see equation \eqref{eq:polyaurn2}.

\section{Proof of lemma \ref{le:expDi}}\label{ap:expDi}

Again, we apply conditioning arguments.  For $\bar{\rho}$,
\begin{align*}
\bar{\rho}  = \E \{ D_i | \bflambda, \bfvarsigma \} & = \E \left\{ \sum_{j \neq i} y_{ij}  \bigg| \bflambda, \bfvarsigma \right\} \\
& = \sum_{j \neq i} \E \left\{ \E( y_{ij} | \xi_i, \xi_j, \bfTheta )  \right\}  \\
 & = [ I - 1 ] \left[ \Pr(\xi_i = \xi_j) \E_{H^{\bflambda_D}} \{ \theta_{k,k} \} + \Pr(\xi_i \ne \xi_j) \E_{H^{\bflambda_O}} \{ \theta_{k,l} \}  \right],
\end{align*}
where the last equality follows from lemma \ref{le:problink}.

There are a few alternative ways to obtain this result that might be of interest.  For example, let $D$ be the number of links in the network $\bfY$ composed of $I$ subjects, $L$ be the number of factions in the network and $m_l$ for $l=1,\ldots,L$ the size of the $l$-th community, then %and $X_{k,l}$ be the number of links between community $k$ and community $l$.
\begin{align*}
\E \{ D | \bflambda, \bfvarsigma \} = \E \left\{ \E \left\{ D | L, \{ m_l \}, \bfTheta \right\} \right\} & = \E\left\{ \sum_{l=1}^{L} \sum_{k \neq l, k=1}^{L} m_{l} m_{k}  \theta_{k,l} + \sum_{l=1}^{L} m_{l} (m_{l}-1)  \theta_{ll} \right \}.
	\end{align*}
	
The number of factions and the sizes of the factions are determined exclusively by the weights $\{ w_k \}$, hence they are independent of $\theta$'s and
\begin{align*}
\E \{ D | \bflambda, \bfvarsigma \}  & = \sum_{l=1}^{L} \sum_{k \neq l, k=1}^{L} \E_{H^{\bflambda_O}} \{ \theta_{k,l} \} \E \left( m_{l} m_{k} \right)  + \sum_{l=1}^{L}  \E_{H^{\bflambda_D}} \{ \theta_{ll} \} \E \left( m_{l}^2 - m_{l} \right)  \\
		& = \E_{H^{\bflambda_O}} \{ \theta_{k,l} \} \E \left\{ \sum_{l=1}^{L} \sum_{k \neq l, k=1}^{L}  m_{l} m_{k} \right\} + \E_{H^{\bflambda_D}} \{ \theta_{ll} \}  \left\{  \E \left(\sum_{l=1}^{L}  m_{l}^2 \right) - E \left( \sum_{l=1}^{L } m_{l} \right) \right\}  \\
		& = \E_{H^{\bflambda_O}} \{ \theta_{k,l} \} \left\{  I^2 -  \E \left( \sum_{l=1}^{L} m_{l}^2 \right)\right\} + \E_{H^{\bflambda_D}} \{ \theta_{ll} \} \left\{ \E \left(\sum_{l=1}^{L}   m_{l}^2 \right) - I \right\}.
	\end{align*}
In the above expression, we make repeated use of the fact that $\sum_{l=1}^{L} m_l = I$ almost surely.  Hence,
\begin{align*} 
\E \{ D | \bflambda, \bfvarsigma \} = I^2 \E_{H^{\bflambda_O}} \{ \theta_{k,l} \} -  I\E_{H^{\bflambda_D}} \{ \theta_{k,k} \}  +  \left( \E_{H^{\bflambda_D}} \{ \theta_{k,k} \} -  \E_{H^{\bflambda_O}} \{ \theta_{k,l} \} \right) {\E \left\{  \sum_{l=1}^{L}  m_{l}^2 \right\}}.
\end{align*}
Finally, since observations are exchangeable the mean number of links per observation is simply
\begin{multline}
\bar{\rho} =  \E \{ D_i | \bflambda, \bfvarsigma \}  =  \E \left\{ \frac{D}{I} \bigg| \bflambda, \bfvarsigma \right\}  =  \\
 I \E_{H^{\bflambda_O}} \{ \theta_{k,l} \} -  \E_{H^{\bflambda_D}} \{ \theta_{k,k} \}  + \frac{1}{I} \left( \E_{H^{\bflambda_D}} \{ \theta_{k,k} \} -  \E_{H^{\bflambda_O}} \{ \theta_{k,l} \} \right) {\E \left\{  \sum_{l=1}^{L}  m_{l}^2 \right\}}.  \label{eq:rhoSm2}
\end{multline}

An alternative way to compute $\bar{\rho}$ is working directly with $D_1$, without lost of generality assume that subject $1$ falls into the first community,
\begin{align*} 
\bar{\rho} = \E \{ D_1 | \bflambda, \bfvarsigma \} &=  \E \left\{ \E \left( D_1 | L, \{ m_l \}, \bfTheta  \right) \right\} = \E\left\{ (m_1 -1)  \theta_{1,1} + \sum_{l=2}^{L} m_{l}  \theta_{1l} \right \}\\
 & = \E \{ m_{1} \} \E_{H^{\bflambda_D}} \{ \theta_{k,k} \} -  \E_{H^{\bflambda_D}} \{ \theta_{k,k} \} + \E\left\{ \sum_{l=2}^{L} m_{l} \right \}  \E_{H^{\bflambda_O}} \{ \theta_{k,l} \}
\end{align*}
After regrouping and using the fact that $\sum_{l=2}^{L} m_l = I - m_1$ almost surely,
\begin{align} \label{eq:rhom}
\bar{\rho}  = I \E_{H^{\bflambda_O}} \{ \theta_{k,l} \} -  \E_{H^{\bflambda_D}} \{ \theta_{k,k} \} + \left( \E_{H^{\bflambda_D}} \{ \theta_{k,k} \} -  \E_{H^{\bflambda_O}} \{ \theta_{k,l} \} \right)  \E \{ m_{1} \}.
\end{align}
Perhaps the simplest expression $\E \{ D_i | \bflambda, \bfvarsigma \}$ was already derived in lemma \ref{le:expDi}, 
rearranging terms
\begin{align}\label{eq:rhothetabar} 
\bar{\rho} &  = I \E_{H^{\bflambda_O}} \{ \theta_{k,l} \} -  \E_{H^{\bflambda_D}} \{ \theta_{k,k} \} + \left( \E_{H^{\bflambda_D}} \{ \theta_{k,k} \} -  \E_{H^{\bflambda_O}} \{ \theta_{k,l} \} \right) \left\{ 1 + (I-1) \Pr(\xi_i = \xi_j) \right\}.
\end{align}
Further, the equivalent expressions (\ref{eq:rhoSm2}), (\ref{eq:rhom}) and (\ref{eq:rhothetabar}) are proof of the following not so evident equalities
\begin{align*}
\E \{ m_{1} \} = \frac{1}{I} {\E \left\{  \sum_{l=1}^{L}  m_{l}^2 \right\}} =   1 + (I-1) \Pr(\xi_i = \xi_j) .
\end{align*}
We are unaware of any proof of this result in the literature on Poisson-Dirichlet processes.

In the case of $\bar{\kappa} = \V( D_i | \bflambda, \bfvarsigma )$, we used similar arguments and the fact that $y_{ij}$ and $y_{ih}$ are conditionally independent to derive the second moment first,
\begin{align*}
\E \{ D^2_i | \bflambda, \bfvarsigma \} & = \E \left\{ \left(\sum_{j \neq i} y_{ij} \right)^2 \Bigg| \bflambda, \bfvarsigma \right\} \\
& = [I-1] \E \left\{ \E( y^2_{ij} | \xi_i, \xi_j, \bfTheta )  \right\}  + \frac{[I-1][I-2]}{2} \E \left\{ \E( y_{ij} | \xi_i, \xi_j, \bfTheta ) \E( y_{ih} | \xi_i, \xi_h, \bfTheta ) \right\}.  \\
%& = \rho + \frac{(I-1)(I-2)}{2} \E \left\{ \E( y_{ij} | \xi_i, \xi_j, \bfTheta ) \E( y_{ih} | \xi_i, \xi_h, \bfTheta ) \right\} 
\end{align*}
Now $\E \left\{ \E( y^2_{ij} | \xi_i, \xi_j, \bfTheta )  \right\} = \Pr(\xi_i = \xi_j) \E_{H^{\bflambda_D}} \{ \theta_{k,k} \} + \Pr(\xi_i \ne \xi_j) \E_{H^{\bflambda_O}} \{ \theta_{k,l} \}$ and
\begin{align*}
\E \left\{ \E( y_{ij} | \xi_i, \xi_j, \bfTheta ) \E( y_{ih} | \xi_i, \xi_h, \bfTheta ) \right\} &= \E\{ \theta_{k,k}^2 \} \Pr(\xi_i = \xi_j = \xi_h) + \E \{ \theta_{k,k}  \theta_{k,l} \}  \Pr(\xi_i = \xi_j \ne \xi_h) + \\
& \;\;\;\;\;\;\;\;  \E\{ \theta_{k,k}  \theta_{k,l} \}  \Pr(\xi_i = \xi_h \ne \xi_j) + \E \{ \theta_{k,l}^2 \}  \Pr(\xi_i \ne \xi_j = \xi_h) + \\
& \;\;\;\;\;\;\;\;  \left( \E\{ \theta_{k,l} \} \right)^2  \Pr(\xi_i \ne \xi_j \ne \xi_h).
\end{align*}
Again, we can use the P{\'o}lya urn representation of the Poisson Dirichlet process to show that $\Pr(\xi_i = \xi_j \ne \xi_h) = \Pr(\xi_i = \xi_h \ne \xi_j) = \Pr(\xi_i \ne \xi_j = \xi_h) = \frac{(1 - \alpha)(\beta + \alpha)}{(\beta + 1)(\beta + 2)}$, $\Pr(\xi_i \ne \xi_j \ne \xi_h) = \frac{(\beta + \alpha)(\beta + 2\alpha)}{(\beta + 1)(\beta + 2)}$ and $\Pr(\xi_i = \xi_j = \xi_h) = \frac{(1-\alpha)(2-\alpha)}{(\beta + 1)(\beta + 2)}$.

\smallskip
Finally, $\bar{\Delta}$ follows again from a simple conditioning argument.  Note that the conditional independence of $y_{ij}$ and $y_{ji}$ implies that 
\begin{align*}
\Delta^N = \Cov(y_{ij}, y_{ji}) &= \Cov \left\{ \E( y_{ij} | \xi_i, \xi_j, \bfTheta ), \E( y_{ji} | \xi_i, \xi_j, \bfTheta ) \right\} %\\ & = \Cov\left\{ \sum_{k=1}^{\infty}\sum_{l=1}^{\infty} w_k w_l \theta_{k,l} , \sum_{k=1}^{\infty}\sum_{l=1}^{\infty} w_k w_l \theta_{l,k} \right\}
\end{align*}
Now
\begin{align*}
\E \left\{ \E( y_{ij} | \xi_i, \xi_j, \bfTheta ) \E( y_{ji} | \xi_i, \xi_j, \bfTheta ) \right\} &= \E_{H^{\bflambda_D}}\{ \theta_{k,k}^2 \} \Pr(\xi_i = \xi_j) + \E_{H^{\bflambda_O}}\{ \theta_{k,l} \theta_{l,k} \} \Pr(\xi_i \ne \xi_j)
%& = \E_{H^{\bflambda_D}}\{ \theta_{k,k}^2 \} \E\left\{ w_1 \right\} + \left( \E_{H^{\bflambda_O}}\{ \theta_{k,l} \} \right)^2 \left( 1 - \E\left\{ w_1 \right\} \right)
%
%
%\E \left\{ \sum_{k=1}^{N}\sum_{l=1}^{N} w_l w_k \theta_{k,l}\theta_{l,k} \right\} \\
%& =  \E_{H^{\bflambda_O}}\{ \theta_{k,k}^2 \}  \left( \sum_{k=1}^{N} \E\{ w_k^2 \} \right) +  \E_{H_{*}^{\bflambda_J}}\{ \theta_{k,l}\theta_{l,k} \}  \left( 1 - \sum_{k=1}^{N} \E\{ w_k^2 \} \right) \\
%& =   \{ \V_{H^{\bflambda_O}}\{ \theta_{k,k}^2 \} + (\E_{H^{\bflambda_O}}\{ \theta_{k,k}^2 \})^2 \}   \left( \sum_{k=1}^{N} \E\{ w_k^2 \} \right)  \\ 
%& \;\;  +  ( \Cov_{H_{*}^{\bflambda_J}}\{ \theta_{k,l}, \theta_{l,k} \} + \E_{H_{*}^{\bflambda_J}} \{ \theta_{k,l}  \} \E_{H_{*}^{\bflambda_J}} \{ \theta_{l,k}  \} )   \left( 1 - \sum_{k=1}^{N} \E\{ w_k^2 \} \right) \\
\end{align*}
and, %because of the symmetry of $H_{*}^{\bflambda_J}$,
\begin{align*}
\E \left\{ \E( y_{ij} | \xi_i, \xi_j, \bfTheta ) \right\} &= \E \left\{ \E( y_{ji} | \xi_i, \xi_j, \bfTheta ) \right\} = \E_{H^{\bflambda_D}}\{ \theta_{k,k} \} \Pr(\xi_i = \xi_j) + \E_{H^{\bflambda_O}}\{ \theta_{k,l} \} \Pr(\xi_i \ne \xi_j).
\end{align*}
Hence,
\begin{align*}
\Delta^N &= \E \left\{ \E( y_{ij} | \xi_i, \xi_j, \bfTheta ) \E( y_{ji} | \xi_i, \xi_j, \bfTheta ) \right\} - \E \left\{ \E( y_{ij} | \xi_i, \xi_j, \bfTheta )\right\}   \E \left\{ \E( y_{ji} | \xi_i, \xi_j, \bfTheta ) \right\} \\
& =  \Pr(\xi_i = \xi_j) \E_{H^{\bflambda_D}}\{ \theta^2_{k,k} \} +\Pr(\xi_i \ne \xi_j) \left( \E_{H^{\bflambda_O}}\{ \theta_{k,l} \} \right)^2 \\
 & \;\;\;\;\;\;\;\;\;\;\;\;\;\;\;\;\;\;\;\;\;\;\;\;\;\;\;\;\;\;\;\;\;\;\;\;\;\;\;\;\;\;\;\;\;\;\;\;\;\;- \left[  \Pr(\xi_i = \xi_j) \E_{H^{\bflambda_D}}\{ \theta_{k,k} \} +\Pr(\xi_i \ne \xi_j) \E_{H^{\bflambda_O}}\{\theta_{k,l}\}  \right]^2.
\end{align*}
Note that $ \E \left\{ \E( y^2_{ij} | \xi_i, \xi_j, \bfTheta ) \right\} = \E \left\{ \E( y^2_{ji} | \xi_i, \xi_j, \bfTheta ) \right\}  $, therefore $\V(y_{ij} | \bflambda, \bfvarsigma) = \V(y_{ji} | \bflambda, \bfvarsigma)$ and
\begin{align*}
\Delta^D & =  \V(y_{ij} | \bflambda, \bfvarsigma) \\
		 & = \E \left\{ \E( y^2_{ij} | \xi_i, \xi_j, \bfTheta ) \right\} - \left[ \E \left\{ \E( y_{ij} | \xi_i, \xi_j, \bfTheta ) \right\} \right]^2 \\
& =  \Pr(\xi_i = \xi_j) \E_{H^{\bflambda_D}}\{ \theta_{k,k} \} + \Pr(\xi_i \ne \xi_j) \E_{H^{\bflambda_O}}\{ \theta_{k,l} \} \\
 & \;\;\;\;\;\;\;\;\;\;\;\;\;\;\;\;\;\;\;\;\;\;\;\;\;\;\;\;\;\;\;\;\;\;\;\;\;\;\;\;\;\;\;\;\;\;\;\;\;\;\;\;\;\;\;\; - \left[  \Pr(\xi_i = \xi_j) \E_{H^{\bflambda_D}}\{ \theta_{k,k} \} + \Pr(\xi_i \ne \xi_j) \E_{H^{\bflambda_O}}\{\theta_{k,l}\}  \right]^2 \\
 & = \bar{\theta}\{ 1 - \bar{\theta} \}.
\end{align*}

\section{Moment generating function for $D_i$}\label{ap:momgenfuncDi}

Since observations are exchangeable we compute the moment generating function for $D_1$, without lost of generality assume that subject 1 is in the first community. Hence,
\begin{align*}
\E \left( \exp\left\{ t D_1 \right\} \right) &=  \E \left( \exp\left\{ t \sum_{j \ne 1} y_{1,j} \right\} \right)  \\
& =  \E \left(  \E \left[ 
\exp\left\{ t \sum_{j \ne 1} y_{1,j} \right\}
\Bigg|  \bfTheta, \xi_1, \ldots, \xi_I \right] \right) \\
& = \E \left( \E \left[ \left\{ \theta_{1,1}(\exp\left\{ t \right\} - 1 ) + 1 \right\}^{m_1 - 1} \prod_{k=2}^{K}
\left\{ \theta_{1,k}(\exp\left\{ t \right\} - 1 ) + 1 \right\}^{m_k}
\bigg|  \bfTheta
\right] \right) \\
&= \E \left\{ 
\left[ 1 + \sum_{s=1}^{m_1 -1} {m_1 -1 \choose s} \E_{H^{\bflambda_{D}}} (\theta_{1,1}^s) (\exp\{t\}-1)^s \right] \right. \\
& \qquad \qquad \times \left. \prod_{k=2}^{K} \left[ 1 + \sum_{s=1}^{m_k} {m_k \choose s} \E_{H^{\bflambda_{O}}} (\theta_{1,k}^s) (\exp\{t\}-1)^s \right]
 \right\}
%
%
%K, m_1, \cdot m_K 
\end{align*}

Note that, even though there is no close-form solution for the outward expectation, the expression can still be used to compute the moments of $D_i$ by relating its moments to those of $m_1, \ldots, m_K$.  Indeed, taking the first derivative with respect to $t$ and evaluating at $t=0$ we get
$$
\E(D_1 | \bflambda, \bfvarsigma) =  I \E_{H^{\bflambda_O}} \{ \theta_{1,l} \} -  \E_{H^{\bflambda_D}} \{ \theta_{1,1} \} + \left( \E_{H^{\bflambda_D}} \{ \theta_{1,1} \} -  \E_{H^{\bflambda_O}} \{ \theta_{1,l} \} \right)  \E \{ m_{1} \},
%\left(\E\{m_1\} -1\right) \E_{H^{\bflambda_D}}\{ \theta_{1,1} \} + \left( E_{H^{\bflambda_D}}\{ \theta_{1,1} \} - E_{H^{\bflambda_D}}\{ \theta_{1,l} \} \right) \E\{m_1\}
$$
as discussed in Appendix \ref{ap:expDi}.

\section{Proof of lemma \ref{le:transitivity}}\label{ap:transitivity}

Note that 
\begin{align*}
\Pr(y_{ij}=1, y_{ih}=1, y_{jh}=1 | \bflambda, \bfvarsigma) &= \E \left\{  \Pr(y_{ij}=1, y_{ih}=1, y_{jh}=1 |  \bfTheta, \xi_i, \xi_j, \xi_h, \bflambda, \bfvarsigma)  \right\} \\
&= \E\{ \theta_{k,k}^3 \} \Pr(\xi_i = \xi_j = \xi_h) + \E \{ \theta_{k,k}  \theta_{k,l}^2 \}  \Pr(\xi_i = \xi_j \ne \xi_h) + \\
& \;\;\;\;\;\;\;\;  \E\{ \theta_{k,k}  \theta_{k,l}^2 \}  \Pr(\xi_j = \xi_h \ne \xi_i) +
  \E\{ \theta_{k,k} \theta_{k,l}^2 \}  \Pr(\xi_i = \xi_h \ne \xi_j) +  \\ 
& \;\;\;\;\;\;\;\;\;\;\;    \left( \E\{ \theta_{k,l} \} \right)^3  \Pr(\xi_i \ne \xi_j \ne \xi_h).
\end{align*}
Again, because of the exchangeability, we can use the P{\'o}lya urn representation of the process to obtain $\Pr(\xi_i = \xi_j \ne \xi_h) = \Pr(\xi_j = \xi_h \ne \xi_i) =  \Pr(\xi_i = \xi_h \ne \xi_j) = \frac{(1-\alpha)(\beta + \alpha)}{(\beta+1)(\beta+2)}$, $\Pr(\xi_i = \xi_j = \xi_h) = \frac{(1-\alpha)(2-\alpha)}{(\beta+1)(\beta+2)}$ and $\Pr(\xi_i \ne \xi_j \ne \xi_h) = \frac{(\beta + \alpha)(\beta + 2\alpha)}{(\beta+1)(\beta+2)}$.

\section{Details on the MCMC sampler}\label{se:MCMCdetails}

%%%%%%%%%
% SM Algorithm %
%%%%%%%%%	
\subsection{Split-Merge MCMC Algorithm}\label{sec:SM}

    \begin{enumerate}
       \item[1.] Uniformly at random select two networks $a$ and $b$. 
       \item[2A.] If $\bfzeta_a = \bfzeta_b$ propose a SPLIT move:
          \begin{enumerate}
             \item[2A.1] Let $S.ab = \{j : \bfzeta_j = \bfzeta_a = \bfzeta_b \}$, $S.a = \{a\}$ and $S.b = \{ b \}$. \\ Assign the rest of $S.ab$'s components to either $S.a$ or $S.b$ at random with equal probability.
             \item[2A.2] Generate $\xi_a$ and $\xi_b$ through a modified Poyla urn scheme. 
             \item[2A.3] For $\xi_{ab}$ use the actual value of $\xi_{\bfzeta_a}$ 
             \item[2A.4] Run a Gibbs sampler to update  $\xi_a$ , $\xi_b$ and $\xi_{ab}$.
          \end{enumerate} 
          \item[2B.] Otherwise, if $\bfzeta_a \neq \bfzeta_b$ propose a MERGE move: 
          \begin{enumerate}
             \item[2B.1] Let $S.a = \{j : \bfzeta_j = \bfzeta_a \}$, $S.b = \{j : \bfzeta_j = \bfzeta_b \}$ and $S.ab = S.a \cup S.b$
             \item[2B.2] Generate $\xi_{ab}$ through a modified Poyla urn scheme. 
             \item[2B.3] For $\xi_{a}$ and $\xi_{b}$ use the actual value of $\xi_{\bfzeta_a}$ and $\xi_{\bfzeta_b}$, respectively.
             \item[2B.4] Run a Gibbs sampler to update  $\xi_a$ , $\xi_b$ and $\xi_{ab}$.
          \end{enumerate}           
      \item[3.] Evaluate the proposal using a Metropolis-Hastings acceptance ratio. If the proposal is accepted, then $\bfzeta$ and $\bfxi$ change.
      \item[4.] Each $\bfxi_r$ is updated using a regular Gibbs sampler, regardless of the result of the M-H evaluation. 
 \end{enumerate}

%%%%%%%%%
% Initialize bfxi %
%%%%%%%%%
\subsection{Modified Poyla urn scheme}\label{sec:Ini.xi}

Initialize each $\bfxi_r$, $r=1,\ldots,R$ as follows, let $\bfpi$ be a random permutation of $\{1,\ldots,I \}$, set $\xi_{r,\pi_1} = 1$, next $\xi_{r,\pi_2} = 1$ with probability $(1 - \alpha_{2,r})/(\beta_{2,r} + 1)$, $\xi_{r,\pi_2} = 2$ otherwise. Then, for $h=3,\ldots,I$, $\xi_{r,\pi_h} = k$ with probability 
\begin{multline*} 
	p\left( \xi_{r,\pi_h} = k ~\Big|~ \{ \bfY_{j}^{(h)} : \zeta_j = r \} \right) \\
 	= \begin{cases}
			\frac{m_{r,k} - \alpha_r}{\beta_{r} + h -1} \prod_{ \{ j:\zeta_j = r \} } p \left( \bfy_{\pi_h,j}^{(h-1)} ~\Big|~ \{ \bfy_{l,j}^{(h-1)} : l \in \bfpi^{(h-1)}, \xi_{r,l} = k \} \right) & k \leq L_{r}^{(h-1)}   \\
			\frac{\beta_r + \alpha_{r} L_{r}^{(h-1)} }{\beta_{r} + h -1} \prod_{ \{ j:\zeta_j =r \} } p \left( \bfy_{\pi_h,j}^{(h-1)} \right) & k = L_{r}^{(h-1)} + 1   
		\end{cases},
\end{multline*}
		where $\bfY_{j}^{(h)} = \{y_{i,i',j} : i,i' \in \bfpi^{(h)} \}$, $\bfy_{l,j}^{(h-1)} = \{ y_{i,i',j} : (i,i') \in \{ i=l, i' \in \bfpi^{(h-1)} \} \cup \{ i \in \bfpi^{(h-1)}, i'=l \} \}$, with $\bfpi^{(h)} = \{ \pi_1, \ldots, \pi_{h} \} $, $L_{r}^{(h-1)}$ is the total number of groups and $m_{r,k}^{(h-1)}$ is the number of actors in the $k$-th group after assigning the first $h-1$ actors. There,
\begin{multline*} 
	p \left( \bfy_{\pi_h,j}^{(h-1)} ~\Big|~ \{ \bfy_{l,j}^{(h-1)} : l \in \bfpi^{(h-1)}, \xi_{r,l} = k \} \right) \\
	 = \begin{cases}
  p_{\theta} \left( \bfy_{(\pi_h),k,j}^{(h-1)} ~\Big|~ \bfy_{k,k,j}^{(h-1)} \right) \displaystyle{ \prod_{l=1,l \neq k}^{L_{r}^{(h-1)}} } p_{\theta} \left( \bfy_{(\pi_h),l,j}^{(h-1)} ~\Big|~ \bfy_{k,l,j}^{(h-1)} \right) p_{\theta} \left( \bfy_{l,(\pi_h),j}^{(h-1)} ~\Big|~ \bfy_{l,k,j}^{(h-1)} \right) & \mbox{if $j$ is directed} \\
 p_{\theta} \left( \bfy_{(\pi_h),k,j}^{(h-1)}, \bfy_{k,(\pi_h),j}^{(h-1)} ~\Big|~ \bfy_{k,k,j}^{(h-1)} \right) \displaystyle{ \prod_{l=1,l \neq k}^{L_{r}^{(h-1)}} } p_{\theta} \left( \bfy_{(\pi_h),l,j}^{(h-1)} , \bfy_{l,(\pi_h),j}^{(h-1)} ~\Big|~ \bfy_{k,l,j}^{(h-1)} , \bfy_{l,k,j}^{(h-1)} \right) & \mbox{if $j$ is undirected}
 \end{cases}, 
\end{multline*} 
	where $ \bfy_{(\pi_h),l,j}^{(h-1)} = \{ y_{\pi_h,i',j} : i' \in \bfpi^{(h-1)}; \xi_{r,i'}^{(h-1)} = l \}$, $ \bfy_{l,(\pi_h),j}^{(h-1)} = \{ y_{i,\pi_h,j} : i \in \bfpi^{(h-1)}; \xi_{r,i}^{(h-1)} = l \}$, $ \bfy_{k,l,j}^{(h-1)} = \{ y_{i,i',j} : i,i' \in \bfpi^{(h-1)}; \xi_{r,i}^{(h-1)} = k; \xi_{r,i'}^{(h-1)} = l \}$ and $p_{\theta} ( \cdot \mid \cdot )$ can be found using $p_{\theta}(\cdot)$ the marginal posterior of $\bfy$ under a given prior for $\theta$. Moreover, 
	\begin{align*}
		p \left( \bfy_{\pi_h,j}^{(h-1)} \right) &= 
		\begin{cases}
			\prod_{l=1}^{L_{r}^{(h-1)}} p_{\theta} \left( \bfy_{(\pi_h),l,j}^{(h-1)} \right) p_{\theta} \left( \bfy_{l,(\pi_h),j}^{(h-1)} \right) & \mbox{if } j \mbox{ is directed} \\
			\prod_{l=1}^{L_{r}^{(h-1)}} p_{\theta} \left( \bfy_{(\pi_h),l,j}^{(h-1)} , \bfy_{l,(\pi_h),j}^{(h-1)} \right) & \mbox{if } j \mbox{ is undirected} 
		\end{cases}.
	\end{align*}
	Let $q_r$ be the product of the probabilities of the assignment made at each step. Hence, $q_r$ is the probability of obtaining $\bfxi_r$, which will be needed for the Metropolis-Hastings ratio.
	
%%%%%%%%%
% MH Ratio      %
%%%%%%%%%
\subsection{Metropolis-Hastings ratio}\label{sec:MHratio}

 For $\bfeta= (\bfzeta, \bfxi)$ if $\bfeta^*$ is a SPLIT move, accept a SPLIT move with probability
	\begin{multline*}
	a \left( \bfeta^* \mid \bfeta \right) \\
	 = min \left\{ \frac{ 2^{n_{ab} - 2} (\beta_1 + R\alpha_1)}{ \Gamma(1-\alpha_1)} \frac{\Gamma(n_a - \alpha_1) \Gamma(n_b - \alpha_1)}  { \Gamma (n_{ab} - \alpha_1) }  \frac{ p \left( \bfY,  \bfxi_a; S.a \right) p \left(\bfY,  \bfxi_b; S.b \right) }{ p \left(\bfY \bfxi_{ab}; S.ab\right)} \frac{Q.ab}{Q.a Q.b} , 1 \right\},
	\end{multline*}
	where $R=$ the total number of classes before the split, $n_r = |S.r |$ and  
	\[ p \left(\bfY, \bfxi_r ; S.r\right) =  \prod_{j \in S.r} p \left( \bfY_j \mid \bfxi_r \right) 
p(\bfxi_r | \alpha_{2,r}, \beta_{2,r}),\] 
	with $p(\bfxi_r | \alpha_{2,r}, \beta_{2,r})$ given in equation (\ref{eq:XIdist}) and 
	\begin{eqnarray*}
		p \left( \bfY_j \mid \bfxi_r \right) & = & 
		\begin{cases}
			\prod_{l=1}^{L_{r}} p_{\theta} \left( \bfy_{l,l,j} \right) \prod_{l' > l}^{L_{r}} p_{\theta} \left( \bfy_{l,l',j} \right) p_{\theta} \left( \bfy_{l',l,j} \right) & \mbox{if } j \mbox{ is directed} \\
			\prod_{l=1}^{L_{r}} p_{\theta} \left( \bfy_{l,l,j} \right) \prod_{l' > l}^{L_{r}} p_{\theta} \left( \bfy_{l,l',j}, \bfy_{l',l,j} \right)  & \mbox{if } j \mbox{ is undirected} 
		\end{cases},
	\end{eqnarray*}
	where $ \bfy_{l,l',j} = \{ y_{i,i',j} : \xi_{r,i} = l; \xi_{r,i'} = l' \}$ and $p_{\theta}(\cdot)$ the marginal posterior of $\bfy$ under a given prior for $\theta$.  	
	
	$Q.r = p(\alpha_{2,r}) q(\xi_r^{(0)}) q(\xi_r^{(1)}) $ where $p(\alpha_{2,r})$ is the prior on $\alpha_{2,r}$, and $q(\xi_r^{(0)})$ and $q(\xi_r^{(1)})$ are the probabilities of obtaining $\xi_r^{(0)}$ and $\xi_r^{(1)}$ respectively, using the method described in section~\ref{sec:Ini.xi}. 

{\small
\bibliographystyle{bka}
\bibliography{reftesis}
}
\end{document}